\documentclass[a4paper,11pt]{article}
\pdfoutput=1
\usepackage{jheppub} 

\usepackage{latexsym}
\usepackage{amsthm}
\usepackage{amsmath}
\usepackage{graphicx}
\usepackage{mathtools}
\usepackage{slashed}
\usepackage{amssymb}
\usepackage{float}
\usepackage{subcaption}
\usepackage{color}
\definecolor{mycolor}{RGB}{0,30,100}
\newcommand*\diff{\mathop{}\!\mathrm{d}}
\newcommand*\Diff[1]{\mathop{}\!\mathrm{d^#1}}
\newcommand{\nn}{\nonumber}

\newcommand{\be}{\begin{eqnarray}}
\newcommand{\ee}{\end{eqnarray}}
\newcommand{\ma}{\mathrm}
\newcommand{\ml}{\mathcal}
\newcommand{\bs}{\boldsymbol}

\newcommand{\trento}{T\raisebox{-0.5ex}{R}ENTo}

\title{Coupled Boltzmann Transport Equations of Heavy Quarks and Quarkonia in Quark-Gluon Plasma}

\author[a,b]{Xiaojun Yao}
\author[c,b]{Weiyao Ke,}
\author[b]{Yingru Xu,}
\author[b]{Steffen A. Bass,}
\author[b]{Berndt M\"uller}

\affiliation[a]{Center for Theoretical Physics, Massachusetts Institute of Technology, Cambridge, MA, 02139, USA}
\affiliation[b]{Department of Physics, Duke University, Durham, NC 27708, USA}
\affiliation[c]{Nuclear Science Division, Lawrence Berkeley National Laboratory, Berkeley, CA 94720, USA}

\emailAdd{xjyao@mit.edu}
\emailAdd{weiyaoke@lbl.gov}
\emailAdd{yx59@phy.duke.edu}
\emailAdd{bass@duke.edu}
\emailAdd{mueller@phy.duke.edu}

\preprint{MIT-CTP/5192}

\abstract{We develop a framework of coupled transport equations for open heavy flavor and quarkonium states, in order to describe their transport inside the quark-gluon plasma. Our framework is capable of studying simultaneously both open and hidden heavy flavor observables in heavy-ion collision experiments and can account for both, uncorrelated and correlated recombination. Our recombination implementation depends on real-time open heavy quark and antiquark distributions. We carry out consistency tests to show how the interplay among open heavy flavor transport, quarkonium dissociation and recombination drives the system to  equilibrium. We then apply our framework to study bottomonium production in heavy-ion collisions. We include $\Upsilon(1S)$, $\Upsilon(2S)$, $\Upsilon(3S)$, $\chi_b(1P)$ and $\chi_b(2P)$ in the framework and take feed-down contributions during the hadronic gas stage into account. Cold nuclear matter effects are included by using  nuclear parton distribution functions for the initial primordial heavy flavor production. A calibrated $2+1$ dimensional viscous hydrodynamics is used to describe the bulk QCD medium. We calculate both the nuclear modification factor $R_{\ma{AA}}$ of all bottomonia states and the azimuthal angular anisotropy coefficient $v_2$ of the $\Upsilon(1S)$ state and find that our results agree reasonably with experimental measurements. Our calculations indicate that correlated cross-talk recombination is an important production mechanism of bottomonium in current heavy-ion experiments. The importance of correlated recombination can be tested experimentally by measuring the ratio of $R_{\ma{AA}}(\chi_b(1P))$ and $R_{\ma{AA}}(\Upsilon(2S))$.}

\begin{document}
\maketitle
\flushbottom

\section{Introduction}
\label{sect:intro}
Heavy quarkonia are bound states of heavy quark-antiquark pairs $Q\bar{Q}$. The mass spectra of the ground and lower excited quarkonium states can be reasonably well described by the nonrelativistic Schr\"odinger equation with a potential model \cite{Quigg:1977dd}. Inside a hot nuclear medium, i.e., the quark-gluon plasma (QGP), the attractive potential can be significantly suppressed due to the  static screening effect in the plasma \cite{Matsui:1986dk}. As a result, a bound $Q\bar{Q}$ can ``melt" at sufficiently high temperature \cite{Karsch:1987pv,McLerran:1981pb,Mocsy:2007jz}. Therefore, quarkonium suppression (compared to a non QGP baseline) can be used as a signal of the formation of a QGP in heavy-ion collisions. In general, shallower bound states melt at lower temperatures and one would expect a ``sequential" suppression pattern.

However, this simple picture is complicated by other in-medium processes such as the dissociation of quarkonium states via dynamical scattering (i.e. the dynamical screening effect, which is related to the imaginary part of the $Q\bar{Q}$  potential \cite{Laine:2006ns,Beraudo:2007ky}) and heavy quark (re)combination \cite{Thews:2000rj,Andronic:2007bi}. To account for all these effects, semi-classical transport equations have been widely applied  \cite{ Grandchamp:2003uw,Grandchamp:2005yw,Yan:2006ve,Zhao:2007hh,Liu:2009nb,Zhao:2010nk,Song:2011xi,Song:2011nu,Emerick:2011xu,Sharma:2012dy,Nendzig:2014qka,Krouppa:2015yoa,Chen:2017duy,Zhao:2017yan,Du:2017qkv,Aronson:2017ymv,Ferreiro:2018wbd,Yao:2018zrg,Hong:2019ade,Chen:2019qzx,Du:2018wsj,Du:2019tjf}. These calculations typically contain three components: The first part is a temperature-dependent potential that is parametrized from lattice calculations of the free energy of a $Q\bar{Q}$ singlet \cite{Kaczmarek:2002mc,Bazavov:2018wmo} or direct lattice calculations of the real part of the $Q\bar{Q}$ potential \cite{Burnier:2014ssa}. The second input is the dissociation rate of each quarkonium state. Perturbative calculations of dissociation rates include both the gluo-dissociation process \cite{Peskin:1979va,Bhanot:1979vb} and the inelastic scattering with the medium (Landau damping). An effective field theory of QCD, potential nonrelativistic QCD (pNRQCD) \cite{Brambilla:1999xf,Brambilla:2004jw,Fleming:2005pd} has been applied to study both the static screening of the potential and the dissociation rate \cite{Brambilla:2008cx,Brambilla:2011sg,Brambilla:2013dpa}. Similar effective theory has also been used to study dark matter bound state formation \cite{Biondini:2018ovz,Biondini:2019int,Binder:2020efn}. Some studies also included viscous effects \cite{Dumitru:2007hy,Dumitru:2009fy,Du:2016wdx} and modifications due to the quarkonium moving with respect to a static medium \cite{Liu:2006nn,Escobedo:2011ie}. The final component is a recombination model, for example, a statistical hadronization model \cite{Andronic:2007bi}, a coalescence model based on Wigner functions \cite{Chen:2017duy} or a model that is based on detailed balance and a finite relaxation rate \cite{Du:2017qkv}. 

Unlike the first two effects, recombination contains significant model-dependencies in most studies. Here, we shall distinguish between two kinds of recombination: uncorrelated and correlated recombination. 
In uncorrelated recombination, the $Q$ and $\bar{Q}$ originate from differential initial hard vertices. In proton-proton collisions, these heavy quarks and antiquarks would almost never (re)combine to form a quarkonium state due to their separation in phase space. However, in heavy-ion collisions, multiple $Q\bar{Q}$ pairs are produced from the initial hard scatterings in one collision event. The momenta of these heavy quarks and antiquarks change continuously inside the QGP due to diffusion and energy loss. Therefore, the chance for an uncorrelated pair to come close to each other in phase space is higher. When they are close, they may combine to form a quarkonium state. The uncorrelated recombination rate rises with the number of open heavy quarks produced in the collision, which is crucial to explain the small amount of suppression observed for the $J/\psi$ with rising collision energy. Naively, one would expect $J/\psi$ to be more suppressed at higher collision energies due to the hotter medium and stronger plasma screening effects. Therefore, uncorrelated recombination is important for the phenomenology of charmonium. However, its effect on bottomonium production is expected to be negligible, since only a few bottom-antibottom quark pairs are produced in one collision.

In correlated recombination, the $Q$ and $\bar{Q}$ originate from the same initial hard vertex. For example, a $Q\bar{Q}$ pair in a pre-quarkonium resonance or emerging from a previous dissociation of a quarkonium state is considered a correlated pair. The phenomenological effect of correlated recombination for both charmonium and bottomonium has not been systematically explored yet. 

Most recombination models depend on the uncorrelated open heavy quark distributions, they can thus only account for uncorrelated recombination. To explore the physical effects of correlated recombination, we develop a set of coupled transport equations for $Q$ and $\bar{Q}$'s as well as for quarkonia, which allows us to study both, uncorrelated and correlated recombination. The transport equation of quarkonium provides consistent treatment of dissociation and recombination, in the sense that both of them are derived from QCD under systematic nonrelativistic expansions \cite{Yao:2018nmy,Yao:2020kqy}. The derivation is based on the combination of the open quantum system framework (in which correlated recombination is taken into account) and the pNRQCD effective field theory. The application of the open quantum system framework to studying quarkonium in-medium dynamics \cite{Young:2010jq,Borghini:2011ms,Akamatsu:2011se,Akamatsu:2014qsa,Blaizot:2015hya,Katz:2015qja,Kajimoto:2017rel,DeBoni:2017ocl,Blaizot:2017ypk,Blaizot:2018oev,Akamatsu:2018xim,Miura:2019ssi,Sharma:2019xum} and its combination with pNRQCD \cite{Brambilla:2016wgg,Brambilla:2017zei} has recently drawn significant theoretical interest. A quarkonium transport coefficient has been defined in Ref.~\cite{Brambilla:2019tpt}. In the open quantum system approach, quarkonium dissociation is caused by the wavefunction decoherence, which also leads to correlated recombination at the same time. For example, if we start with the $1S$ state: $|\psi(t=0)\rangle = |1S\rangle$, the wavefunction decoherence leads to $|\langle 1S|\psi(t)\rangle|^2<1$ for $t>0$. But at the same time, if the $2S$ state exists as a well-defined bound state, we will also have $|\langle 2S|\psi(t)\rangle|^2>0$, i.e., some $2S$ state is regenerated. Since our transport equation for quarkonium is derived from the open quantum system approach, it can handle correlated recombination.

The coupled transport equations allow us to study the in-medium transport of both open and hidden heavy flavor states. In this paper, we will use this framework to study bottomonium production in heavy-ion collisions. We will include $\Upsilon(1S)$, $\Upsilon(2S)$, $\Upsilon(3S)$, $\chi_b(1P)$ and $\chi_b(2P)$ states in the transport network. By solving the coupled transport equations via test particle Monte Carlo simulations, we will explore the importance of correlated recombination in bottomonium phenomenology. This paper is organized as follows: in Sect.~\ref{sect:transport}, we will introduce the set of coupled transport equations. Then in Sect.~\ref{sect:test}, some simulation tests inside a QGP box will be shown and compared to the system properties in thermal equilibrium. Details on applying the transport equations to the study of heavy-ion collisions will be explained in Sect.~\ref{sect:real}. Results on the nuclear modification factor ($R_\ma{AA}$) and the azimuthal angular anisotropy coefficient $v_2$ of bottomonia will be discussed later in Sect.~\ref{sect:results}. Finally, we will draw conclusions in Sect.~\ref{sect:conclusions}.

\section{Coupled Transport Equations}
\label{sect:transport}

The set of coupled Boltzmann transport equations for the distribution functions of unbound heavy quark-antiquark pairs $Q\bar{Q}$ and each quarkonium state with the quantum number $nls$ ($n$ is for the radial excitation, $l$ the orbital angular momentum and $s$ the spin) is given by
\be
\label{eq:LBE}
(\frac{\partial}{\partial t} + \dot{{\bs x}} _Q\cdot \nabla_{{\bs x}_Q} + \dot{{\bs x}} _{\bar{Q}}\cdot \nabla_{{\bs x}_{\bar{Q}}} ) f_{Q\bar{Q}}({\bs x}_Q, {\bs p}_Q, {\bs x}_{\bar{Q}}, {\bs p}_{\bar{Q}}, t) &=& \ml{C}_{Q\bar{Q}}  -  \ml{C}_{Q\bar{Q}}^{+} +  \ml{C}_{Q\bar{Q}}^{-}\\
(\frac{\partial}{\partial t} + \dot{{\bs x}}\cdot \nabla_{\bs x})f_{nls}({\bs x}, {\bs p}, t) &=& \ml{C}_{nls}^{+}-\ml{C}_{nls}^{-}\,,
\ee
where $\dot{{\bs x}} = \frac{\partial \bs x}{\partial t}$.
The left-hand sides of these equations describe the free streaming of distribution functions in phase space while the right-hand sides contain collision terms of the heavy particles interacting with the plasma. The collision terms with $\pm$ superscripts represent the quarkonium dissociation ($-$) and recombination ($+$) while the term without any superscript describes the energy and momentum changes of the open heavy quark-antiquark pairs. In the following, we will explain these collision terms in detail.

\subsection{Transport of Open Heavy Quark-Antiquark Pairs}
If we neglect the interaction between the heavy quark-antiquark pair, we can write
\be
\label{eqn:collision_HQ}
\ml{C}_{Q\bar{Q}} = \ml{C}_{Q} + \ml{C}_{\bar{Q}}\,,
\ee
i.e., the heavy quark and antiquark interact independently with the medium. In potential models, the interaction between the $Q\bar{Q}$ pair is attractive for the color singlet and repulsive for the color octet. For a Coulomb potential, the color-averaged potential vanishes if we assume the color is in thermal equilibrium. Furthermore, in-medium $Q\bar{Q}$ potentials are significantly suppressed, so we expect the potential interaction between the $Q\bar{Q}$ pair to be weak. Therefore, throughout this paper, we will neglect the interaction between the $Q\bar{Q}$ pair in the open heavy flavor transport equations. As a remark, we note that a factorized $Q\bar{Q}$ distribution $f_{Q\bar{Q}}({\bs x}_Q, {\bs p}_Q, {\bs x}_{\bar{Q}}, {\bs p}_{\bar{Q}}, t) = f_{Q}({\bs x}_Q, {\bs p}_Q, t) f_{\bar{Q}}({\bs x}_{\bar{Q}}, {\bs p}_{\bar{Q}}, t)$
indeed leads to Eq.~(\ref{eqn:collision_HQ}). But the opposite is not true in general. Most recombination models implicitly assume the factorization of the $Q\bar{Q}$ distributions and thus cannot study the correlated recombination. But here we only assume Eq.~(\ref{eqn:collision_HQ}) in this work.

We will use a weak coupling picture for the transport of open heavy quarks \cite{ Gossiaux:2008jv,Gossiaux:2009mk,Uphoff:2014hza,Cao:2016gvr}. The interaction of open heavy quarks (and antiquarks) with the medium is described by scattering between open heavy quarks and medium partons (which include both light (anti)quarks and gluons, abbreviated as $q$ and $g$ respectively). The collision term $\ml{C}_Q$ includes three types of scattering processes: the elastic ${2\rightarrow2}$ scattering $q+Q\rightarrow q+Q$ and $g+Q\rightarrow g+Q$, the inelastic ${2\rightarrow3}$ scattering $q+Q\rightarrow q+Q+g$ and $g+Q\rightarrow g+Q+g$ and the inelastic ${3\rightarrow2}$ scattering $q+Q+g\rightarrow q+Q$ and $g+Q+g\rightarrow g+Q$, and similarly for $\ml{C}_{\bar{Q}}$. In this work, we will use the Monte Carlo simulations in the Lido package \cite{Ke:2018tsh} to solve the transport equations of open heavy quark-antiquark pairs. The Lido package contains both a linearized Boltzmann transport description and a model that is based on the Langevin equation with radiation corrections. The latter description has been reported in Ref.~\cite{Xu:2017obm}. We will only use the linearized Boltzmann description in this work.

\subsection{Transport of Quarkonia}
For the dissociation and recombination terms in the transport equations, we will use the expressions in Ref.~\cite{Yao:2018sgn}. Detailed expressions of the relevant collision terms can be found in Appendix~\ref{app:rates}. The calculations therein are based on a version of pNRQCD under the hierarchy of scales $M \gg Mv \gg Mv^2 \gtrsim T \gtrsim m_D$. Here $M$ is the heavy quark mass, $v$ the typical relative velocity between the $Q\bar{Q}$ pair inside the bound state, $T$ the temperature of the QGP and $m_D$ the Debye mass. The typical size of quarkonium is roughly give by $r\sim \frac{1}{Mv}$ and the typical binding energy is about $Mv^2$. The last inequality $T \gtrsim m_D$ means the QGP is weakly-coupled. For charmonium, we have $v^2\sim 0.3$ while for bottomonium, $v^2\sim 0.1$ \cite{Bodwin:1994jh}. They both give $Mv^2\sim 500$ MeV.

One may worry that the hierarchy is not always true in real heavy-ion collisions. In the early time of the QGP expansion, the temperature can be $ \gtrsim 450 $ MeV. Due to the static screening effect, the in-medium binding energies of quarkonium states can be much smaller than their vacuum binding energies, especially for excited states. Thus, our assumed hierarchy indeed breaks down in the early stage and in principle one has to use a different version of pNRQCD if there still exists a hierarchy of scales. However, even before the breakdown of the assumed hierarchy, as the temperature increases, the dissociation rates of excited quarkonium states such as $\Upsilon$(2S) and $\chi_b$(1P) blow up rapidly. What happens in our calculations is that after a very short time period in the early stage, the excited quarkonium states have dissociated and evolve then as an unbound, correlated $Q\bar{Q}$ pair. This is the right physics: When our hierarchy of scales breaks down, we can either have $ T\gtrsim Mv  \gg Mv^2 $ or $Mv \gg T \gg Mv^2$. We do not consider $Mv \gg T \gg Mv^2$ here because the real values of $v$ do not allow this hierarchy to happen for both charmonium and bottomonium. (For $\gg$ to be valid, one at least needs a factor of three in the ratio.) So we only need to consider $ T\gtrsim Mv  \gg Mv^2 $ here. Whenever this occurs, one expects both the plasma screening effects to be extremely strong ($\frac{1}{Mv}$ gives the rough size of the quarkonium state) and no bound states can be well-defined. In this case, quarkonium, even if it binds, binds very weakly and it is reasonable to assume that it behaves more like an unbound $Q\bar{Q}$ pair. Then the correct description is the transport of an unbound $Q\bar{Q}$ pair. Therefore, in our calculations, whenever the temperature is high enough to break our assumed hierarchy, the quarkonium state dissociates after a tiny time step and then evolves as an unbound, correlated $Q\bar{Q}$ pair. This is a good approximation of the correct physics. The transport of excited quarkonia as well-defined bound states is valid again in the later stage of the evolution, when the temperature drops and our hierarchy is resumed. In this sense, most excited quarkonium states are probably generated via recombination in the later stage of the QGP evolution. Their suppression mechanism is mainly the decorrelation of the $Q\bar{Q}$ pair in coordinate and momentum space (or decoherence of their wavefunction). Those excited states that are observed may probably come from recombination of $Q\bar{Q}$ pairs that are still correlated. We will discuss this in more detail in Sect.~\ref{sect:results}.

For the current calculations, we work to the leading order in the nonrelativistic expansion (the multipole expansion under this hierarchy of scales is equivalent to a nonrelativistic expansion). At this order, the higher Fock state $| Q\bar{Q}g \rangle$ of quarkonium (in which the $Q\bar{Q}$ is a color octet) is suppressed at least by two powers of $v$ with respect to the leading Fock state $|Q\bar{Q} \rangle$ (in which the $Q\bar{Q}$ is a color singlet) \cite{Bodwin:1994jh}. Therefore, in our calculations, quarkonium is always a $Q\bar{Q}$ pair in the color singlet. Unbound $Q\bar{Q}$ pairs can be in the color singlet or octet states. At the order of the nonrelativistic expansion that we are studying right now, the transition between a quarkonium and an unbound pair only occurs via a dipole interaction between the color singlet and octet states. In other words, the dissociation product is a $Q\bar{Q}$ in the color octet state and only color octet $Q\bar{Q}$ pairs can recombine as quarkonia via the dipole interaction. Keeping only the dipole interaction in the calculation works here because of the hierarchy $Mv \gg T$. When the quarkonium or the $Q\bar{Q}$ pair is at rest with respect to the local medium, the typical energy of a medium parton is $E_g\sim T$. The typical size of quarkonium is given by $r\sim \frac{1}{Mv}$. So the quarkonium size is small in the sense that $r E_g \ll 1$ and thus the dipole vertex is a weak-coupling interaction. However, if the quarkonium or the $Q\bar{Q}$ is moving with respect to the local medium (for example, when the quarkonium state has a finite transverse momentum), the typical energy of medium partons in the quarkonium (or $Q\bar{Q}$) rest frame is boosted, $E_g\sim \gamma T$ where $\gamma$ is the boost factor that depends on the relative velocity between the quarkonium (or the center-of-mass of the $Q\bar{Q}$) and the local medium. The condition $rE_g \ll 1$ may no longer be true and our calculations would break down then. Rigorously speaking, our calculations only apply to quarkonium states at low transverse momentum. In practice, we should be careful when interpreting our results in the mid and high transverse momentum regions. We will come back to this issue later in Sect.~\ref{sect:results}.

The potentials used in the calculations are Coulombic: $V_s = -C_F\frac{\alpha_{s}^{\ma{pot}}}{r}$ for color singlet and $V_o = \frac{1}{2N_c}\frac{\alpha_{s}^{\ma{pot}}}{r}$ where $C_F = \frac{N_c^2-1}{2N_c}$ and $N_c = 3$. We will take $\alpha_{s}^{\ma{pot}}$ in the potentials to be a parameter and choose its value as $\alpha_{s}^{\ma{pot}} = 0.36$. The coupling constant in the scattering vertices will be taken to be constant $\alpha_s=0.3$. We will vary these two coupling constants and discuss the calculation uncertainties in Section~\ref{sect:results}. The effects of running coupling and nonperturbative potentials will be left to future studies\footnote{The different values of $\alpha_{s}^{\ma{pot}}$ and $\alpha_s$ chosen here already hint at the importance of nonperturbative potentials.}. Since the octet potential is non-zero here, the wavefunction of the unbound octet pair is a Coulomb scattering wave rather than a plane wave. In this way, we re-sum an infinite number of Coulomb exchanges between the octet pair in the initial (for recombination) or final (for dissociation) state. At the leading order in the nonrelativistic expansion, the potential is independent of the orbital angular momentum and spin. Thus the dissociation rates are the same for states separated by fine and hyperfine splittings. The recombination rates are also the same up to a spin degeneracy factor $g_s=\frac{1}{4}$ or $\frac{3}{4}$.

We include the following scattering channels for the dissociation and recombination: $g+H\leftrightarrow Q + \bar{Q}$, $q+H\leftrightarrow q+ Q + \bar{Q}$ and $g+H\leftrightarrow g+ Q + \bar{Q}$ where $H$ can indicate any quarkonium state. The first process is induced by real gluon absorption (for dissociation) and emission (for recombination). The last two processes are mediated by virtual gluons (inelastic scattering). The elastic scattering between quarkonia and medium gluons is neglected here because it occurs at higher orders in the multipole expansion (or equivalently here, nonrelativistic expansion) \cite{Yao:2018sgn}. The direct transitions between different quarkonia species are also omitted due to the same reason. Expressions of the reaction rates of the $1S$, $2S$ and $1P$ states can be found in Appendix~\ref{app:rates}. For $3S$ and $2P$ states, we assume they cannot exist inside the QGP. In other words, they dissociate immediately after entering the QGP and cannot be (re)generated inside the hot medium.

One important feature of our framework is the inclusion of correlated recombination (because we keep track of the evolution of the two-particle distribution of the $Q\bar{Q}$ pair rather than the distribution of a singlet heavy quark). The correlated recombination leads to cross-talk between different quarkonium states. When an excited quarkonium state such as $\Upsilon(2P)$ or $\Upsilon(3S)$ dissociates, the produced $Q\bar{Q}$ can form a lower excited state or the ground state $\Upsilon(1S)$. As will be discussed later, this is an important production mechanism for the ground state. When an excited state dissociates at high temperature, the dissociated $Q\bar{Q}$ pair may form the ground state and then survive the subsequent evolution. When the temperature drops and a ground state dissociates occasionally (the dissociation rate is still non-vanishing even if the static screening effect is small at low temperature), the dissociated $Q\bar{Q}$ pair may form an excited state. The time evolution of the whole system is a network of reactions among unbound heavy quarks, antiquarks and all quarkonia states. Our framework can handle this reaction network and study the physical impacts of the cross-talk recombination.

\subsection{Monte Carlo Simulations}
We solve the coupled transport equations by test particle Monte Carlo simulations. We sample a certain number of $Q\bar{Q}$ pairs and quarkonia according to the distribution functions at the initial time. Mathematically, the distribution function for each particle species is represented by
\be
f({\bs x}, {\bs p}, t) = (2\pi)^3\sum_i \delta^3({\bs x} - {\bs x}_i) \delta^3({\bs p} - {\bs p}_i) \,,
\ee
and similarly for the two-particle distribution functions.
Here ${\bs x}_i$ and ${\bs p}_i$ are the position and momentum of the $i$-th sampled test particle.
The integral of a distribution function over the whole phase space gives the total number of the particle associated with that distribution. The positions and momenta of the sampled particles obey their initial distribution functions. Then we evolve the positions and momenta of all particle species step by step. The time step size is chosen as $\Delta t = 0.01$ fm/c in the laboratory frame. At each time step, we consider each of the following processes.

\subsubsection{Free Streaming}
The position of the particle changes according to
\be
{\bs x}(t+\Delta t) = {\bs x}(t) + \Delta t \frac{{\bs p}(t)}{E(t)}\,,
\ee
where $E(t)$ and ${\bs p}(t)$ are the energy and momentum of the particle at the current time step.

\subsubsection{Momentum Change}
This process is implemented via the Lido package. For given heavy quark (antiquark) momentum and local temperature, the package calculates the scattering rate of the heavy quark (antiquark) with the medium in each scattering channel. If a certain scattering process occurs, the package generates a light scattering partner utilizing local medium properties (temperature and flow field). Its outgoing momentum is sampled from the differential scattering rate. The outgoing momentum of the heavy quark (antiquark) can be obtained from energy and momentum conservation. Then we can update the momentum of the heavy quark (antiquark).

\subsubsection{Dissociation}
For a quarkonium state $nls$ with a certain momentum (velocity) and a position at some local temperature, we calculate its dissociation rate in the laboratory frame. The method to obtain the dissociation rate from the collision term $\ml{C}^-$ can be found in Ref.~\cite{Yao:2018sgn}. The rate times the time step size leads to the dissociation probability in this step. If it is determined (by Monte Carlo sampling) that the quarkonium state dissociates, we sample the incoming and/or outgoing momenta of the relevant light particles and obtain the momenta of the outgoing $Q\bar{Q}$ pair from energy and momentum conservation. The positions of the unbound $Q$ and $\bar{Q}$ are given by the position of the quarkonium before dissociation. Then we remove this quarkonium state from the relevant particle list and add the produced $Q\bar{Q}$ pair to the list of heavy quarks and antiquarks.

\subsubsection{Recombination}
For each unbound $Q\bar{Q}$ pair, we need to calculate their recombination rate. In the Monte Carlo simulation, the two-particle distribution function of the unbound $Q\bar{Q}$ pair is represented by
\be
f({\bs x}_Q, {\bs p}_Q, {\bs x}_{\bar{Q}}, {\bs p}_{\bar{Q}}, t) = (2\pi)^6 \sum_{i,j} \delta^3({\bs x}_Q - {\bs x}_i) \delta^3({\bs p}_Q - {\bs p}_i) \delta^3({\bs x}_{\bar{Q}} - \tilde{\bs x}_j) \delta^3({\bs p}_{\bar{Q}} - \tilde{\bs p}_j) \,.
\ee
The recombination rate of a specific pair can be obtained by the following replacement
\be \nn
&&\delta^3({\bs x}_Q - {\bs x}_i) \delta^3({\bs p}_Q - {\bs p}_i) \delta^3({\bs x}_{\bar{Q}} - \tilde{\bs x}_j) \delta^3({\bs p}_{\bar{Q}} - \tilde{\bs p}_j) \\
&\to&  \delta^3 \Big({\bs x}_{\ma{cm}} - \frac{{\bs x}_i+\tilde{\bs x}_j}{2} \Big) \delta^3 \big({\bs p}_{\ma{cm}} - ( {\bs p}_i + \tilde{\bs p}_j) \big) \delta^3\Big({\bs p}_{\ma{rel}} - \frac{{\bs p}_i-\tilde{\bs p}_j}{2} \Big) \frac{1}{(2\pi\sigma^2)^{\frac{3}{2}}}e^{-\frac{({\bs x}_i-\tilde{\bs x}_j)^2}{2\sigma^2}} \,,
\ \ \ \ \ 
\ee
where ${\bs x}_{\ma{cm}}$, ${\bs p}_{\ma{cm}}$, ${\bs x}_{\ma{rel}}$ and ${\bs p}_{\ma{rel}}$ are the center-of-mass (cm) and relative positions and momenta. This Gaussian ansatz is motivated by the recombination formula derived in Ref.~\cite{Yao:2018nmy}, in which the recombination term for a $Q\bar{Q}$ far away from each other is suppressed exponentially by the bound state wavefunction. The width of the Gaussian is chosen to be the typical size of the bound state. More specifically, for $1S$, $\sigma = a_B$ where $a_B = \frac{2}{\alpha_s C_F M}$ is the Bohr radius. For $2S$ and $1P$ state, we have $\sigma = 2a_B$. The recombination rate of this specific pair can be obtained by doing the following replacement in the recombination term $\ml{C}^+$ in the Boltzmann equation (The expression of $\ml{C}^+$ can be found in Appendix~\ref{app:rates}.)
\be
f({\bs x}_Q, {\bs p}_Q, {\bs x}_{\bar{Q}}, {\bs p}_{\bar{Q}}, t) \to
(2\pi)^3\delta^3\Big({\bs p}_{\ma{rel}} - \frac{{\bs p}_i-\tilde{\bs p}_j}{2} \Big) \frac{1}{(2\pi\sigma^2)^{\frac{3}{2}}} e^{-\frac{({\bs x}_i-\tilde{\bs x}_j)^2}{2\sigma^2}} \,.
\ee

In practice, for each unbound $Q\bar{Q}$ pair with positions ${\bs x}_i$, $\tilde{\bs x}_j$ and momenta ${\bs p}_i$, $\tilde{\bs p}_j$, we first boost the pair into their cm frame. Then we compute their relative momentum and calculate their recombination rate in the cm frame and then boost the rate back into the laboratory. If the recombination into a specific quarkonium state occurs, we sample the momentum of the outgoing quarkonium state based on the differential rate and energy-momentum conservation. The position of the quarkonium state is given by the cm position of the unbound pair before recombination. Finally we remove the unbound pair from the list of heavy quarks and antiquarks and add a new quarkonium state to the relevant list of quarkonium states.

\section{Test Simulation inside QGP Box}
\label{sect:test}
Before we study quarkonium production in heavy-ion collisions, we validate our test particle Monte Carlo simulations for the coupled transport equations. We solve the equations inside a cubic volume of QGP matter at fixed temperature with a side length $L=10$ fm. The box has periodic boundary conditions, i.e., when a particle reaches the boundary of the box, it appears on the opposite side. In other words, the medium behaves as an infinitely large QGP with a finite heavy flavor density (which include both open and hidden heavy flavor states).

We focus on the bottom system, since the nonrelativistic expansion works better for bottom than for charm. We will sample a fixed number of unbound $b\bar{b}$ pairs initially. Their positions are randomly distributed in the volume while their momenta obey thermal or uniform distributions (other distributions can also be specified). In the mode of uniform momentum distribution, each component of the bottom quark's momentum, $p_i$, $i=x,y,z$, is sampled from a uniform distribution between $0$ and $3$ GeV. In the uniform momentum distribution mode, if we turn off the transport equation of unbound pairs and only simulate the transport equations of quarkonia, we find that the system does not properly thermalize. It is the transport of open heavy flavors that drives the kinematic thermalization of all heavy quark states.  These findings have been reported in Refs.~\cite{Yao:2017fuc,Yao:2018zrg}. Here we extend the previous studies to the case of excited states. For simplicity, we will simulate all the cases with the open heavy flavor transport equations turned on. The lessons we learn by comparing the case with open heavy flavor transport equations turned on and that off have been discussed before. We focus on demonstrating the consistency of the numerical implementation of dissociation and recombination here.

We initialize $N_{b,\ma{tot}}=50$ bottom quarks and $N_{\bar{b},\ma{tot}}=50$ antibottom quarks in the QGP volume. Their positions are random and their momenta are sampled in the uniform distribution mode, as described above. We consider the following three cases:
\begin{enumerate}
\item The temperature of the QGP box is fixed to be $300$ MeV throughout. We only simulate the $\Upsilon(1S)$ channel for quarkonium, i.e., only the dissociation and recombination of $\Upsilon(1S)$ are allowed.

\item Only the $\Upsilon(2S)$ channel is turned on. Here, the temperature is fixed at $180$ MeV.\footnote{In principle, since we use a Coulomb potential here, excited bottomonia states exist at high temperature. In the simulation tests, one can choose a higher temperature. But at high temperature, the hidden bottom fraction from excited bottomonia is tiny in thermal equilibrium and one requires large statistics to obtain reasonable results. Furthermore, reaction rates become bigger as temperature increases, so a smaller time step is required.}

\item Same as case 2, but only $\chi_b(1P)$ is studied.

\end{enumerate}

\begin{figure}[h]
    \centering
    \begin{subfigure}[t]{0.48\textwidth}
        \centering
        \includegraphics[height=2.0in]{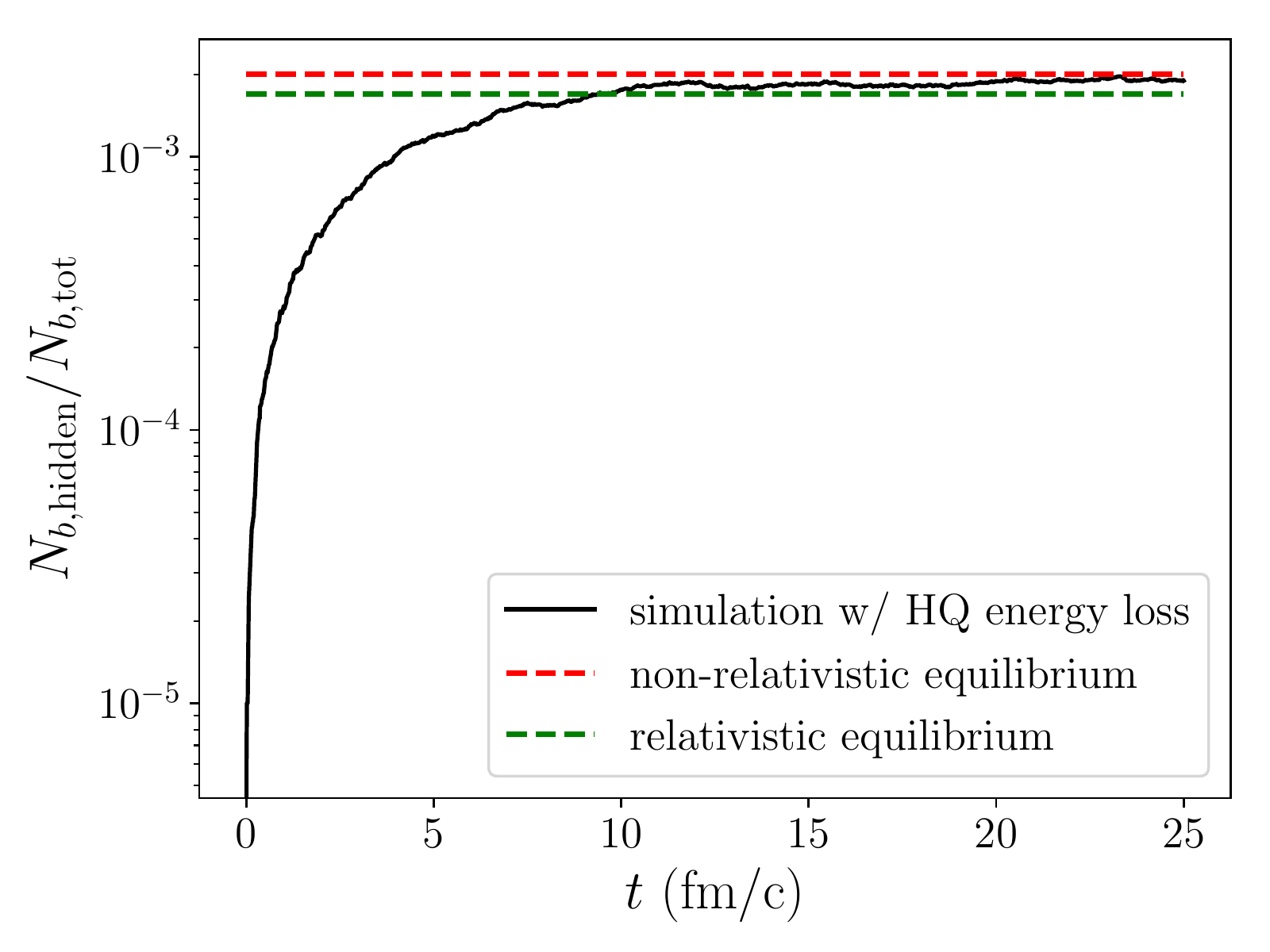}
        \caption{Case 1: $\Upsilon(1S)$ in a $300$ MeV QGP box.}
    \end{subfigure}%
    ~
    \begin{subfigure}[t]{0.48\textwidth}
        \centering
        \includegraphics[height=2.0in]{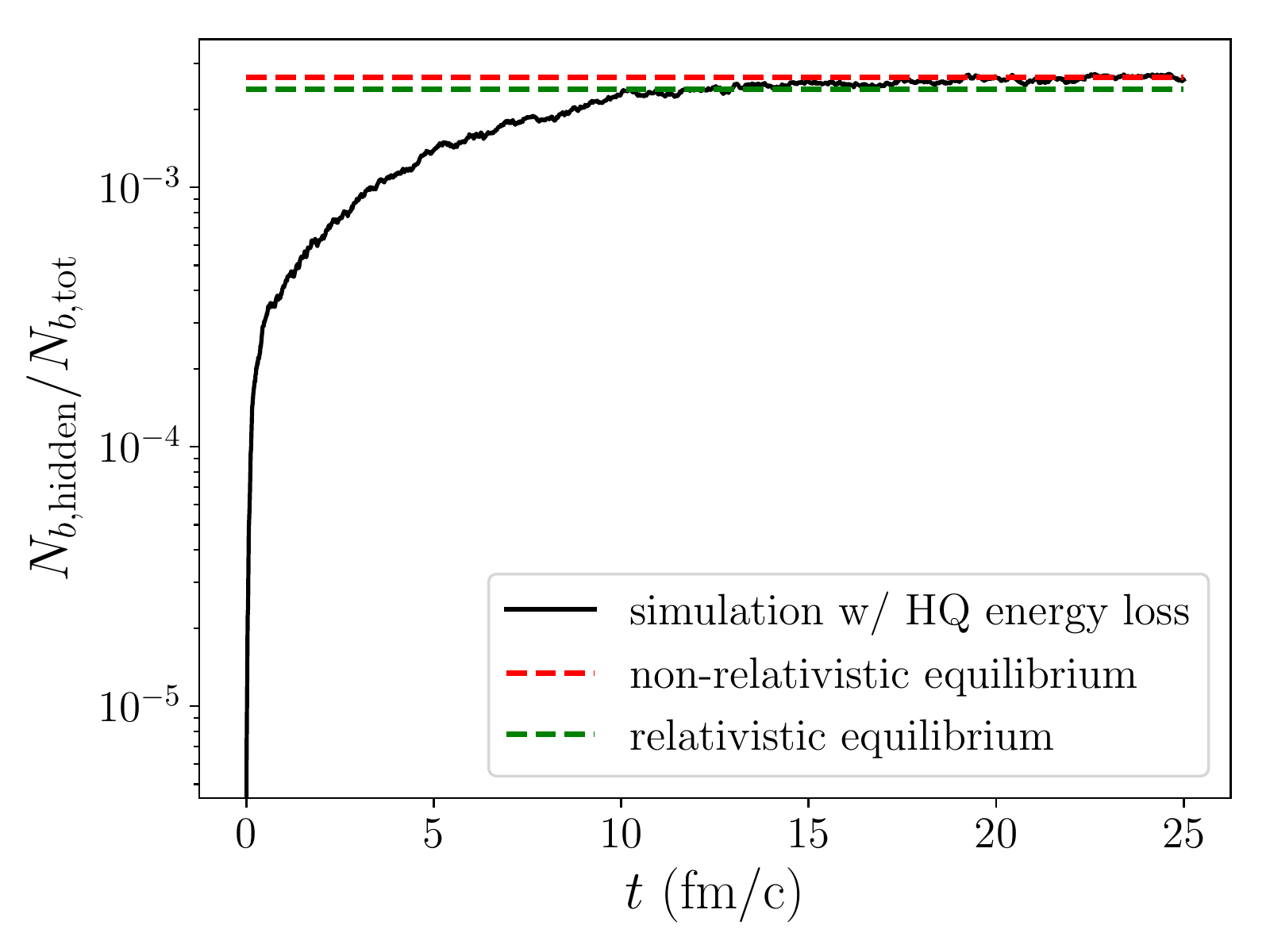}
        \caption{Case 2: $\Upsilon(2S)$ in a $180$ MeV QGP box.}
    \end{subfigure}
    
    \begin{subfigure}[t]{0.48\textwidth}
        \centering
        \includegraphics[height=2.0in]{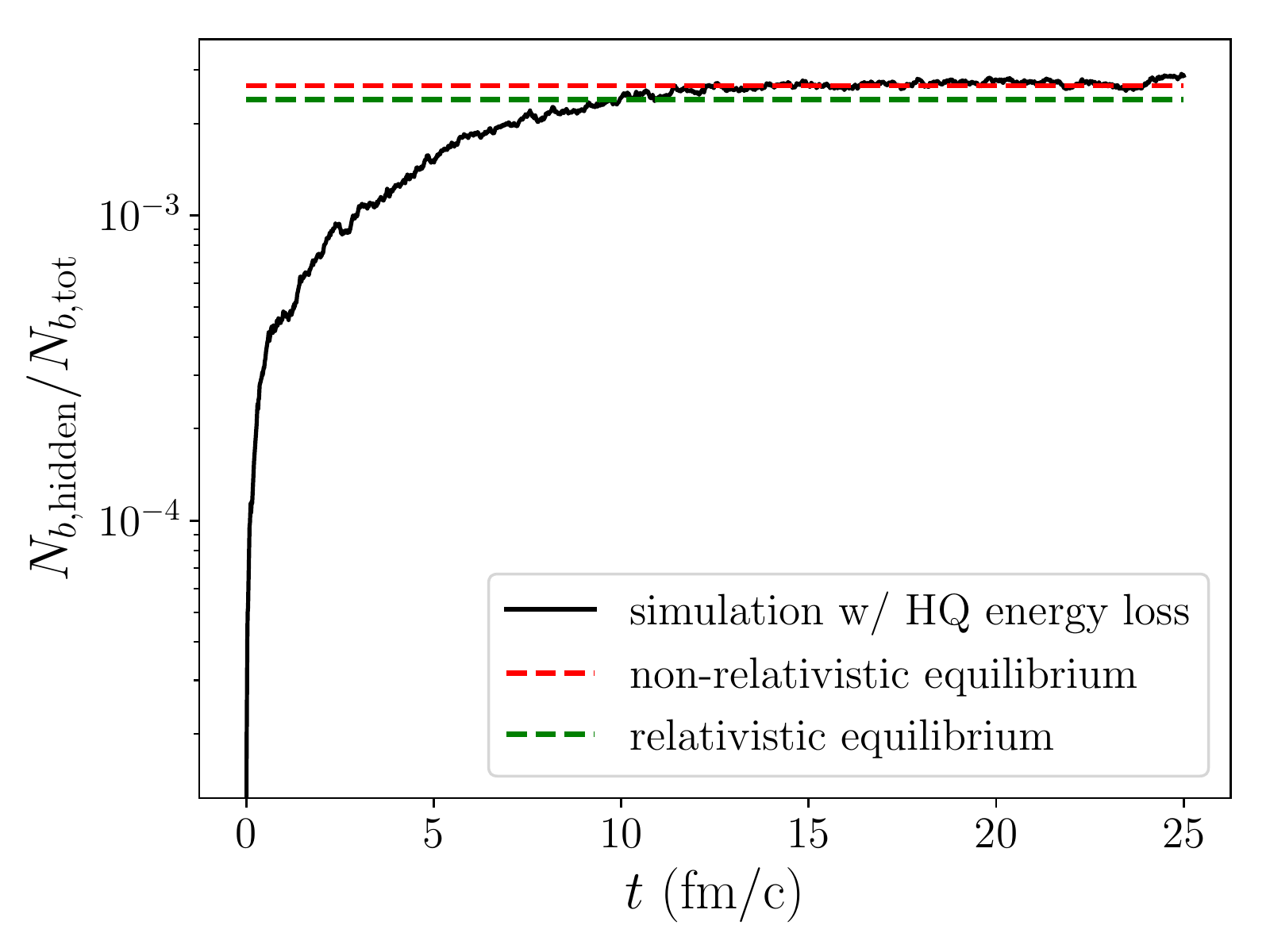}
        \caption{Case 3: $\chi_b(1P)$ in a $180$ MeV QGP box.}
    \end{subfigure}    
\caption{Comparison of the hidden bottom fraction between simulations and properties in thermal equilibrium. The system approaches equilibrium due to the interplay between quarkonium dissociation and recombination.}
\label{fig:balance}
\end{figure}

We will compute the hidden bottom fraction as a function of time, which is defined by
\be
\frac{N_{b,\ma{hidden}}}{N_{b,\ma{tot}}} = \frac{N_{b,\ma{hidden}}}{N_{b,\ma{open}} + N_{b,\ma{hidden}}} = \frac{N_H}{50}\,,
\ee
where $N_H$ is the total number of all bottomonium states. For the three cases listed above, only one bottomonium state contributes to $N_H$. We will compare simulation results of the hidden bottom fraction with that in thermal equilibrium, which can be calculated as follows. In thermal equilibrium, the numbers of open bottom quarks and a specific quarkonium state $H$ are given by
\be
N^\ma{eq}_{i} &=& g_i \ma{Vol} \int\frac{\diff^3p}{(2\pi)^3}\lambda_{i}e^{-E_i(p)/T}\,,
\ee 
with $i=b$, $\bar{b}$ or $H$, $E_i(p)=\sqrt{M_i^2+p^2}$ relativistically and $M_i+\frac{p^2}{2M_i}$ nonrelativistically. Here $g_i$ is the spin and color degeneracy factor and $\lambda_i$ is the fugacity. We have $g_b=6$ for the bottom quark, $g_H=3$ for $\Upsilon(nS)$ and averaged $\chi_b(nP)$. Since the total number of bottom quark is equal to that of antibottom, $\lambda_b = \lambda_{\bar{b}}$. In detailed balance, we have $\lambda_H = \lambda_b \lambda_{\bar{b}} = \lambda_b^2$. With this relation, one can solve the fugacities from the balance equation with a given volume:
\be
N_{b,\ma{tot}} = N^\ma{eq}_{b} + N^\ma{eq}_{H}\,.
\ee
Once we obtain the fugacities, we can compute the hidden bottom fraction in thermal equilibrium:
\be
\frac{N^\ma{eq}_{H}}{N_{b,\ma{tot}}}\,.
\ee
The comparison between our simulation results and the system properties in thermal equilibrium is depicted in Fig.~\ref{fig:balance} for the three cases listed above. The results are obtained from averaging 10000 simulation events. As can be seen from the plots, the interplay between dissociation and recombination can drive the system to equilibrium. The relaxation rate of the system is about $7-8$ fm/c for the system conditions used here. This number is on the order of the lifetime of QGP in real collisions. But in general, the relaxation rate depends on the initial $p_T$ spectrum and density of the heavy particles, as well as the temperature of the medium. Our simulations can reproduce the correct limit of the hidden bottom fraction in thermal equilibrium. These tests serve as consistency checks in our studies and we are now ready to move on and simulate real collision systems.

\section{Treatment of Relativistic Heavy-Ion Collisions}
\label{sect:real}
In this section we will describe our numerical setup to study bottomonium production in heavy-ion collisions. We include $\Upsilon(1S)$, $\Upsilon(2S)$, $\Upsilon(3S)$, $\chi_b(1P)$ and $\chi_b(2P)$ states in the quarkonium transport equations. To solve the coupled transport equations, we need an initial condition and a bulk QGP medium. For the calculations of $R_\ma{AA}$, we also need to include all feed-down contributions from excited states to the ground and lower excited states in the hadronic gas stage. We will explain the calculation of the initial phase space distribution, the bulk QGP evolution and the feed-down network in a sequence below. 

\subsection{Initial Conditions}
\label{sect:cnm}
\begin{figure}
    \centering
    \begin{subfigure}[t]{0.48\textwidth}
        \centering
        \includegraphics[height=2.0in]{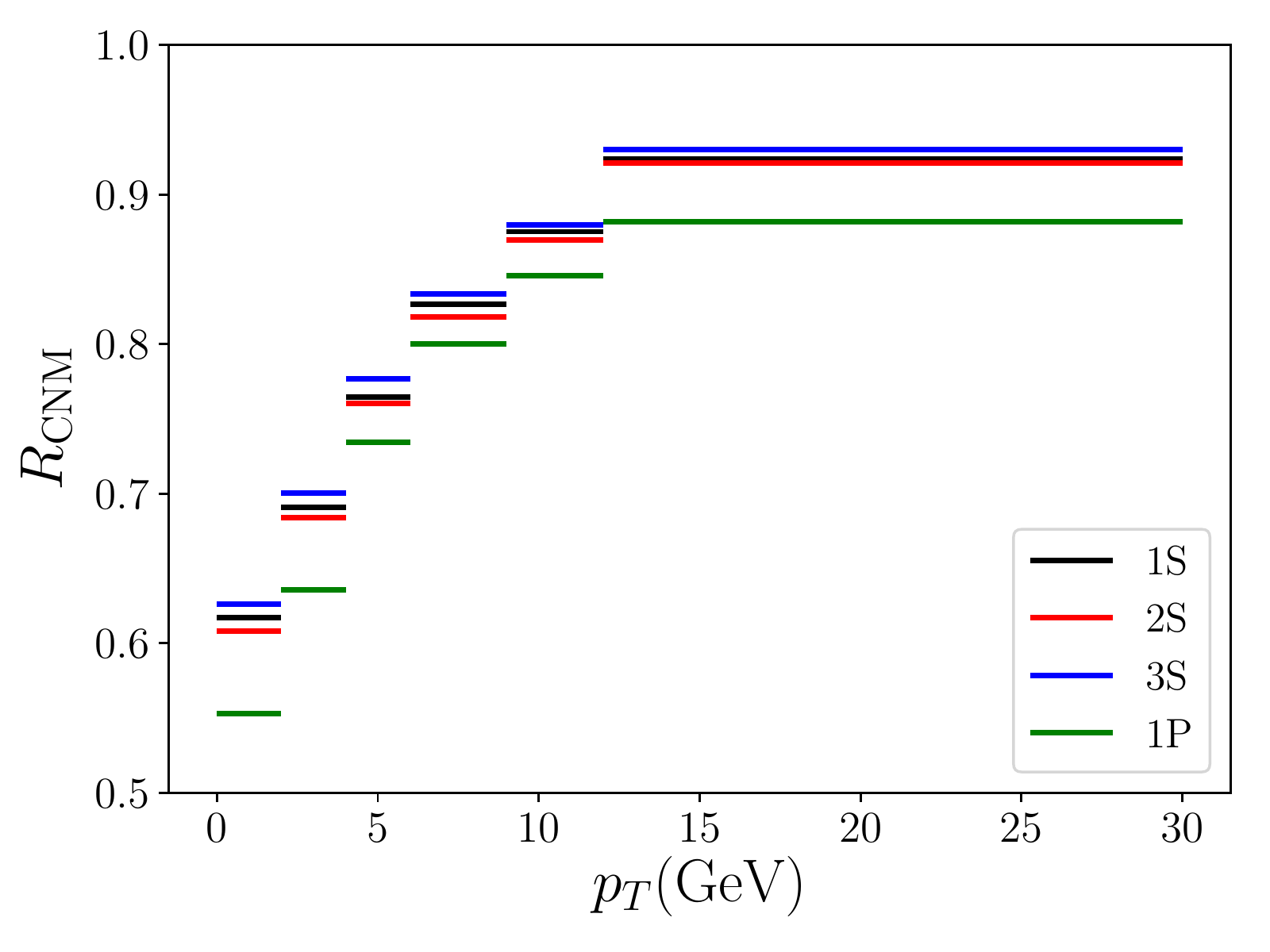}
        \caption{Central value of the CNM effect as a function of transverse momentum.}
    \end{subfigure}%
    ~
    \begin{subfigure}[t]{0.48\textwidth}
        \centering
        \includegraphics[height=2.0in]{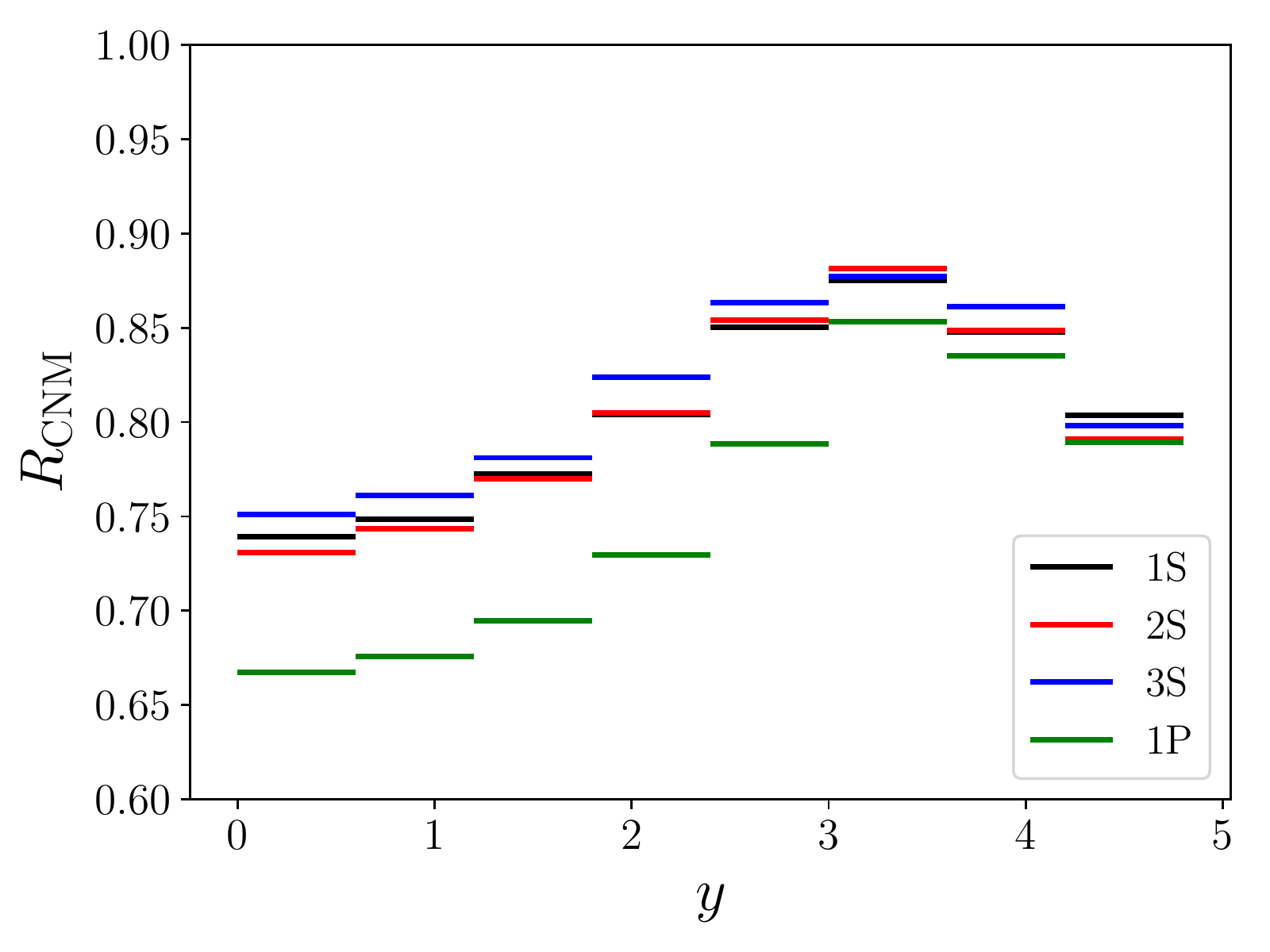}
        \caption{Central value of the CNM effect as a function of rapidity.}
    \end{subfigure}
    
    \begin{subfigure}[t]{0.25\textwidth}
        \centering
        \includegraphics[height=1.1in]{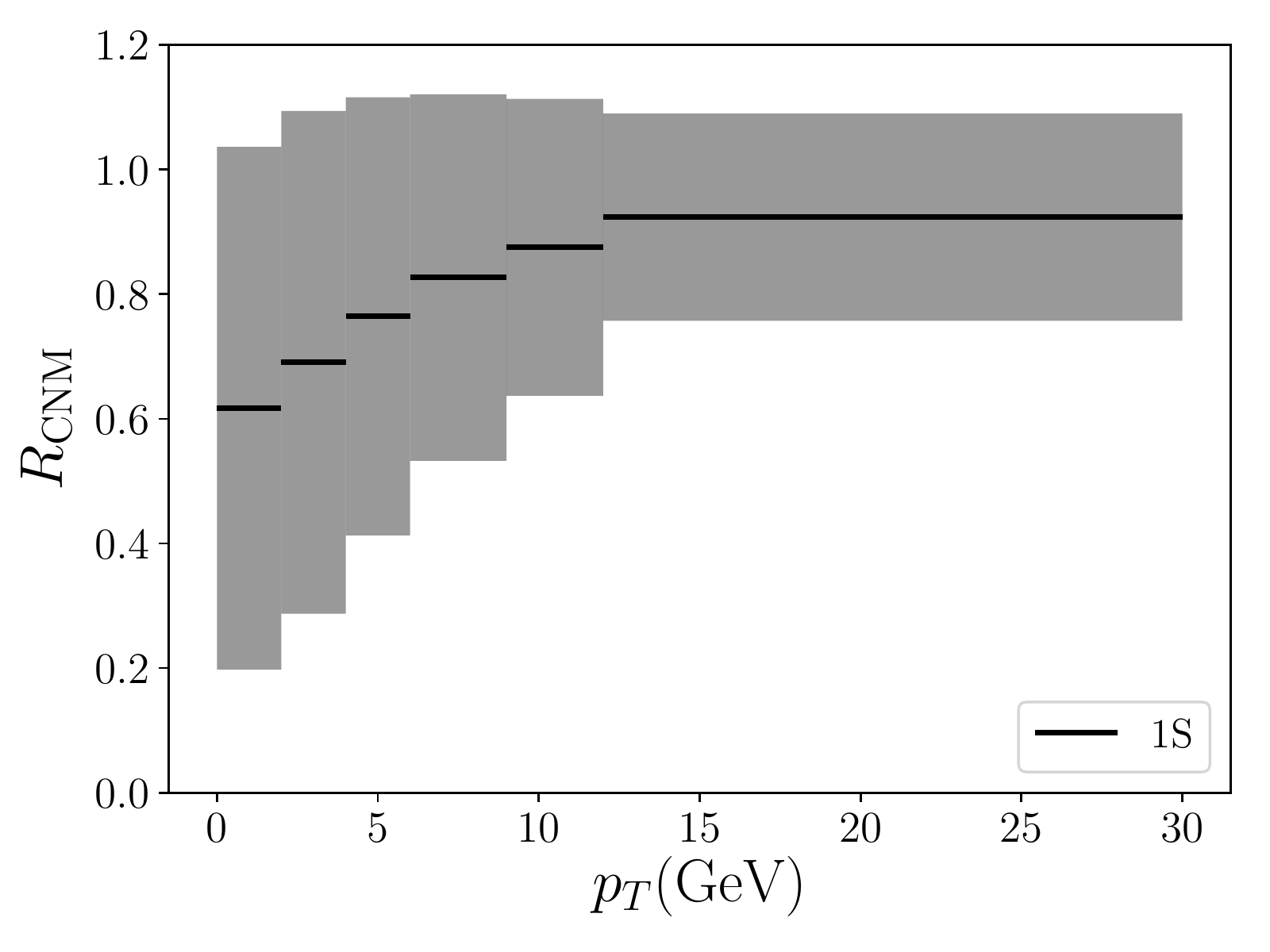}
        \caption{Uncertainty band of $R_{\ma{CNM}}(p_T)$ for $1S$.}
    \end{subfigure}%
    \begin{subfigure}[t]{0.25\textwidth}
        \centering
        \includegraphics[height=1.1in]{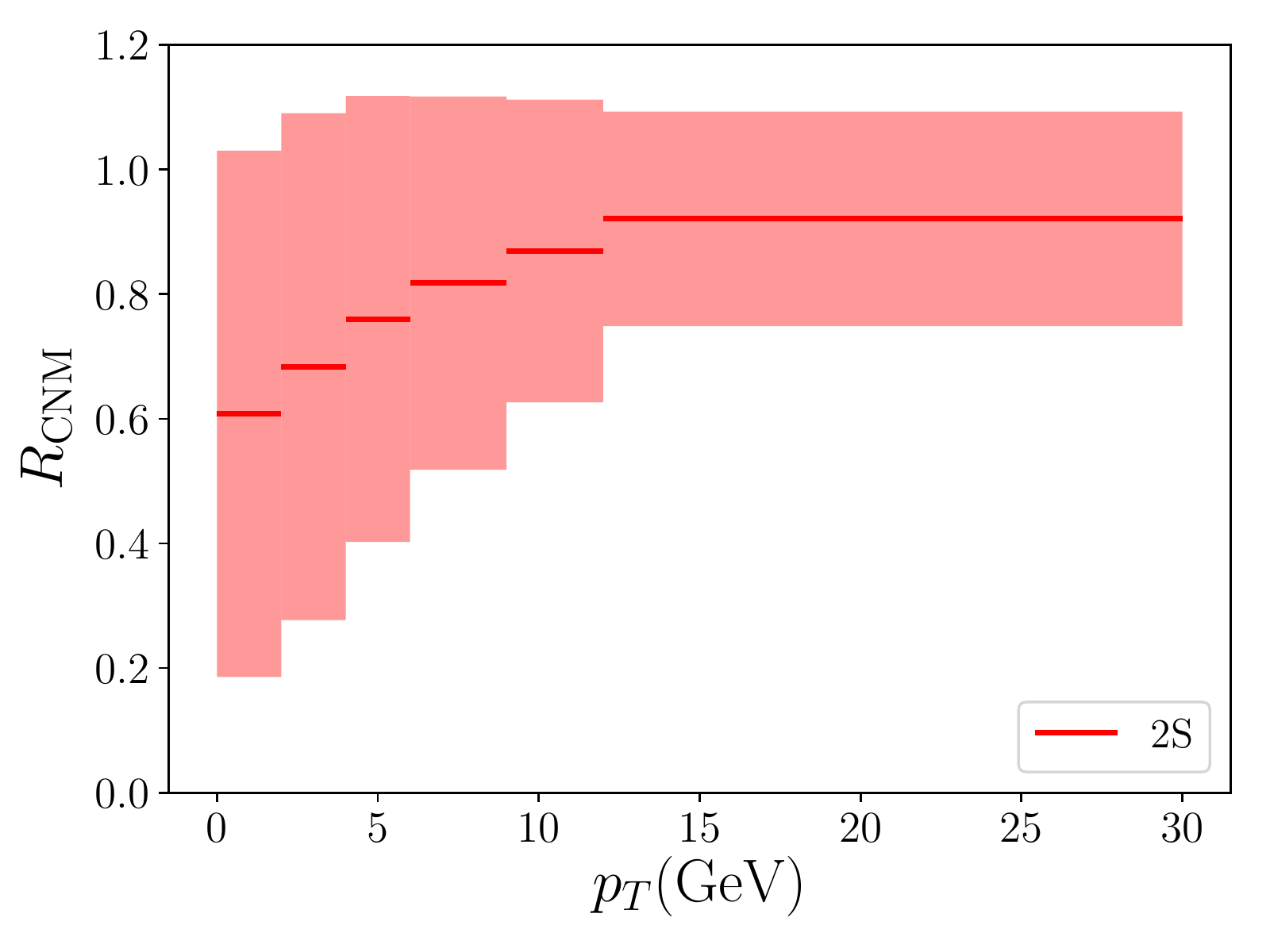}
        \caption{Uncertainty band of $R_{\ma{CNM}}(p_T)$ for $2S$.}
    \end{subfigure}%
    \begin{subfigure}[t]{0.25\textwidth}
        \centering
        \includegraphics[height=1.1in]{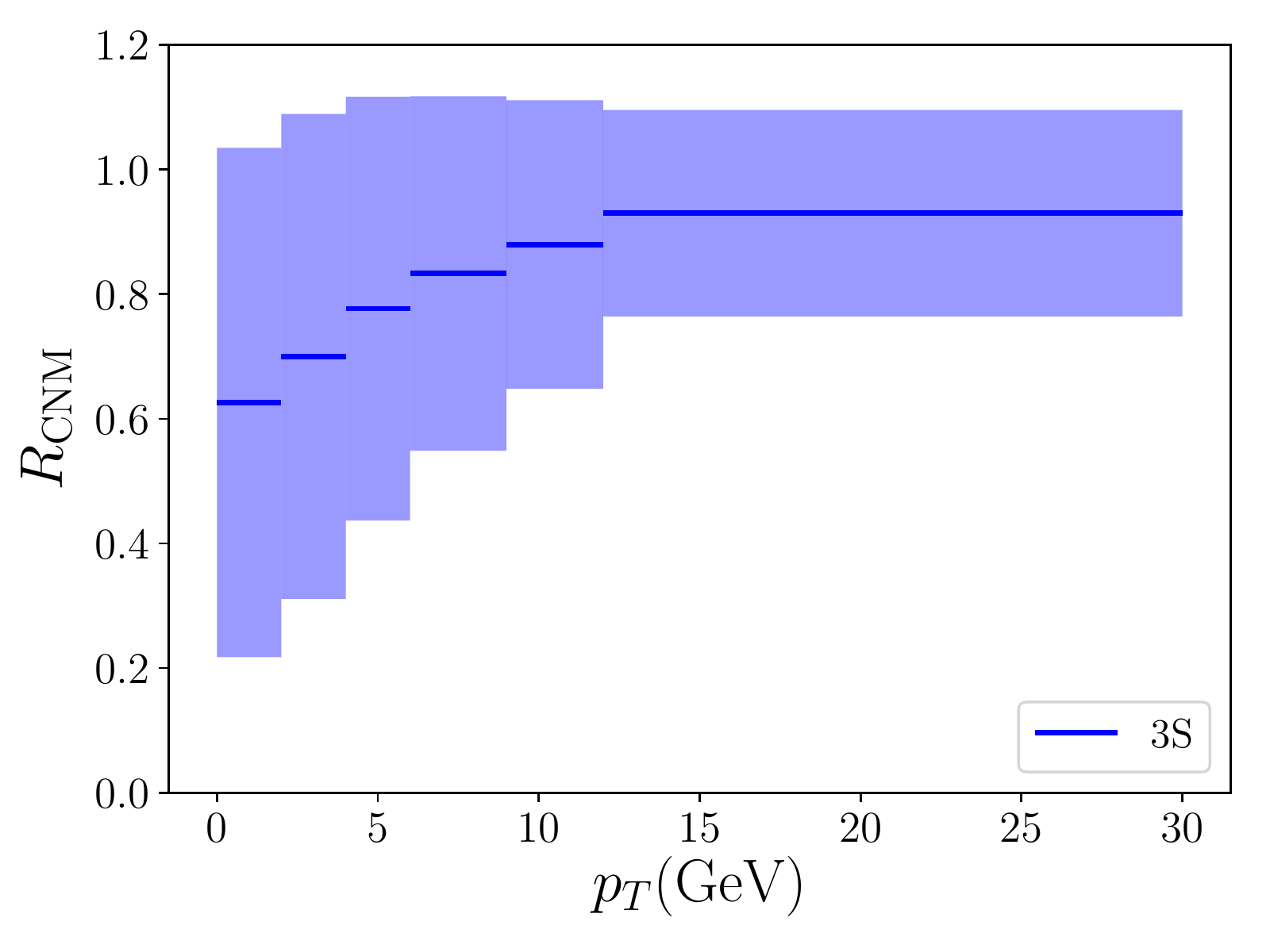}
        \caption{Uncertainty band of $R_{\ma{CNM}}(p_T)$ for $3S$.}
    \end{subfigure}%
        \begin{subfigure}[t]{0.25\textwidth}
        \centering
        \includegraphics[height=1.1in]{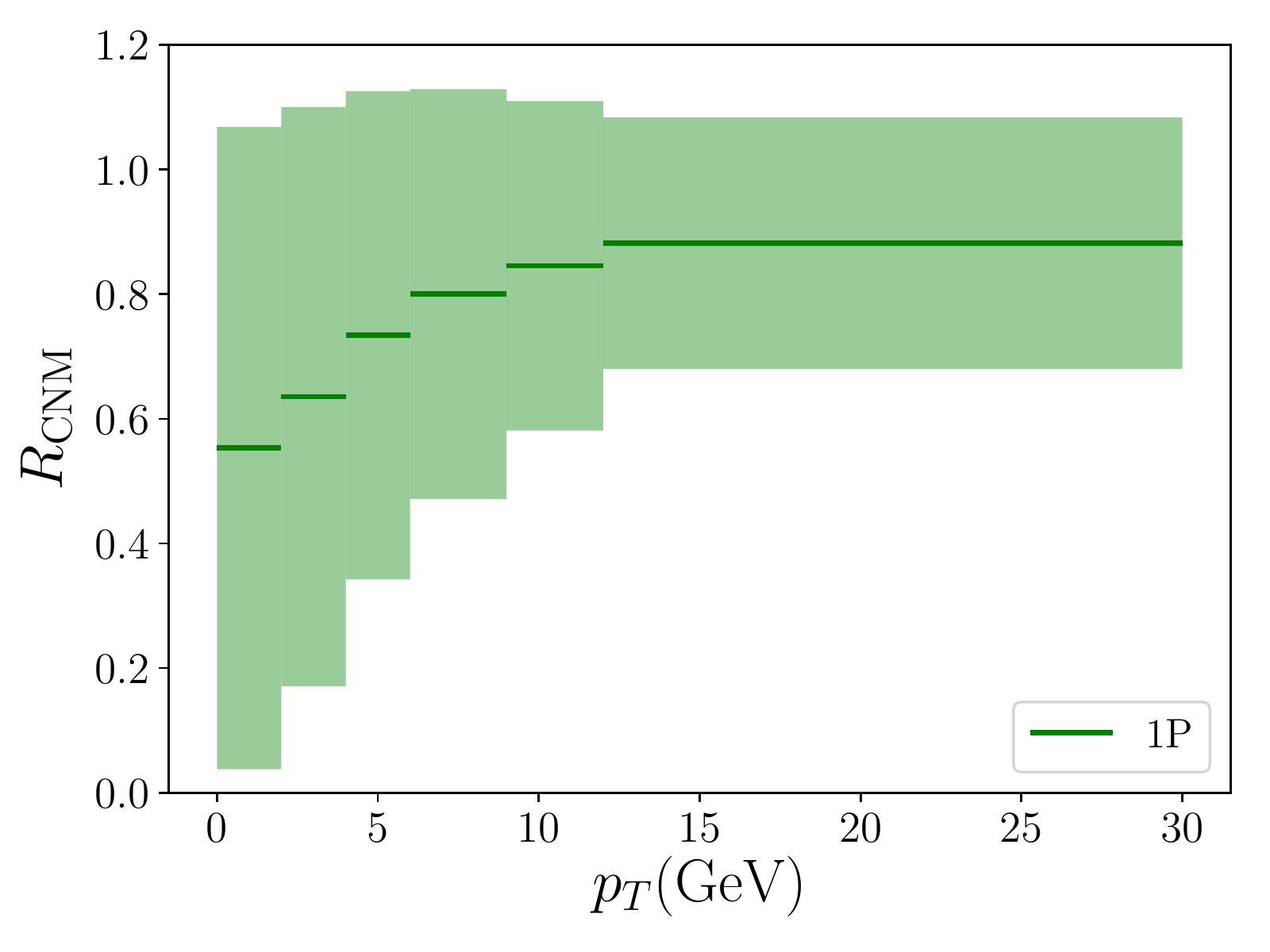}
        \caption{Uncertainty band of $R_{\ma{CNM}}(p_T)$ for $1P$.}
    \end{subfigure}%
    
    \begin{subfigure}[t]{0.25\textwidth}
        \centering
        \includegraphics[height=1.1in]{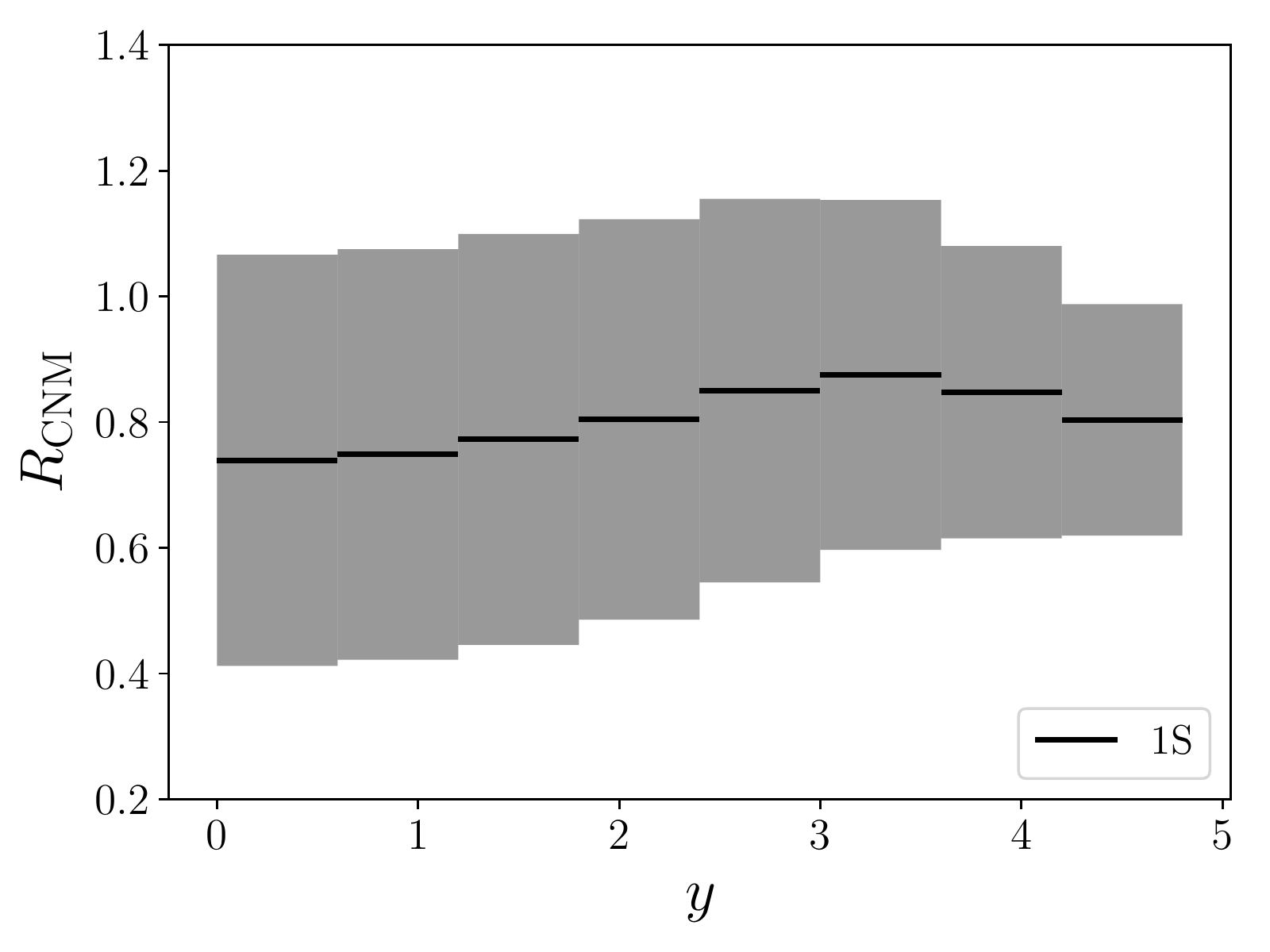}
        \caption{Uncertainty band of $R_{\ma{CNM}}(y)$ for $1S$.}
    \end{subfigure}%
    \begin{subfigure}[t]{0.25\textwidth}
        \centering
        \includegraphics[height=1.1in]{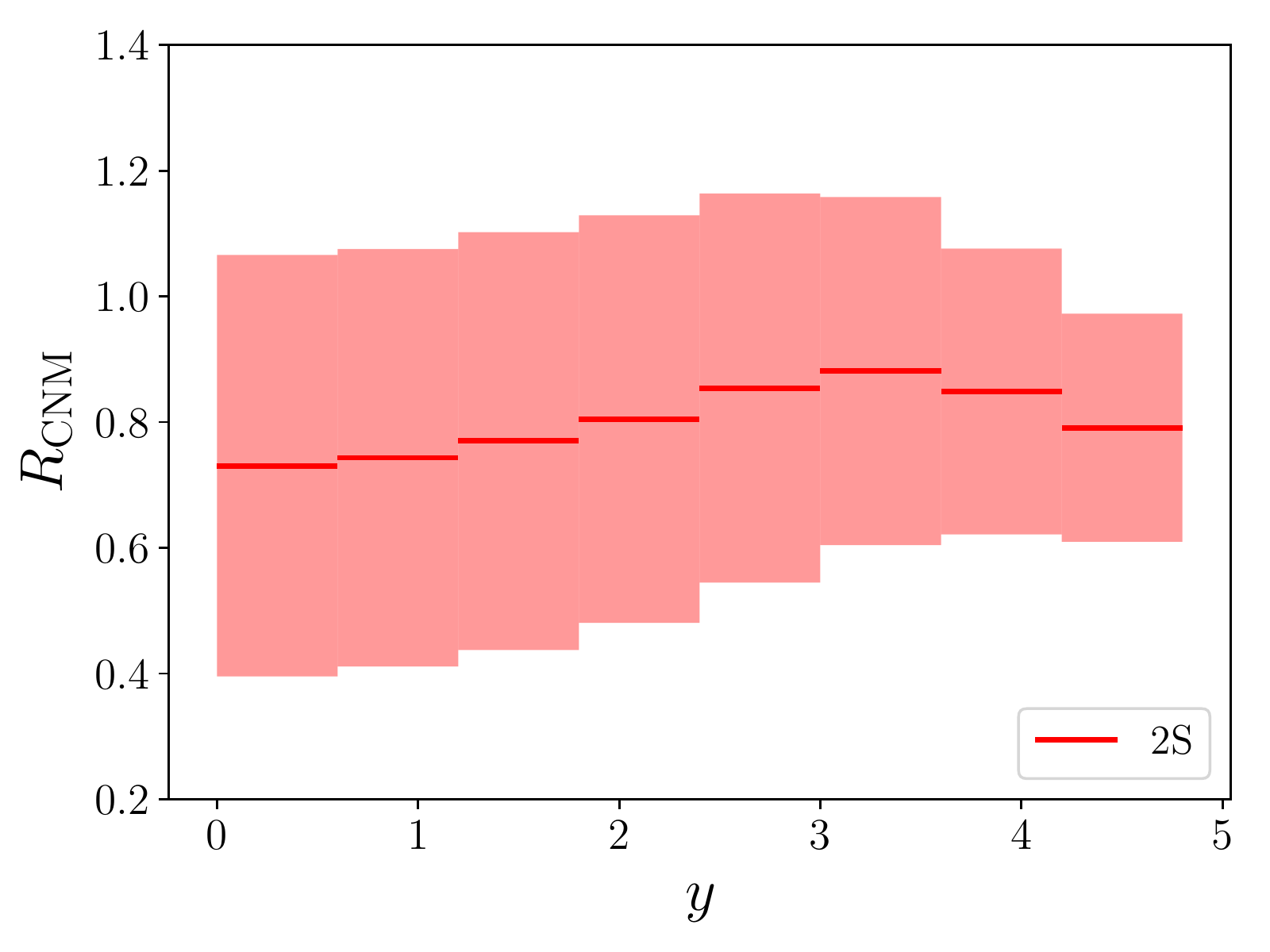}
        \caption{Uncertainty band of $R_{\ma{CNM}}(y)$ for $2S$.}
    \end{subfigure}%
    \begin{subfigure}[t]{0.25\textwidth}
        \centering
        \includegraphics[height=1.1in]{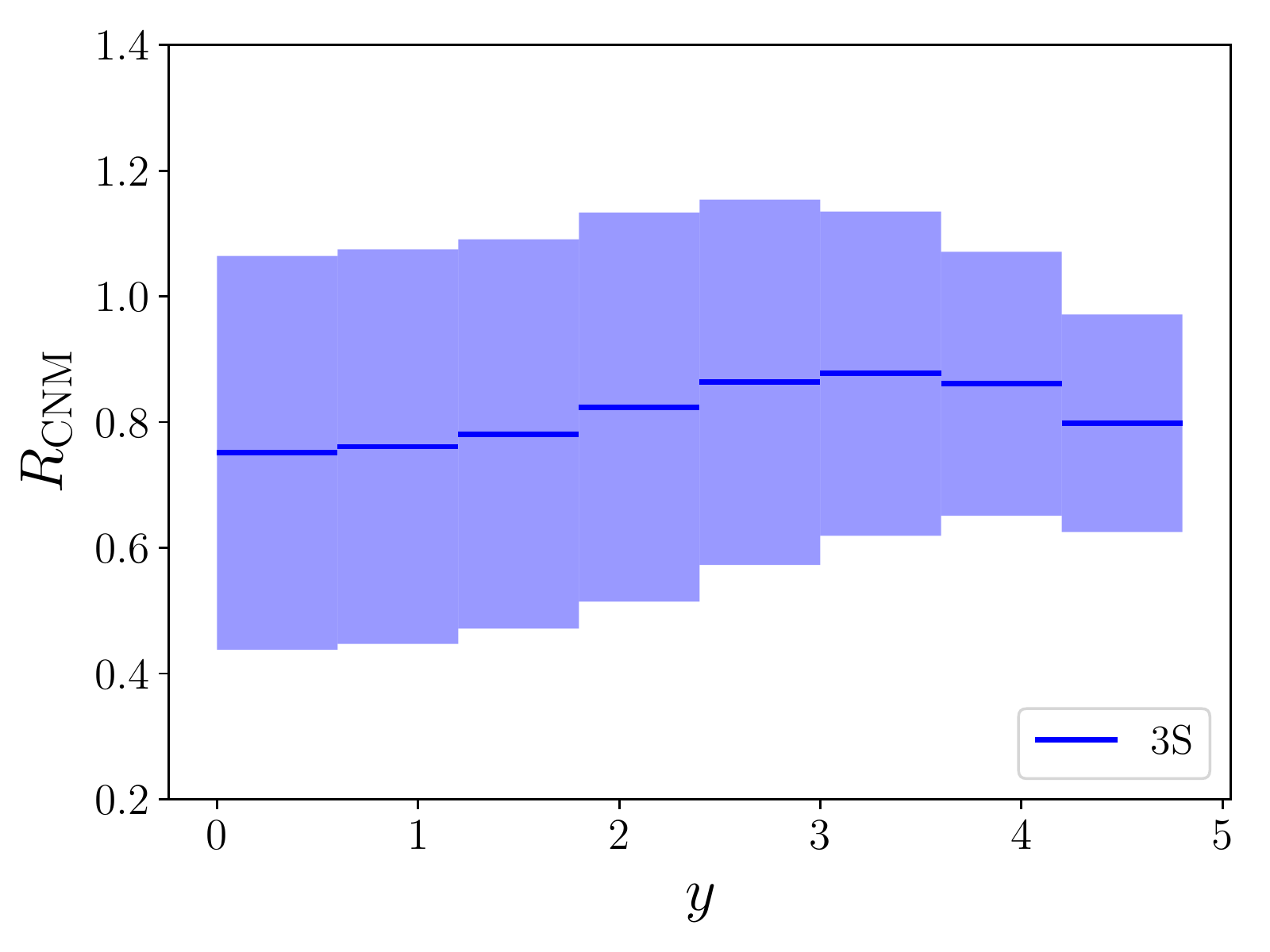}
        \caption{Uncertainty band of $R_{\ma{CNM}}(y)$ for $3S$.}
    \end{subfigure}%
        \begin{subfigure}[t]{0.25\textwidth}
        \centering
        \includegraphics[height=1.1in]{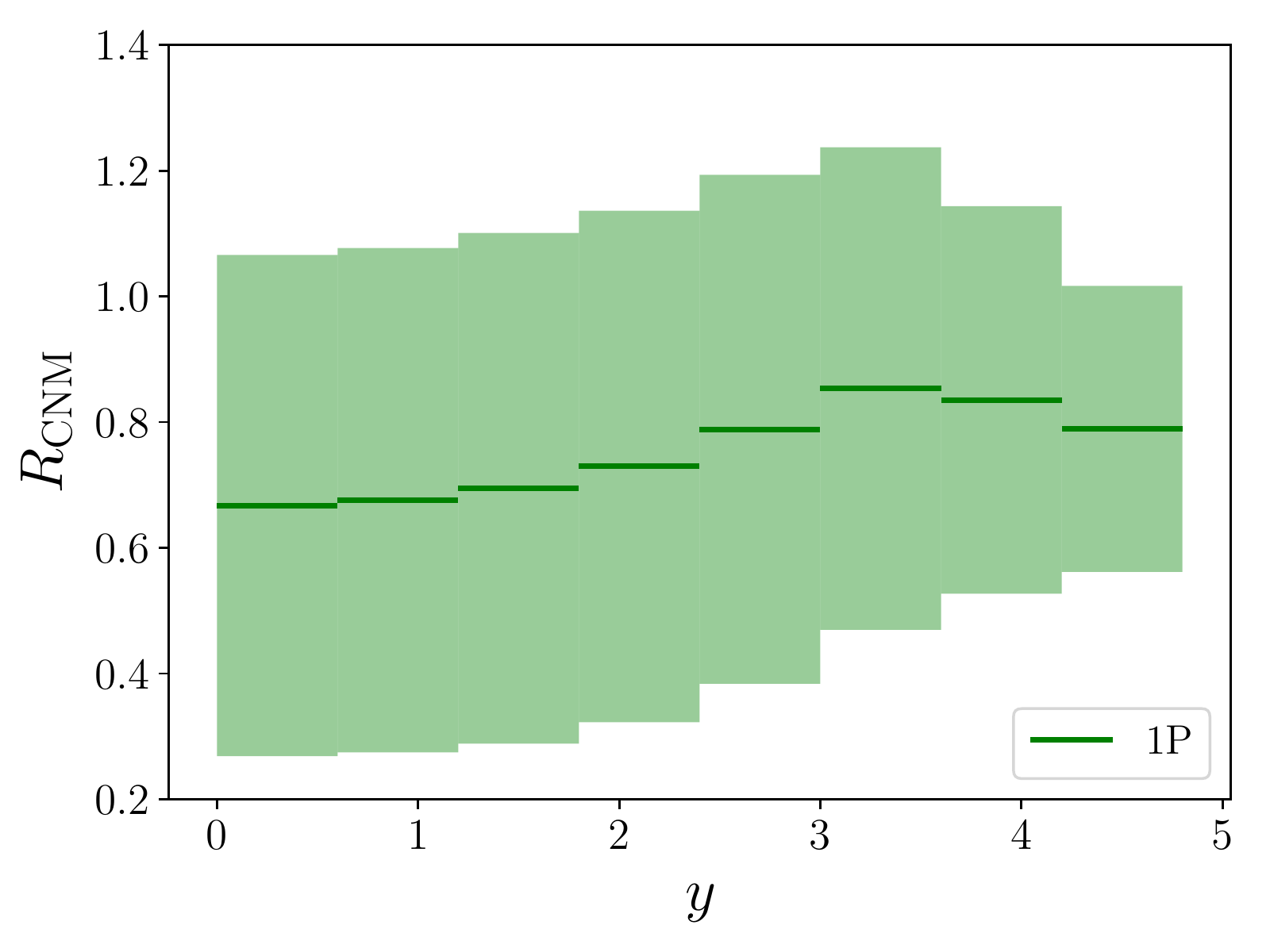}
        \caption{Uncertainty band of $R_{\ma{CNM}}(y)$ for $1P$.}
    \end{subfigure}%
\caption{CNM effects on bottomonia originated from nPDF at $5.02$ TeV Pb-Pb collisions as functions of transverse momentum and rapidity.}
\label{fig:cnm}
\end{figure}

The initial phase space distributions are determined as follows. The momenta of heavy quark-antiquark pairs and each quarkonium state are calculated and sampled utilizing \textsc{Pythia} \cite{Sjostrand:2014zea}. The quarkonium production calculation in \textsc{Pythia} is based on the NRQCD factorization \cite{Bodwin:1994jh}. We use the nuclear parton distribution function (nPDF) parametrized by EPPS16 \cite{Eskola:2016oht}. The nPDF is the only cold nuclear matter (CNM) effect we include. The suppression factors caused solely by the CNM effects $R_\ma{CNM}$ for the Pb-Pb collisions at $5.02$ TeV are shown in Fig.~\ref{fig:cnm} for different bottomonia states. We use the same bins in the transverse momentum $p_T$ and rapidity $y$ as the CMS measurements, which will be shown later. The uncertainty bands are estimated by the following method: We calculate the CNM effect for each set of the nPDF's in the EPPS16 parametrization, by using the LHAPDF interface \cite{Buckley:2014ana}. The set $1$ corresponds to the central value while the sets $2-41$ are error sets. (In the LHAPDF interface, the set $0$ corresponds to the central value while the sets $1-40$ are error sets.) The uncertainty is given by (41) of Ref.~\cite{Eskola:2016oht}:
\be
\Delta R_{\ma{CNM}} = \frac{1}{2} \sqrt{\sum_{i=1}^{20} \Big[ R_{\ma{CNM}}(2i) - R_{\ma{CNM}}(2i+1) \Big]^2} \,,
\ee
where $R_{\ma{CNM}}(i)$ denotes the CNM effect calculated from the $i$-th error set.

The production of $\chi_b(2P)$ is not available yet in \textsc{Pythia}, so we will assume the CNM effect on $\chi_b(2P)$ is the same as that on $\chi_b(1P)$. Furthermore, \textsc{Pythia} with nPDF EPPS16 cannot describe the nuclear modification factor $R_{\ma{pAu}}= 0.82\pm0.10(\ma{stat})\pm0.08(\ma{syst})$ measured by the STAR collaboration \cite{Wang:2019vau}. So we use $R_\ma{CNM} = (R_{\ma{pAu}})^2 = 0.67$ as the central value of the CNM effect for the $200$ GeV Au-Au collision. We estimate the uncertainty in the $R_{\ma{pAu}}$ as $\Delta R_{\ma{pAu}} = \sqrt{0.11^2+0.08^2}\approx 0.128$, and the uncertainty of $R_\ma{CNM}$ in Au-Au collisions as $2\times R_{\ma{pAu}}\times \Delta R_{\ma{pAu}} = 0.21$. We assume the $R_\ma{CNM}$ here is $p_T$-independent since the $p_T$ range covered by the STAR measurement is limited. The $y$-dependence of $R_\ma{CNM}$ is not needed here because the STAR measurements are carried out in the mid-rapidity region ($|y|<0.5$).

The position distribution of the initial production vertices of heavy quark-antiquark pairs and quarkonia are calculated and sampled using the \trento\ 
model \cite{Moreland:2014oya}. The \trento\ 
model assumes the entropy density deposited by the collision at mid rapidity is given by
\be
s(\tau_0, {\bs x}_\perp) \propto \bigg( \frac{(T_A)^p+(T_B)^p}{2}  \bigg)^{1/p}\,,
\ee
where $T_A = T_A({\bs x}_\perp)$ and $T_B = T_B({\bs b}_\perp - {\bs x}_\perp)$ are the nuclear thickness functions of the two projectiles separated by the impact parameter ${\bs b}_\perp$, $p$ is a parameter that has been calibrated on bulk observables of the QGP and $\tau_0$ is the thermalization time of the system after the initial collision. We choose $\tau_0 = 0.6$ fm/c, before which the system is just free streaming. The parametrized entropy density will be used as the initial condition of the hydrodynamic equation, which will be explained in the next subsection. The production of heavy quark-antiquark pairs and quarkonia are thought to be hard processes because of the large mass scale. Therefore, their initial production probability is assumed to be proportional to the binary collision density in the transverse plane, which is proportional to $T_AT_B$. Since the nuclear thickness function is used in both the initial entropy density and the initial heavy quark-antiquark production, the corona effect is taken into account in our calculations.

All heavy particles are assumed to be produced at $\tau = 0$ and they propagate via free streaming until $\tau=\tau_0 = 0.6$ fm/c. Production at $\tau = 0$ is considered a valid assumption for heavy quark-antiquark pairs because their production time is about $\frac{1}{M}$ in their rest frame. But this may no longer be true for quarkonia, whose formation time is estimated as $\frac{1}{Mv^2}$ in their rest frame, which is the time for the heavy quark-antiquark pair to develop the quarkonium wavefunction of the relative motion\footnote{For heavy quark-antiquark pairs and quarkonia with finite transverse momenta, their formation times in the laboratory frame will be boosted by a $\gamma$-factor. So even the ground quarkonium state may form inside the thermal medium. But it should be noted that the total dissociation rate in the laboratory frame increases with the transverse momentum. The effect of the finite formation time is small. See the following main text for the arguments.}. For bottom quarks and bottomonia, $\frac{1}{M} \sim 0.05$ fm/c while $\frac{1}{Mv^2} \sim 0.4$ fm/c. For excited quarkonia states, the formation time would be even longer because of the smaller binding energies. For the ground state, it is probably formed before $\tau_0$. Since we assume particles are free streaming before $\tau_0$, it does not matter whether the ground state is free streaming as a fully formed quarkonium state or an unbound pre-resonant heavy quark-antiquark pair. But it matters for excited quarkonium states, because they may be formed inside the thermal QGP. In other words, at $\tau_0$, we may only have pre-resonant excited quarkonium states. But this is just a tiny effect in our calculations. The excited states have very large dissociation rates when the temperature is high. They will dissociate in one time step after entering the QGP and becoming correlated unbound pair to evolve further. It really does not matter if we have pre-resonant or completely formed excited quarkonium states, because they evolve as unbound $Q\bar{Q}$ pair when entering the thermal QGP. Improvements can be done by including the relative momentum broadening of the $Q\bar{Q}$ pair in the pre-thermalization stage, which will be left to future studies.

For our calculation on bottomonia, we will only initialize quarkonia states but not unbound $b\bar{b}$ pairs, because the number of such pairs produced in current heavy-ion collision experiments is very small (for central Pb-Pb collisions at $5.02$ TeV, the average number of unbound $b\bar{b}$ pairs is less than one per rapidity). So unlike charmonium production, uncorrelated recombination is negligible for bottomonium production. Since most of the unbound heavy quark-antiquark pairs are produced back-to-back in the transverse plane initially, correlated recombination of these unbound pairs is also negligible. Due to the even smaller production cross section of bottomonium, we will assume in our calculations at most one bottomonium state is produced initially in one heavy-ion collision event.

\subsection{Medium Evolution}
We use a $2+1$ dimensional viscous hydrodynamic model VISHNew \cite{Song:2007ux,Shen:2014vra} to describe the evolution of bulk QGP matter. The package numerically solves
\be
\partial_{\mu} T^{\mu\nu} &=& 0
\ee
with the energy-momentum tensor
\be
T^{\mu\nu} &=& eu^{\mu}u^{\nu} - (P+\Pi)(g^{\mu\nu}-u^{\mu}u^{\nu}) + \pi^{\mu\nu}, \\
\Pi &=& -\zeta\nabla\cdot u, \\
\pi^{\mu\nu} &=& 2\eta\nabla^{\langle\mu}u^{\nu\rangle}
\ee
for given initial conditions.
Here $e$ and $P$ are the local energy density and pressure, $g^{\mu\nu}$ is the metric and $u^{\mu}$ is the local four-velocity of the medium. $\Pi$ is the bulk stress with the bulk viscosity $\zeta$, and $\pi^{\mu\nu}$ is the shear stress tensor with the shear viscosity $\eta$. The angle bracket means traceless symmetrization.

Both the bulk and shear viscosities are parameters here and can be temperature-dependent. We will use the parametrizations and calibrations in Ref.~\cite{Bernhard:2016tnd}. Ref.~\cite{Bernhard:2016tnd} uses the \trento\ model to calculate the initial entropy density $s$ and then obtain the energy density and pressure with an equation of state calculated by lattice QCD. To obtain the initial stress-energy tensor at $\tau_0$, Ref.~\cite{Bernhard:2016tnd} further assumes the initial flow velocity and the viscous terms vanish at $\tau_0$. All parameters in the \trento\ model and the hydrodynamic equations are calibrated to experimental observables on light hadrons. For $2.76$ TeV Pb-Pb collisions, we will use the parameters calibrated to charged particles shown in Table~III of Ref.~\cite{Bernhard:2016tnd} with the exception of the parameter $p$ in the \trento\ model. We choose $p=0$, which is consistent with the calibration in Ref.~\cite{Bernhard:2016tnd} (the calibrated value of $p$ is $0.03_{-0.17}^{+0.16}$). For the parameter calibrations in other collision energies and systems, we will follow Ref.~\cite{Xu:2017obm}.

\subsection{Feed-Down Network}
We terminate the transport evolution when the local medium temperature drops to $T_c=154$ MeV, i.e., we neglect the dissociation of quarkonium states in the hadronic gas. 
The final nuclear modification factor $R_\ma{AA}(i)$ for $i = \Upsilon(1S)$, $\Upsilon(2S)$, $\Upsilon(3S)$, $\chi_b(1P)$ and $\chi_b(2P)$ includes both the CNM effect $R_\ma{CNM}$ and the hot medium effect. The evaluation of the CNM effect has been discussed in Sect.~\ref{sect:cnm}. We will now discuss how we compute the hot medium effect.

Our calculation includes, unlike many others, correlated recombination and we find that this effect plays a crucial role. We need to take into account the following situation: Some quarkonium state produced initially may end up as a different quarkonium state. For example, a $2P$ state produced initially, melts inside the QGP and recombines as a $1S$ state which survives the following in-medium evolution. To this end, for each centrality bin, we simulate $N^{\ma{init}}_i$ events in each of which one $i$-quarkonium state is initialized. After the in-medium evolution, among these $N^{\ma{init}}_i$ events, there are $N_{i\to j}$ events that have a $j$-quarkonium state in the end. We note in general $\sum_j N_{i\to j} \leq N^{\ma{init}}_i$. For example, $N_{1S\to 2S}$ is the number of $\Upsilon(2S)$ states generated from those $N^{\ma{init}}_{1S}$ events where the initial quarkonium state is a $\Upsilon(1S)$. $N_{i\to i}$ is the ``surviving" number whose physical meaning is well-known (it also has contribution from the correlated recombination), and has been calculated in many studies. $N_{i\to j}$ with $j\neq i$ is the contribution from correlated, cross-talk recombination.

As explained in Sect.~\ref{sect:cnm}, we only initialize one bottomonium state in each relevant collision event. For simplicity of the expressions, we will assume all $N^{\ma{init}}_i$'s are the same and equal to $N^{\ma{init}} = 30000$ in our calculations of $R_\ma{AA}$. The total final number of the $i$-quarkonium state produced from $N^{\ma{init}}$'s collision events ($N^{\ma{init}}$ events for each state we consider) is given by (we include CNM effect here)
\be
\label{eqn:final_num}
N_i^\ma{final} = \sum_j R_\ma{CNM}(j) \frac{\sigma_j}{\sigma_i}  N_{j\to i}\,,
\ee
where $R_\ma{CNM}(j)$ is the CNM effect on the primordial production of the $j$-quarkonium state and $\sigma_i$ is the initial primordial production cross section of the $i$-quarkonium state (without any feed-down contributions). The ratios of the initial production cross sections are listed in Appendix~\ref{app:fd}. Now we include the feed-down contributions for $\Upsilon(1S)$ and $\Upsilon(2S)$:
\be
N^{\ma{init,fd}}_{2S} &=& N^{\ma{init}} + \sum_{j = 3S, 2P}  \frac{\sigma_j}{\sigma_{2S}} N^{\ma{init}} \ma{Br}[j\to 2S] \\
N^{\ma{init,fd}}_{1S} &=& N^{\ma{init}} + \frac{\sigma_{2S}}{\sigma_{1S}}N^{\ma{init,fd}}_{2S} \ma{Br}[2S \to 1S]  + 
\sum_{j = 1P, 3S, 2P}  \frac{\sigma_j}{\sigma_{1S}} N^{\ma{init}} \ma{Br}[j\to 1S] \\
N^{\ma{final,fd}}_{2S} &=& N^{\ma{final}}_{2S} + \sum_{j = 3S, 2P}  N^{\ma{final}}_j \ma{Br}[j\to 2S] \\
N^{\ma{final,fd}}_{1S} &=& N^{\ma{final}}_{1S} + N^{\ma{final,fd}}_{2S} \ma{Br}[2S \to 1S]  + 
\sum_{j = 1P, 3S, 2P}   N^{\ma{final}}_j \ma{Br}[j\to 1S]\,,
\ee
where we have used $N^{\ma{init}}_i=N^{\ma{init}}$ for all $i$'s. The branching ratios in vacuum are listed in Appendix~\ref{app:fd}. Finally the nuclear modification factors for $i=1S$, $2S$ are given by
\be
R_\ma{AA}(i) = \frac{N^{\ma{final,fd}}_i}{N^{\ma{init,fd}}_i}\,.
\ee
For $j=3S$, $1P$ and $2P$, we do not consider any feed-down contributions. Then their nuclear modification factors are given by
\be
R_\ma{AA}(j) = \frac{N_j^\ma{final}}{N^{\ma{init}}_j} = \frac{N_j^\ma{final}}{N^{\ma{init}}}\,.
\ee

\section{Results}
\label{sect:results}
\subsection{Uncertainty Estimates}
We will discuss three uncertainty sources here. The first source is the uncertainty in the EPPS16 parametrizations of the nPDF. The uncertainty bands of the CNM effect caused by the nPDF have been estimated in Section~\ref{sect:cnm} and will be included in the plots to be shown in the next subsection.

The second uncertainty originates in the values of $\alpha_s=0.3$ and $\alpha_s^{\ma{pot}}=0.36$. We will vary these two coupling constants by $\pm10\%$ to estimate the uncertainty caused by the parameter values. In the next subsection, we will show three curves for the $R_{\ma{AA}}$ calculation results, where the middle curve corresponds to the central values of the parameters $\alpha_s=0.3$ and $\alpha_s^{\ma{pot}}=0.36$. The lower curve corresponds to $\alpha_s=0.27$ and $\alpha_s^{\ma{pot}}=0.32$ while the upper one represents $\alpha_s=0.33$ and $\alpha_s^{\ma{pot}}=0.4$.

The final source of uncertainty is in the experimental measurements of cross sections and branching ratios, which are listed in Appendix~\ref{app:fd} and used as inputs in the calculations. We expect the first two uncertainty sources will dominate over the last one, so we exclude the last uncertainty source in the following analysis.

\subsection{Results for $\Upsilon(nS)$}
\begin{figure}
    \centering
    \begin{subfigure}[t]{0.45\textwidth}
        \centering
        \includegraphics[height=2.0in]{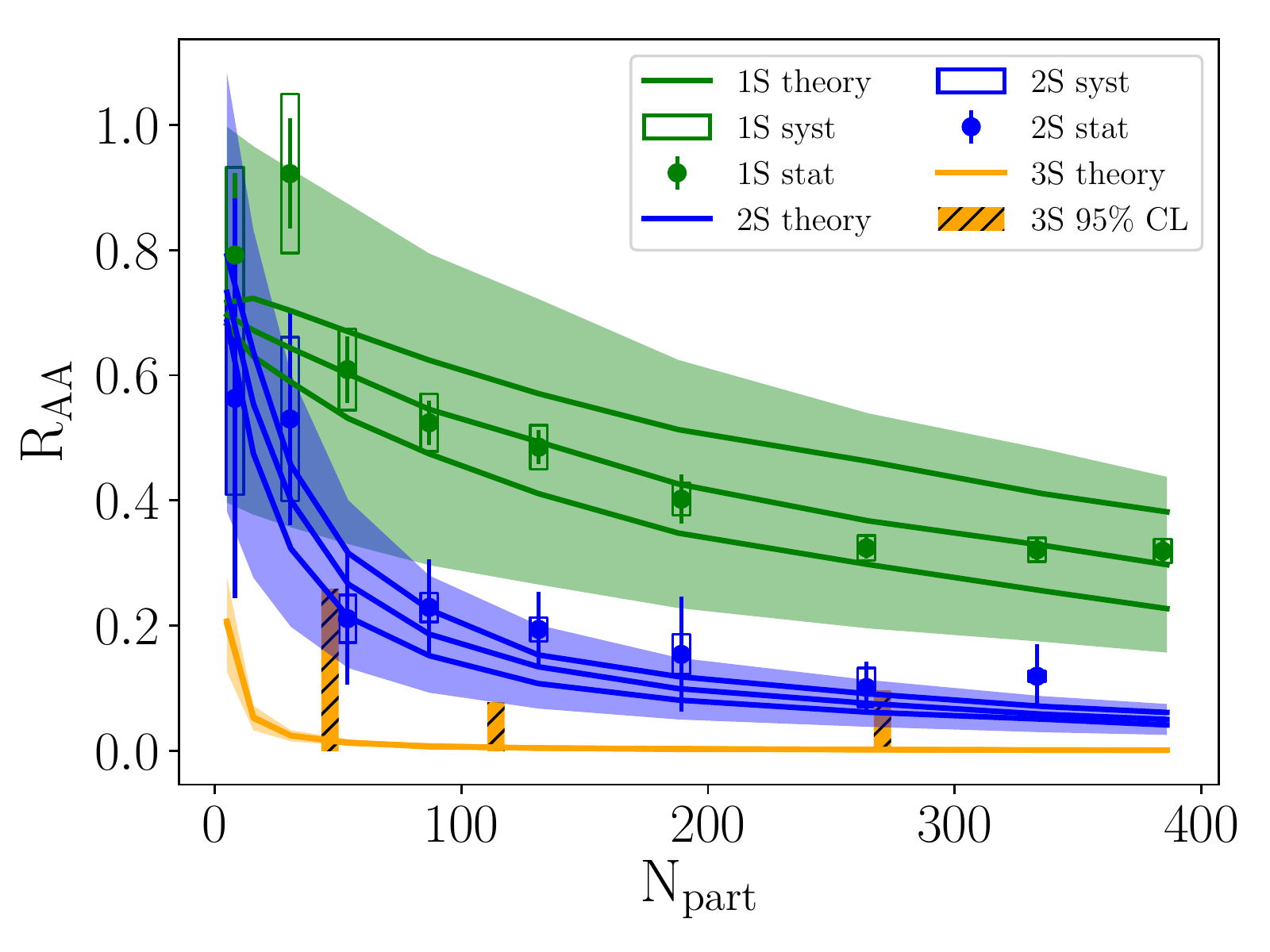}
        \caption{With cross-talk recombination.}
        \label{fig:502Npart_corr}
    \end{subfigure}%
    ~
    \begin{subfigure}[t]{0.45\textwidth}
        \centering
        \includegraphics[height=2.0in]{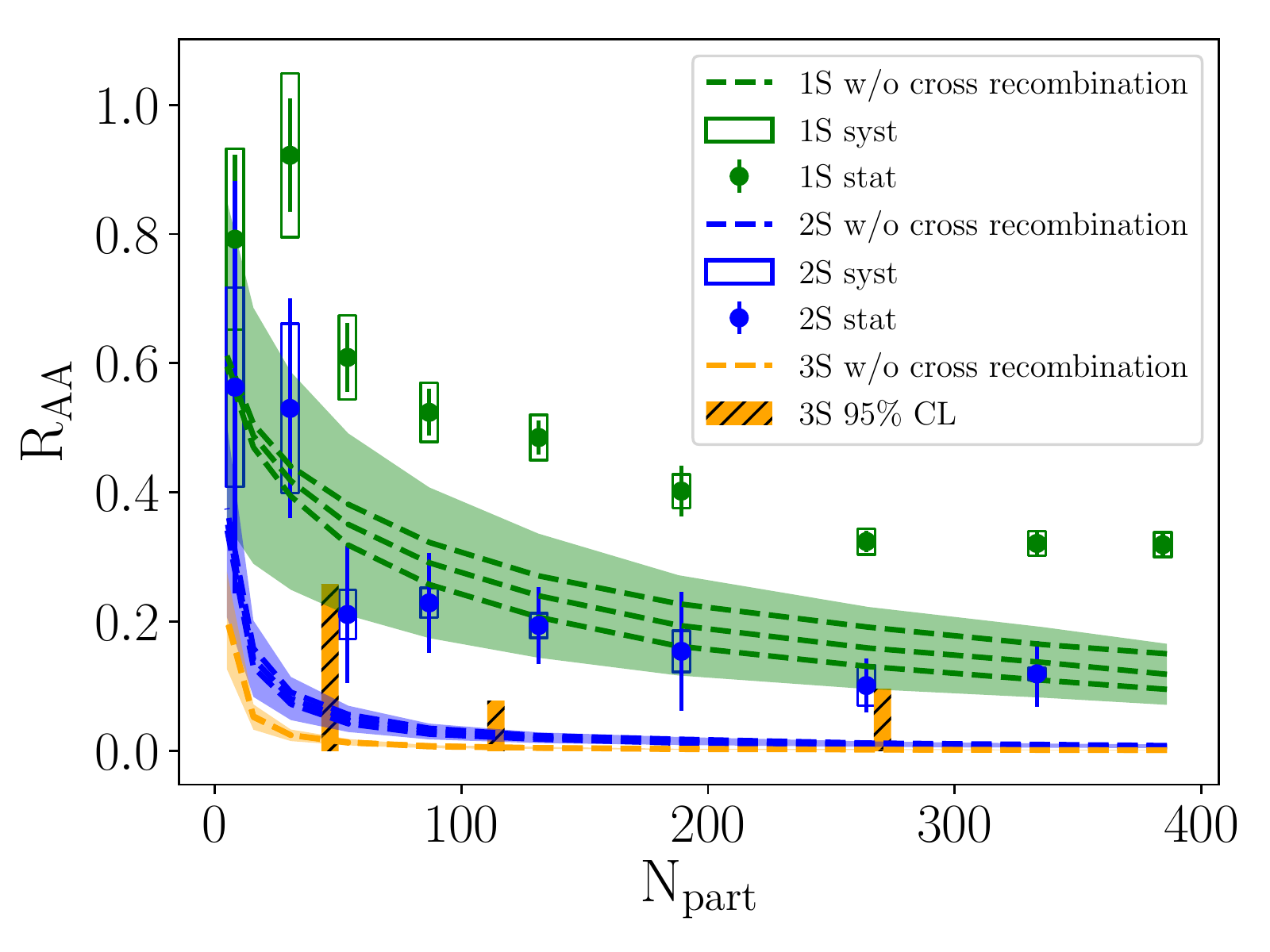}
        \caption{Without cross-talk recombination.}
        \label{fig:502Npart_nocorr}
    \end{subfigure}
    \caption{Bottomonia $R_\ma{AA}$ as functions of centrality at $5.02$ TeV Pb-Pb collisions. The upper and lower curves correspond to calculations with parameters that differ by $\pm10\%$ respectively from the parameters used in the middle curve. The band indicates the nPDF uncertainty that is centered at the middle curve. Experimental data are taken from Ref.~\cite{Sirunyan:2018nsz}.}
    \label{fig:502Npart}
\end{figure}

We first show the results of $R_\ma{AA}$ as a function of centrality at $5.02$ TeV Pb-Pb collisions. The comparison between our calculation and the measurements by the CMS collaboration is shown in Fig.~\ref{fig:502Npart}. The three curves in the plot correspond to the three sets of parameters, as discussed in the previous subsection. The uncertainty band is solely from the uncertainty of the nPDF and is centered at the middle curve, which corresponds to the central values of the parameters $\alpha_s=0.3$ and $\alpha_s^{\ma{pot}}=0.36$. As can be seen in the plot, the nPDF uncertainty dominates over the uncertainty of the parameters. To explicitly demonstrate the importance of correlated recombination, we show the full calculation results in Fig.~\ref{fig:502Npart_corr} and the results without any contribution from the cross-talk recombination in Fig.~\ref{fig:502Npart_nocorr}, i.e., in the latter case, the contribution from an initial $i$-quarkonium state ending up as a final $j$-quarkonium state ($j\neq i$) is excluded. Remaining same-species ($i\to i$) correlated recombination is still included in the latter case, though we expect its contribution to be much smaller than the cross-talk recombination. As can be seen from the comparison of the two figures, the cross-talk recombination is crucial to describe the data, even if one takes into account the uncertainties from the nPDF and parameter values. Furthermore, it seems unlikely to describe the data for $1S$ and $2S$ simultaneously without including the cross-talk recombination, by further increasing the coupling constant in the potential, since the change of $R_{\ma{AA}}(2S)$ from the parameter variation is tiny. For the $1S$ state production, most excited states dissociate quickly when entering the QGP because of the initial high temperature. These melted correlated $b\bar{b}$ pairs may form $1S$ state after the dissociation because the $1S$ state can exist at high temperature. For the $2S$ production, most of the primordially produced $2S$ states cannot survive the in-medium evolution in most centrality bins. They are regenerated when the temperature cools down and allows their existence in the medium. At low temperature, a $2S$ state can be formed from a correlated unbound $b\bar{b}$ pair (which may come from a dissociated quarkonium state such as a $1S$ state). For the $3S$ production, since we assume they cannot exist inside QGP (melting temperature is below $154$ MeV), correlated recombination does not affect its production. The $3S$ state is only produced in very peripheral collisions due to the corona effect. 

\begin{figure}
    \centering
    \begin{subfigure}[t]{0.45\textwidth}
        \centering
        \includegraphics[height=2.0in]{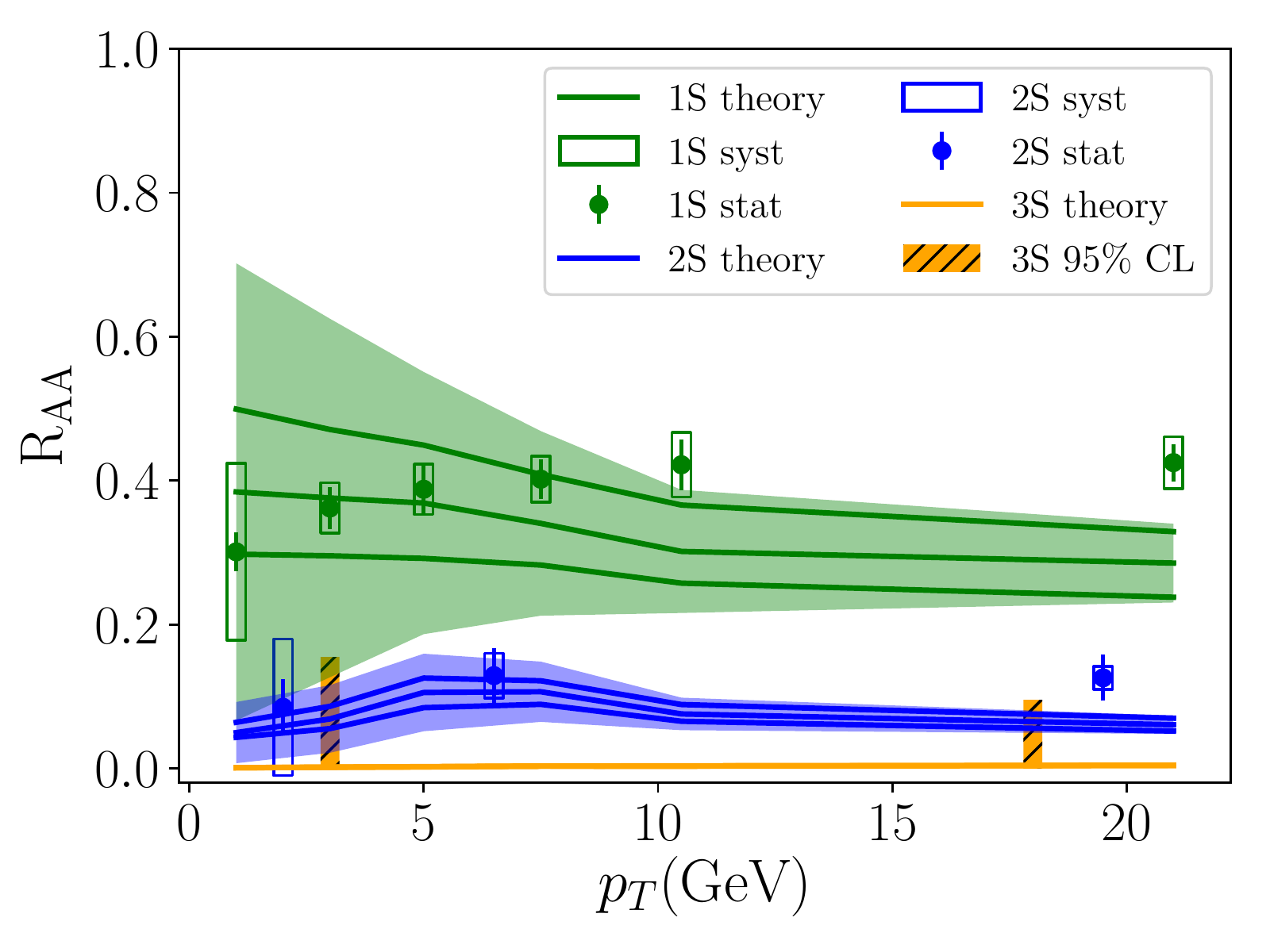}
        \caption{$R_\ma{AA}(p_T)$ with CNM effects and with cross-talk recombination.}
        \label{fig:502pTa}
    \end{subfigure}%
    ~
    \begin{subfigure}[t]{0.45\textwidth}
        \centering
        \includegraphics[height=2.0in]{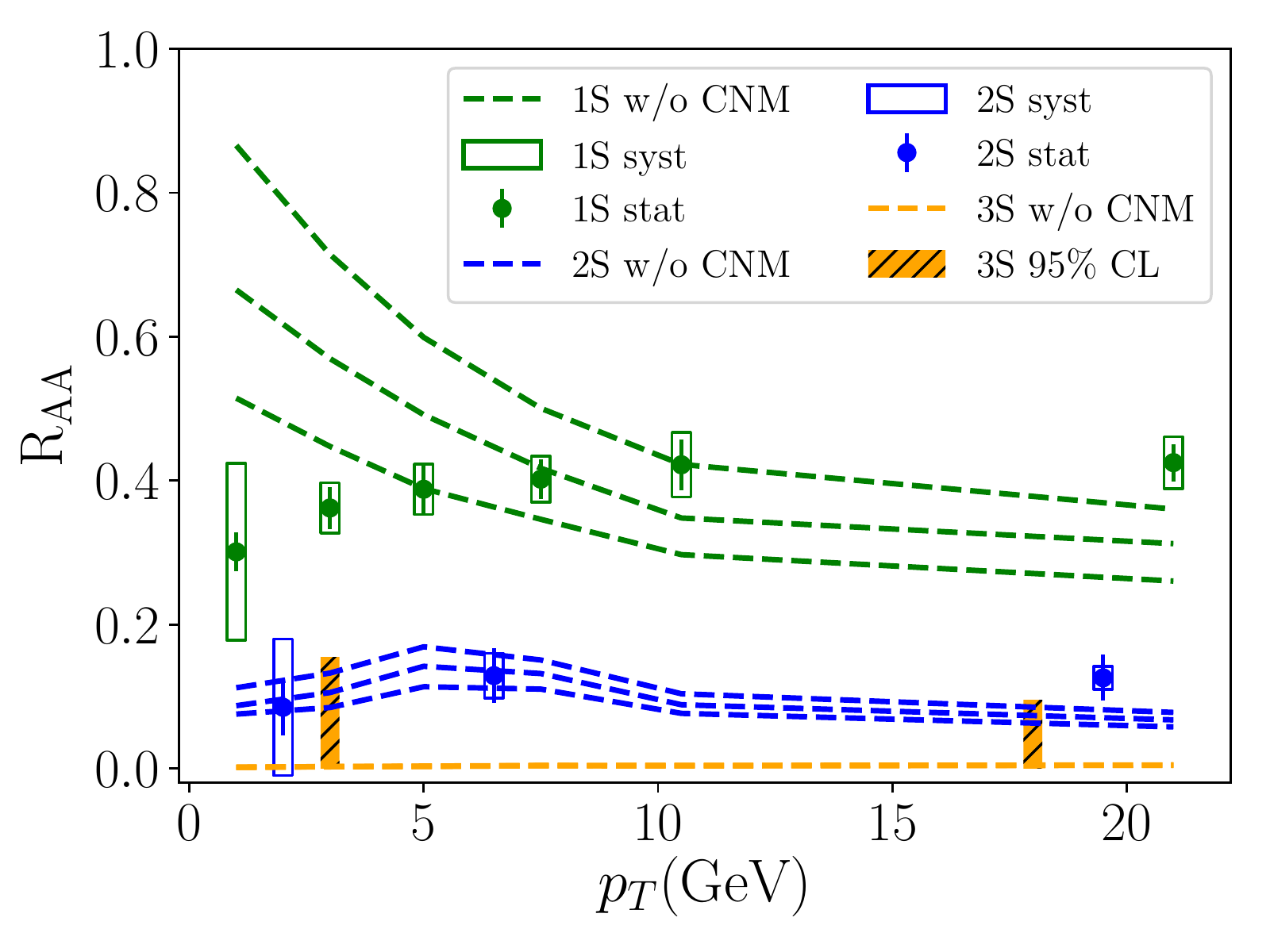}
        \caption{$R_\ma{AA}(p_T)$ without CNM effects but with cross-talk recombination.}
        \label{fig:502pTb}
    \end{subfigure}
    
    \begin{subfigure}[t]{0.45\textwidth}
        \centering
        \includegraphics[height=2.0in]{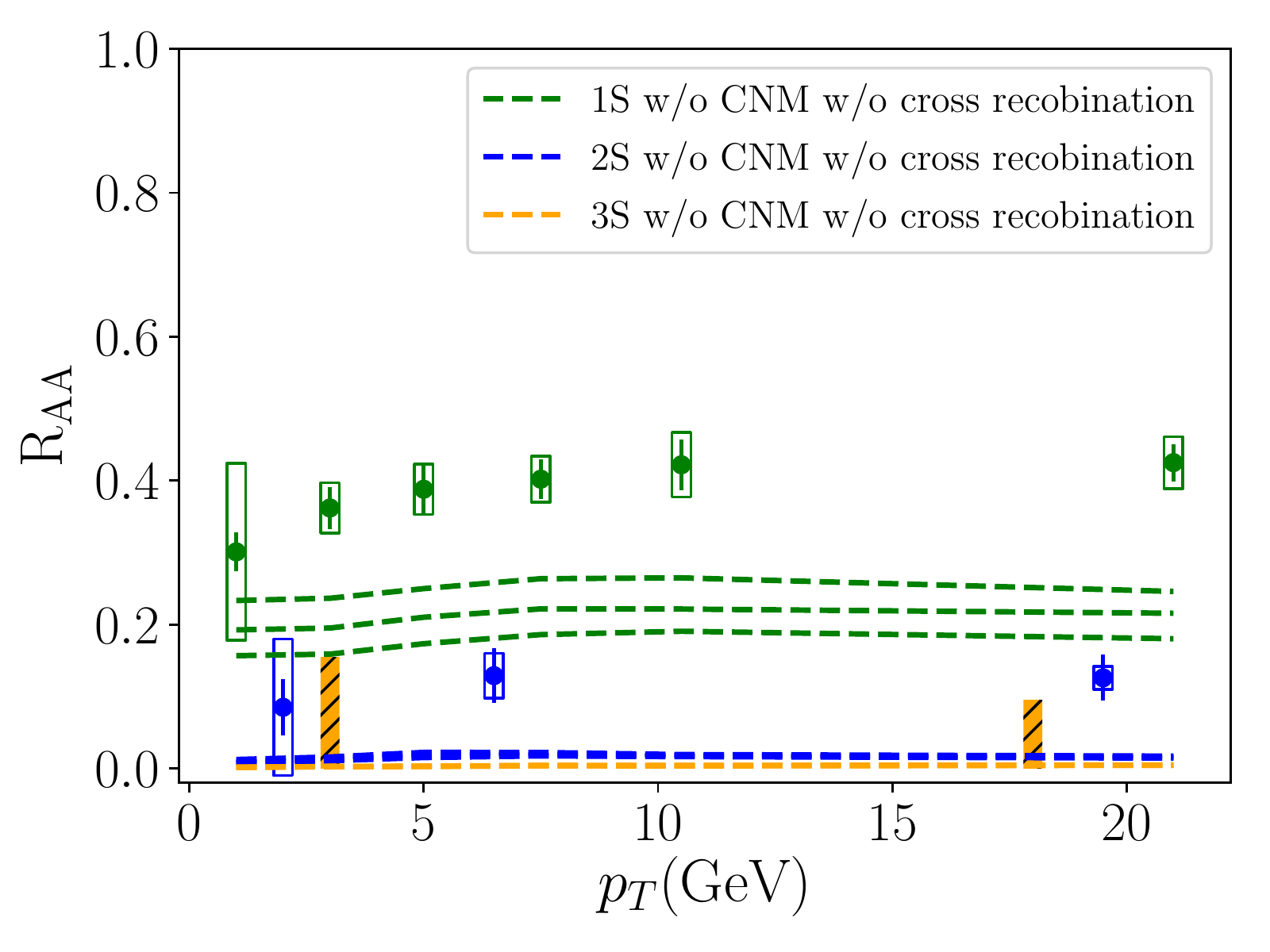}
        \caption{$R_\ma{AA}(p_T)$ without CNM effects and without cross-talk recombination.}
        \label{fig:502pTc}
    \end{subfigure}
    ~
    \begin{subfigure}[t]{0.45\textwidth}
        \centering
        \includegraphics[height=2.0in]{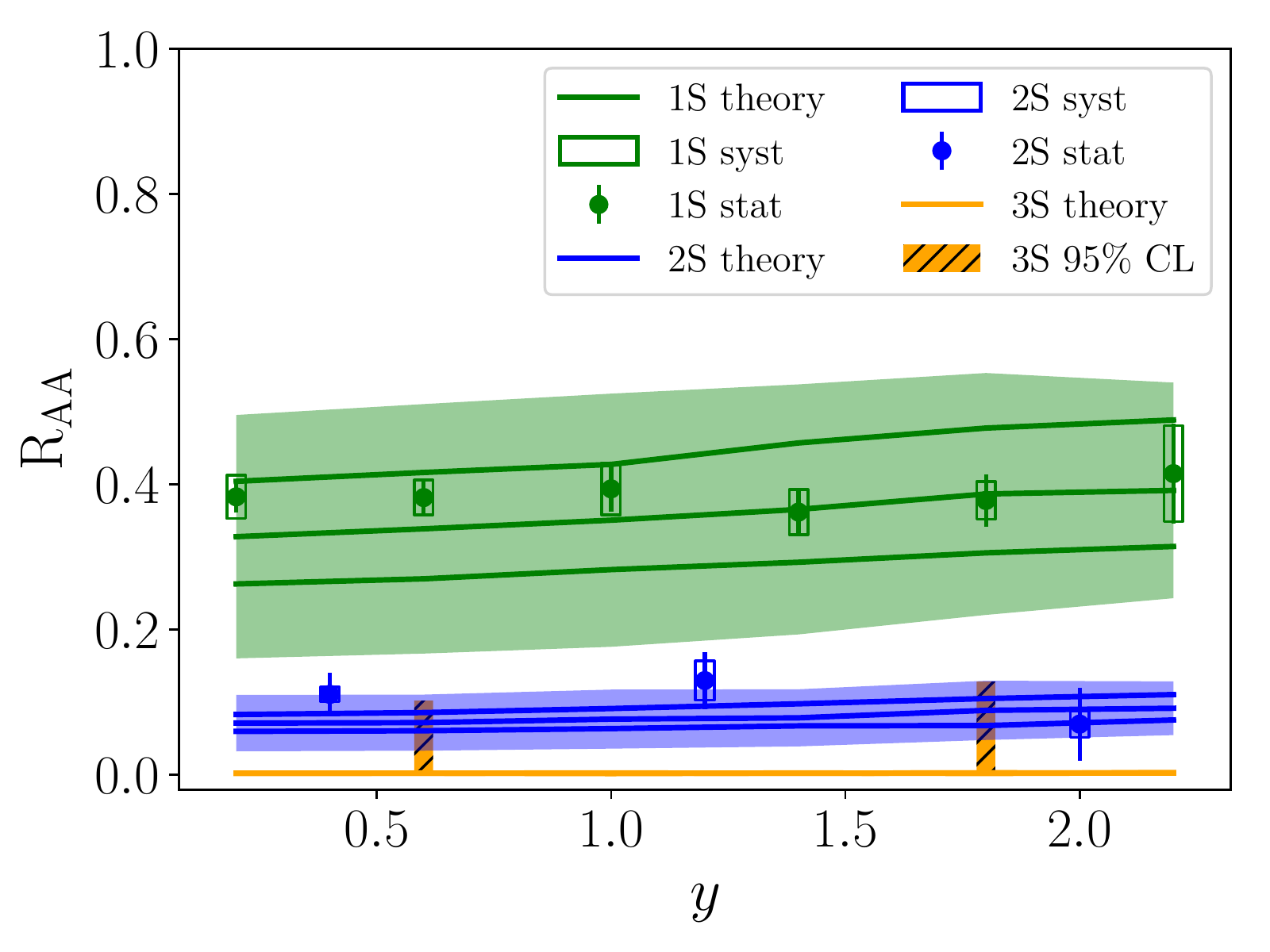}
        \caption{$R_\ma{AA}(y)$ with CNM effects and with cross-talk recombination.}
        \label{fig:502y}
    \end{subfigure}
    \caption{Bottomonia $R_\ma{AA}$ as functions of transverse momentum and rapidity at $5.02$ TeV Pb-Pb collisions. The upper and lower curves correspond to calculations with parameters that differ by $\pm10\%$ respectively from the parameters used in the middle curve. The band indicates the nPDF uncertainty that is centered at the middle curve. Experimental data are taken from Ref.~\cite{Sirunyan:2018nsz}.}
\label{fig:502pTy}
\end{figure}

Next we discuss the results of $R_\ma{AA}$ as a function of the transverse momentum, shown in Fig.~\ref{fig:502pTa}. First we notice that the experimental results of $R_\ma{AA}$ are almost flat as a function of $p_T$. This is a highly non-trivial result because the CNM effect has a dramatic dependence on $p_T$, as shown by the central values in Fig.~\ref{fig:cnm}, though the uncertainty band is quite large. The $p_T$ variation of the CNM effect can be understood as follows: At mid-rapidity, the energy fraction carried by the two gluons that fuse to produce quarkonium (in the collinear factorization, for a non-zero $p_T$, a recoil particle is produced back-to-back to the quarkonium state in the transverse plane), is given by $x=\frac{2m_T}{\sqrt{s}}$ where $m_T=\sqrt{p_T^2+M_H^2}$. Bottomonium mass is about $M_H\approx10$ GeV. So at $5.02$ TeV Pb-Pb collisions, $x$ goes from $0.004$ to $0.009$ as $p_T$ increases from $0$ to $20$ GeV. As $x$ increases, the nuclear modification factor $R_g(x)$ of the gluon PDF increases. The CNM effect on the production cross section grows as $[R_g(x)]^2$. This growing trend is true for every fitting set in the EPPS16. Furthermore, the scale $m_T$ used in the perturbative calculation also increases as $p_T$ increases. At higher energy scales, the nPDF modification factor $R_g$ is expected to be closer to unity, which further increases its $p_T$ dependence.

The final flat $R_{\ma{AA}}$ means for bottomonium, the $p_T$-dependent CNM effect is washed out by the hot medium effect. To show this more explicitly, we plot the calculation results of $R_{\ma{AA}}$ without the CNM effects in Fig.~\ref{fig:502pTb}. We see the hot medium effect leads to a raise of $R_{\ma{AA}}$ at low transverse momentum. The reason behind is related with correlated recombination: Suppose a quarkonium state dissociates and then recombines later (via correlated recombination). After dissociation but before recombination, the correlated $Q\bar{Q}$ pair interacts with the medium and loses energy. When the pair recombines, it has a smaller $p_T$ than that right after the dissociation. This leads to an increase of quarkonium states at low $p_T$. Furthermore, low $p_T$ quarkonium states stay inside the medium for a longer time and probably go through the low temperature part of the QGP expansion. Since recombination at high temperature is ineffective, quarkonium states at low $p_T$ have a higher chance to recombine (via correlated recombination) if they dissociate than those at high $p_T$. To demonstrate this point more clearly, we plot the calculation results without any CNM effects and without any cross-talk recombination in Fig.~\ref{fig:502pTc}, where the suppression has only a mild $p_T$ dependence.

Our calculations can describe the $R_\ma{AA}$ for $p_T<10$ GeV and start to deviate from the data as $p_T$ becomes bigger. This may indicate the breakdown of the nonrelativistic expansion that our calculations are heavily replied on. For a finite $p_T$, the typical energy of medium light particles in the rest frame of quarkonium is given by $\frac{T}{\sqrt{1-v_T^2}}$ rather than $T$, where $v_T$ is the transverse velocity of the quarkonium with respect to the local medium. Our calculations are valid if $\frac{rT}{\sqrt{1-v_T^2}} \ll 1$ where $r\sim\frac{1}{Mv}$ is the typical size of quarkonium. Under our assumed hierarchy $rT\ll 1$, the condition $\frac{rT}{\sqrt{1-v_T^2}} \ll 1$ is valid when $v_T$ is small. Higher order contributions in the nonrelativistic expansion become important and have to be included consistently as $p_T$ increases\footnote{Under our assumed hierarchy, the multipole expansion is equivalent to the nonrelativistic expansion. One should keep in mind that the dipole interaction vertex grows linearly with the quarkonium size and thus the reaction rate grows with the size. But physically we know there is an upper limit of the rate when the size is large: The rate cannot be bigger than twice the reaction rate of one heavy quark interacting with the medium. When the quarkonium size is large enough, the heavy quark and antiquark interact with the medium almost independently. Therefore, we expect that including higher order terms in the nonrelativistic expansion will reduce the rate and thus the $R_{\ma{AA}}$ at large $p_T$ will go up.}. Our calculations probably indicate that this happens when $p_T>10$ GeV. One should note that for charmonium this can happen at a much smaller $p_T$ because what matters in the argument is the transverse velocity rather than the transverse momentum. Since most bottomonia are produced at low transverse momentum, the deviation at mid and high $p_T$ does not affect the centrality and rapidity dependence of $R_\ma{AA}$ in our calculations. 

Then we show the results of $R_\ma{AA}$ as a function of the rapidity in Fig.~\ref{fig:502y}. Our calculations are consistent with the experimental measurements. Since we use a $2+1$ dimensional viscous hydrodynamics, the hot medium effect is independent of $y$. Furthermore, we see in Fig.~\ref{fig:cnm} that the $y$ dependence of the CNM effect is also mild. Therefore, the final $R_\ma{AA}$ is almost flat in the rapidity.

\begin{figure}
    \centering
    \begin{subfigure}[t]{0.45\textwidth}
        \centering
        \includegraphics[height=2.0in]{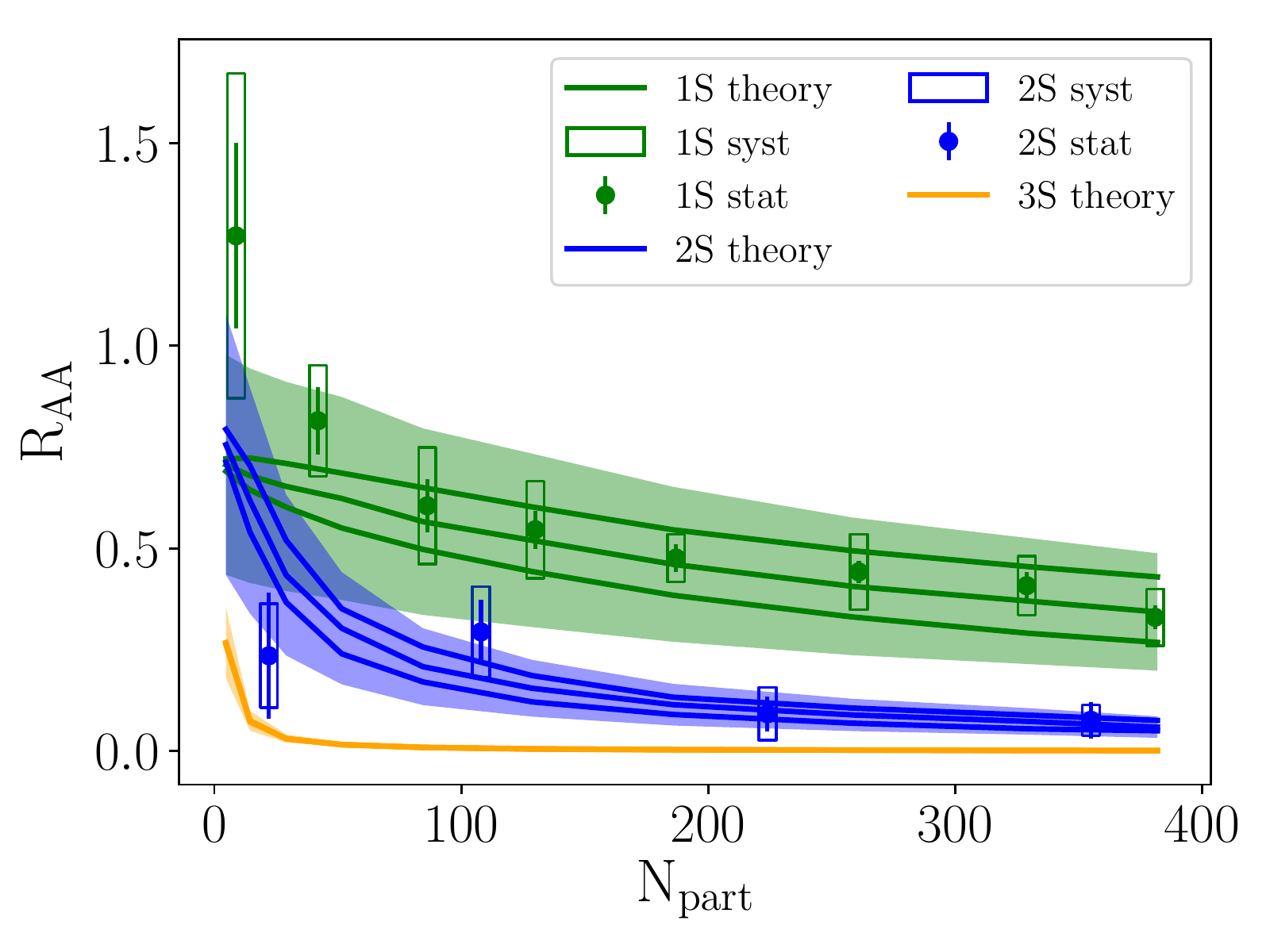}
        \caption{$R_\ma{AA}(N_\ma{part})$ at $2.76$ TeV Pb-Pb collisions.}
    \end{subfigure}%
    ~
    \centering
    \begin{subfigure}[t]{0.45\textwidth}
        \centering
        \includegraphics[height=2.0in]{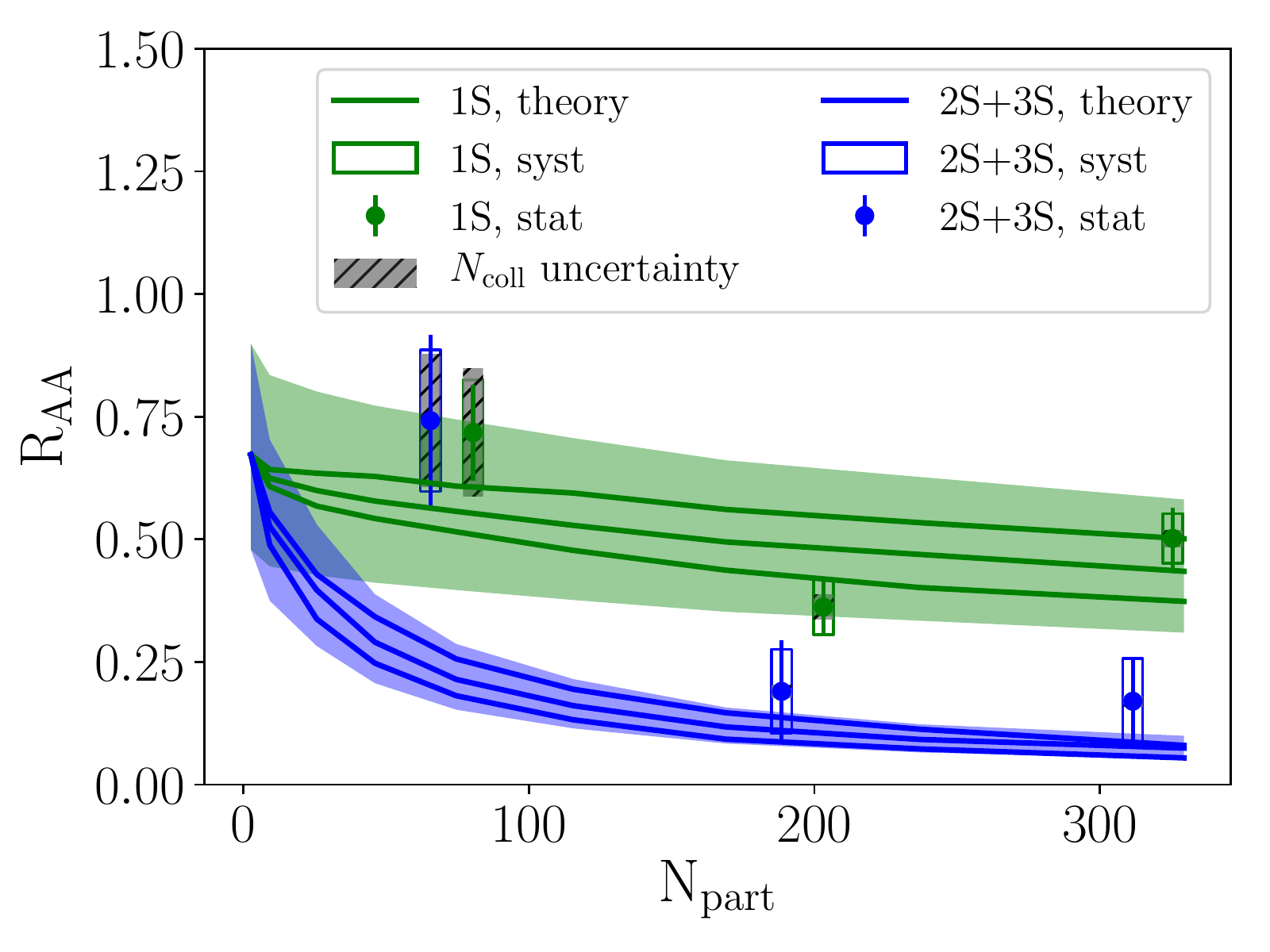}
        \caption{$R_\ma{AA}(N_\ma{part})$ at $200$ GeV Au-Au collisions.}
        \label{fig:star_Npart}
    \end{subfigure}%
    
    \begin{subfigure}[t]{0.45\textwidth}
        \centering
        \includegraphics[height=2.0in]{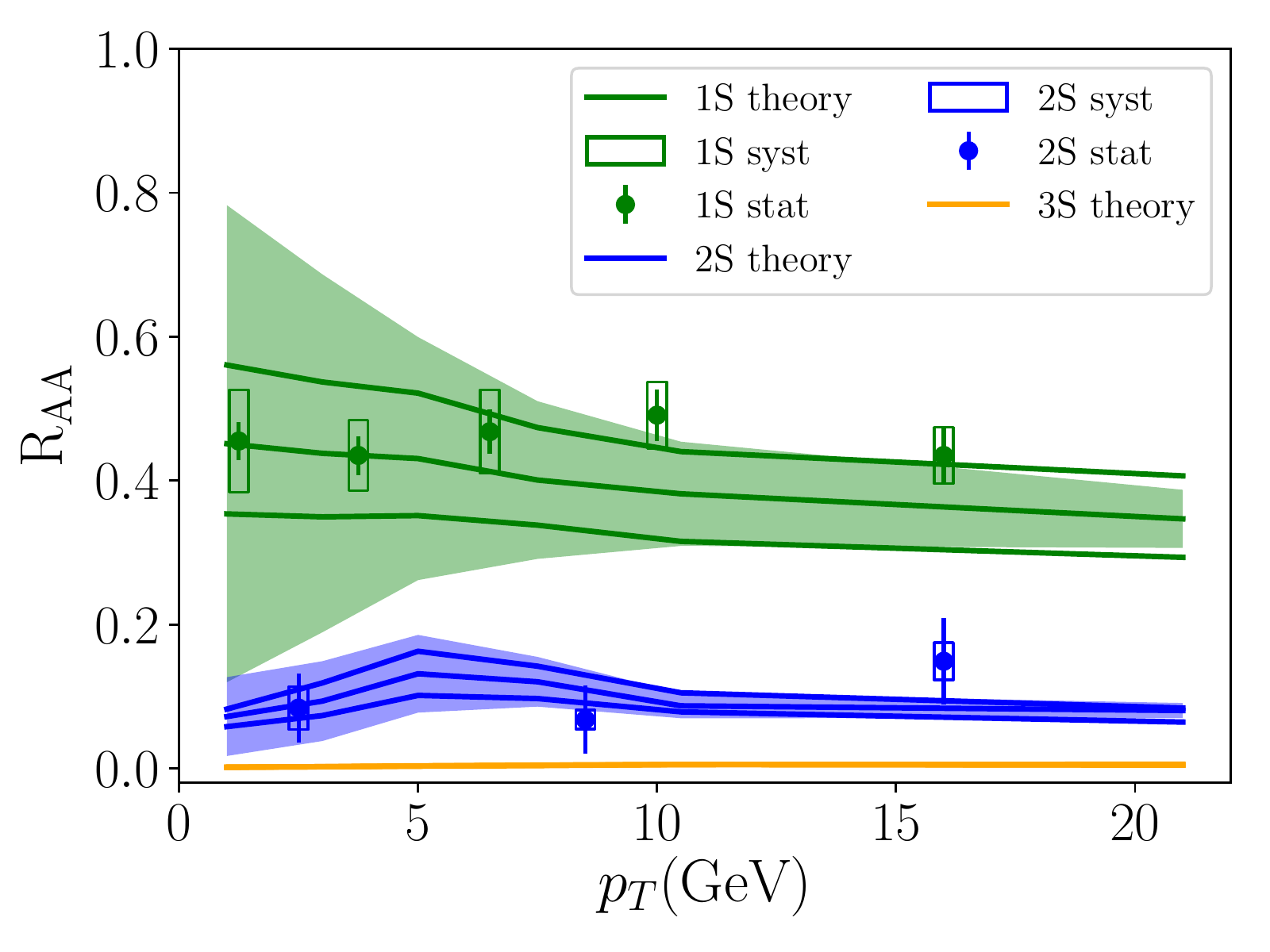}
        \caption{$R_\ma{AA}(p_T)$ at $2.76$ TeV Pb-Pb collisions.}
    \end{subfigure}
    ~
    \begin{subfigure}[t]{0.45\textwidth}
    \centering
    \includegraphics[height=2.0in]{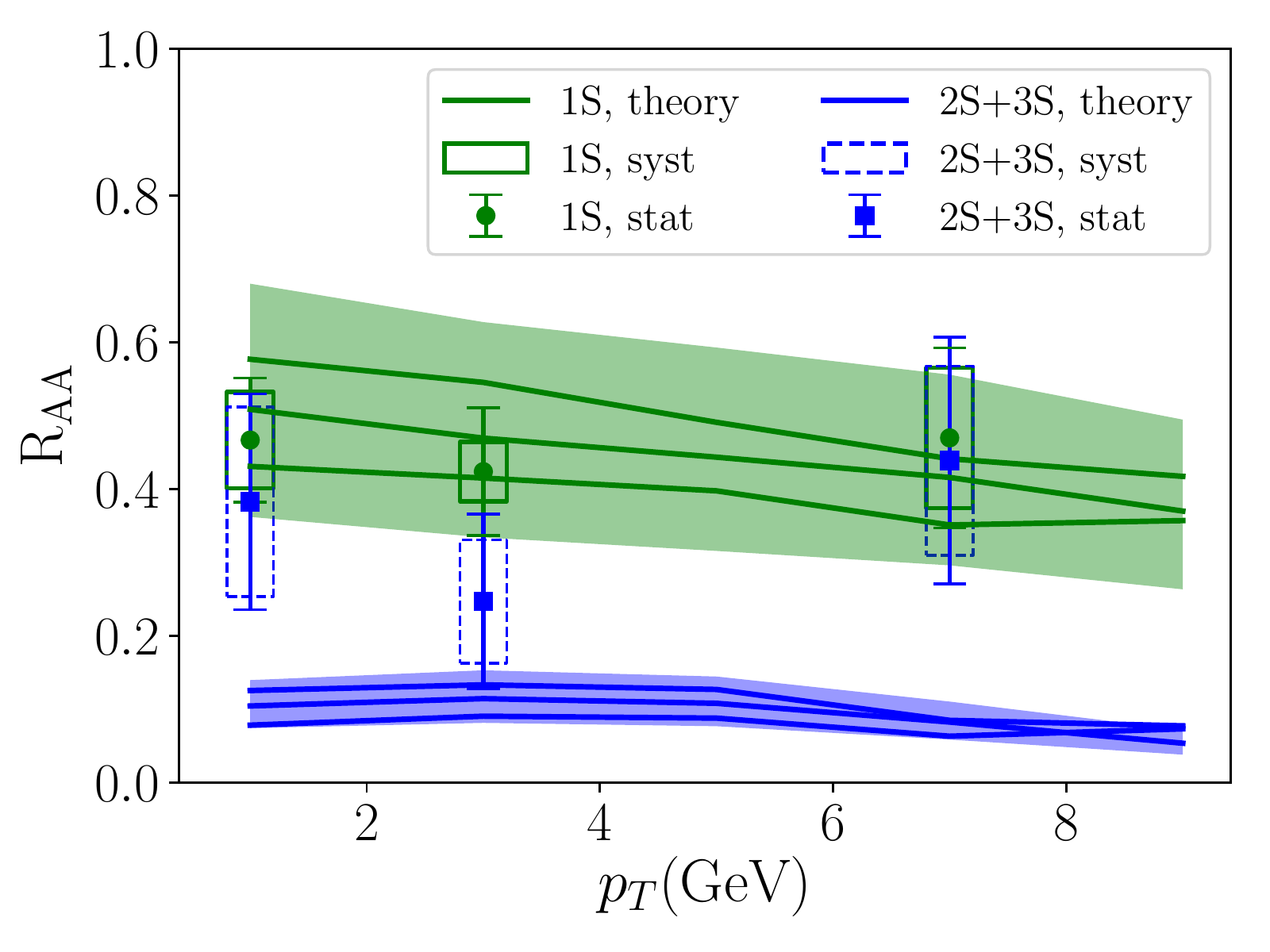}
    \caption{$R_\ma{AA}(p_T)$ at $200$ GeV Au-Au collisions.}
    \label{fig:star_pT}
    \end{subfigure}  

    \begin{subfigure}[t]{0.45\textwidth}
    \centering
    \includegraphics[height=2.0in]{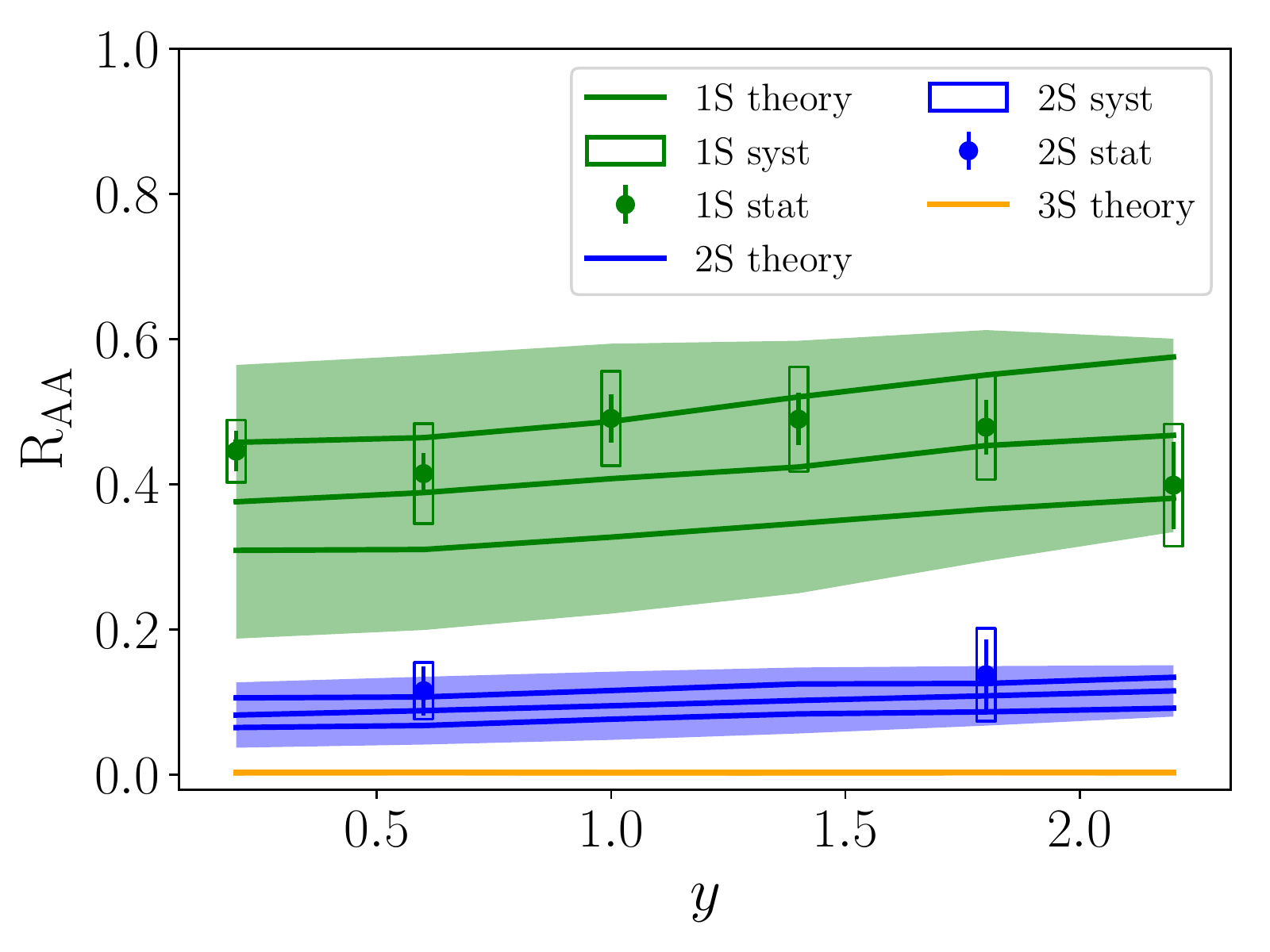}
    \caption{$R_\ma{AA}(y)$ at $2.76$ TeV Pb-Pb collisions.}
    \end{subfigure}
\caption{Bottomonia $R_\ma{AA}$ as functions of centrality, transverse momentum and rapidity at $2.76$ TeV Pb-Pb and $200$ GeV Au-Au collisions. The upper and lower curves correspond to calculations with parameters that differ by $\pm10\%$ respectively from the parameters used in the middle curve. The band indicates the nPDF uncertainty that is centered at the middle curve. Data of the CMS and STAR measurements are taken from Refs.~\cite{Khachatryan:2016xxp} and \cite{Wang:2019vau} respectively. The $p_T$ dependence of $R_{\ma{AA}}$ for the $200$ GeV Au-Au collisions is calculated and compared for the centrality range $0-60\%$.}
\label{fig:2760_200}
\end{figure}

The comparison between our calculations and experimental measurements for $2.76$ TeV Pb-Pb collisions and $200$ GeV Au-Au collisions is shown in Fig.~\ref{fig:2760_200}. Our calculations can describe most of the experimental data except $R_\ma{AA}(2S+3S)$ in peripheral Au-Au collisions at $200$ GeV. In Fig.~\ref{fig:star_Npart}, our calculations of $R_\ma{AA}(2S+3S)$ are consistent with the measurements at central collisions, though are slightly lower than the central values. But our calculation result has a large discrepancy with the data point at the peripheral collision. The uncertainty of the experimental measurement there is quite large due to the large uncertainty in the determination of $N_{\ma{coll}}$. The uncertainty associated with $N_{\ma{coll}}$ is small for central collisions. This discrepancy with the single data point in peripheral collisions leads to the discrepancy in the $p_T$ dependence of $R_\ma{AA}(2S+3S)$, as depicted in Fig.~\ref{fig:star_pT}. We expect that the future sPHENIX collaboration will provide data with high precision and then the origin of the current discrepancy issue will be more clear: Either the calculations are consistent with improved measurements or the calculations need improving. Improvements of the calculations can be carried out in understanding the CNM effects at $200$ GeV Au-Au collisions. Since we assume nPDF is the only CNM effect, we use a constant $R_\ma{CNM} = 0.67$ for all centrality bins. It is possible that the real CNM effect is centrality-dependent and $R_\ma{CNM}$ goes up as the collision becomes more peripheral. An increase of $R_\ma{CNM}$ in peripheral collisions will lessen the discrepancy we see here. 

\begin{figure}
\centering
\includegraphics[width=0.67\textwidth]{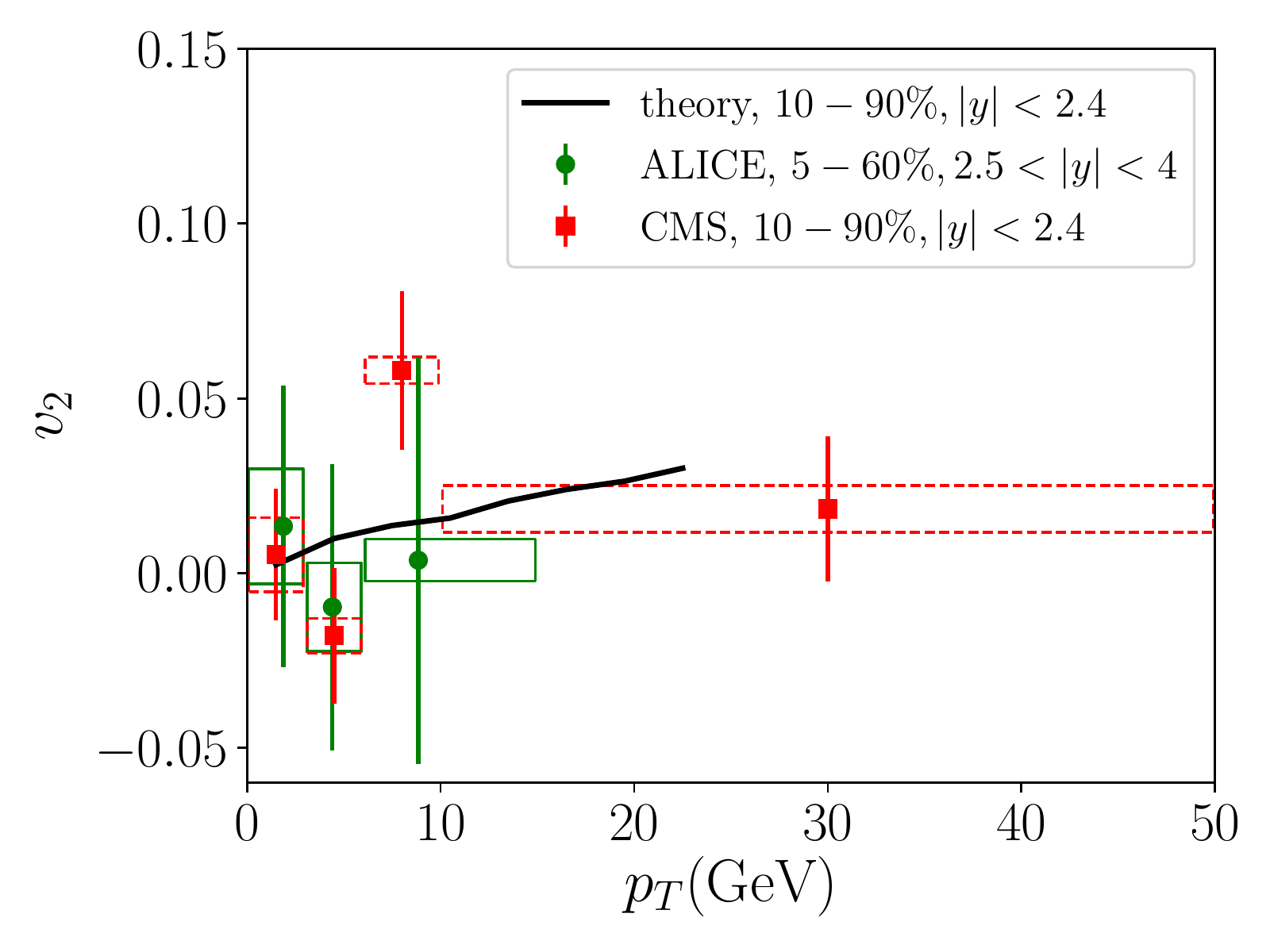}
\caption{Azimuthal angular anisotropy coefficient $v_2$ of $\Upsilon(1S)$ at $5.02$ TeV Pb-Pb collisions. The experimental results of the ALICE and CMS collaborations are taken from Refs.~\cite{Acharya:2019hlv} and~\cite{Sirunyan:2020qec}.}
\label{fig:v2}
\end{figure}

Finally, we study the azimuthal angular anisotropy of $\Upsilon(1S)$. In particular, we compute the $v_2$ coefficient which is defined by
\be
E\frac{\diff^3N_{1S}}{\diff p^3} =\frac{1}{2\pi} \frac{\diff^2N_{1S}}{p_T\diff p_T\diff y}\big(1+2v_2\cos[2(\phi-\Psi_\ma{RP})]\big)\,,
\ee
in which we neglect higher order harmonics. Here $\phi-\Psi_\ma{RP}$ is the azimuthal angle of the $1S$ state with respect to the reaction plane, which is defined event-by-event. Our calculation result for the $5.02$ Pb-Pb collision in the centrality range\footnote{For the $v_2$ observable, the CMS collaboration defines the centrality using hard probe triggers. So in the $v_2$ calculation, we define the centrality bin in \trento\ by using the number of binary collisions $N_{\ma{coll}}$ rather than the multiplicity.} $10-90\%$ is shown in Fig.~\ref{fig:v2}. We only show the result from the central values of the parameters. We also plot the experimental results measured by the ALICE collaboration which are in a different rapidity and centrality regions, just to show the state-of-art of the current bottomonium $v_2$ measurements. We stop our calculations at $p_T=24$ GeV because as we have seen in the $R_{\ma{AA}}$ comparison, higher order corrections in the nonrelativistic expansion start to become important as $p_T$ increases, which are neglected in our current setup. We can calculate $v_2$ at higher $p_T$ values in our current setup but their physical meaning is less robust and we cannot learn much from doing that. Our calculation result is consistent with the current experimental data, though current measurements have large statistical errorbars. The last CMS data point has a quite large $p_T$ range: $10-50$ GeV. Our nonrelativistic expansion calculation would definitely break down at $50$ GeV. Furthermore, at such a high $p_T$, the fragmentation production would start to dominate and the suppression mechanism would mainly be jet energy loss. Due to the steep $p_T$ spectrum in the primordial production of bottomonium, most of the contribution to the last data point comes from $p_T \sim 10-20$ GeV. So the comparison we show still has some physical meaning for the last data point.

Several physical processes contribute to the development of the quarkonium $v_2$. The first contribution is the path dependence. In peripheral collisions, the QGP has a elliptic shape. Quarkonia moving along the longer axis will be more suppressed. Also, the reaction rates of quarkonium in the medium depend on the relative velocity between the quarkonium and the local medium, which has a flow velocity. This also influences the shape of $v_2$ as a function of $p_T$. Finally, after the dissociation of quarkonium, the unbound $Q\bar{Q}$ pair can develop some $v_2$ by interacting with the medium. Later if they recombine, the $v_2$ will be partly or fully inherited by the regenerated quarkonium. Uncorrelated recombination can also contribute to $v_2$ and is crucially important for charmonium production in heavy-ion collision. Open charm quarks that develop $v_2$ during the in-medium evolution will contribute to the charmonium $v_2$ if they recombine. But this $v_2$ generation mechanism is negligible for bottomonium since the number of unbound $b\bar{b}$ pairs produced in one collision event is smaller than one per rapidity in mid-rapidity. Future precise measurements on the azimuthal angular anisotropy will greatly help us understand the in-medium dynamics of quarkonium, especially how quarkonia with finite transverse momenta interact with an expanding medium. These non-equilibrium transport properties of quarkonium are not easy to study via finite temperature lattice QCD calculations. The interplay between theory and phenomenology will help deepen our understanding on these, in particular once experimental data with high precision are available.

\subsection{Prediction: $\chi_b(1P)$ More Suppressed than $\Upsilon(2S)$}
One consequence of the important contribution from correlated recombination in bottomonium production would be that the $\chi_b(1P)$ state is more suppressed than the $\Upsilon(2S)$ state. If there were no correlated recombination, one would expect $R_{\ma{AA}}(\chi_b(1P))$ to be similar to $R_{\ma{AA}}(\Upsilon(2S))$, since the two states have similar binding energies and sizes. However, correlated recombination can alter dramatically this naive expectation. Since $\chi_b(1P)$ and $\Upsilon(2S)$ have similar binding energies and sizes, their recombination rates from a correlated $b\bar{b}$ pair are also close. In particular, the probability of an initial $\chi_b(1P)$ state ending up as a $\Upsilon(2S)$ state (via first dissociation and then correlated recombination) is similar to that of an initial $\Upsilon(2S)$ state ending up as a $\chi_b(1P)$ state. But the primordial production cross section of $\chi_b(1P)$ is $4-5$ times that of $\Upsilon(2S)$. Much more $\chi_b(1P)$ states are produced initially and thus, the number of $\Upsilon(2S)$ states regenerated from initial $\chi_b(1P)$ states is much larger than the number of $\chi_b(1P)$ states regenerated from initial $\Upsilon(2S)$ states. Therefore, $\Upsilon(2S)$ is less suppressed than $\chi_b(1P)$.

\begin{figure}
    \centering
    \begin{subfigure}[t]{0.45\textwidth}
        \centering
        \includegraphics[height=2.0in]{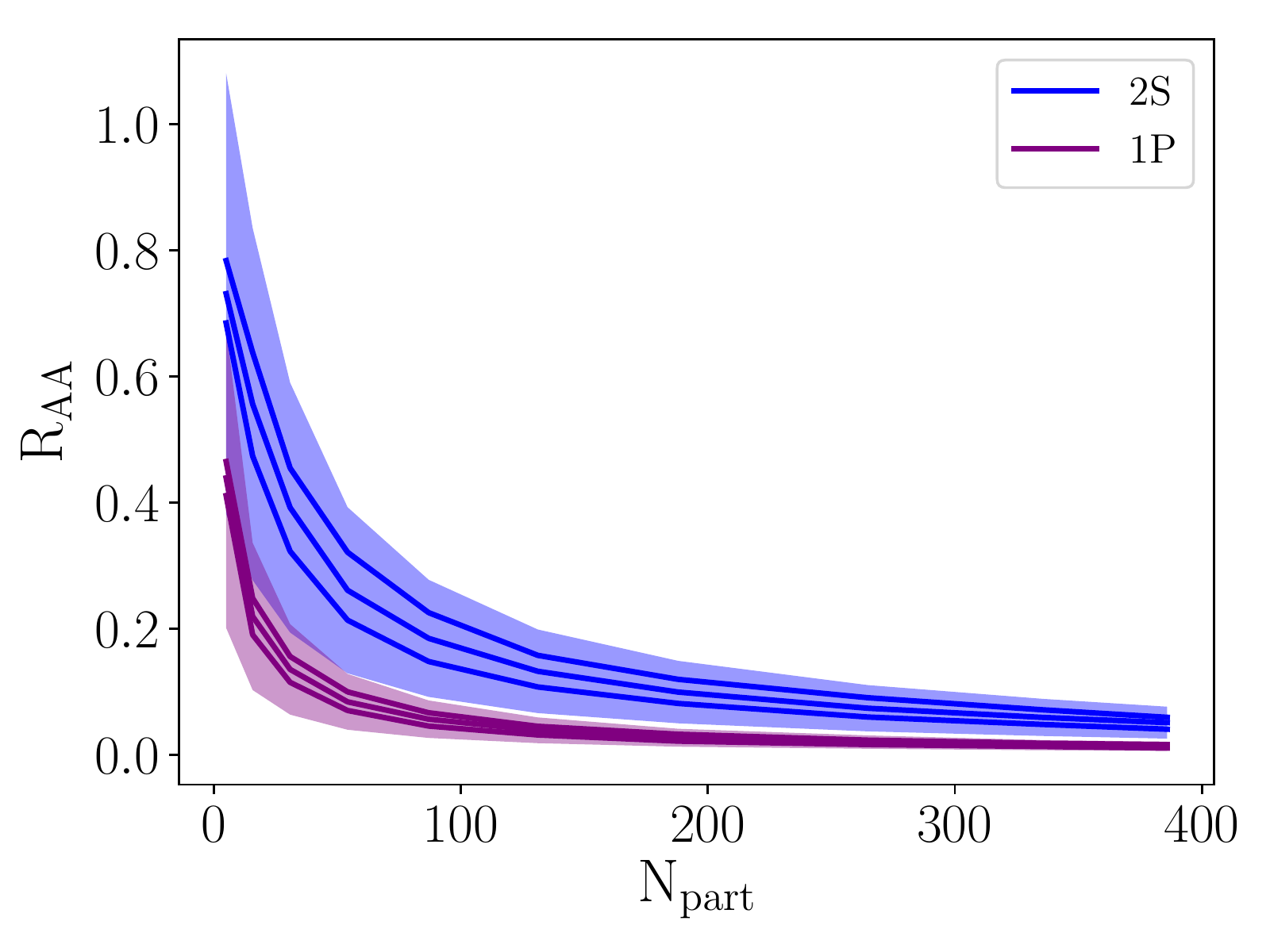}
        \caption{With cross-talk recombination.}
    \end{subfigure}%
    ~
    \centering
    \begin{subfigure}[t]{0.45\textwidth}
        \centering
        \includegraphics[height=2.0in]{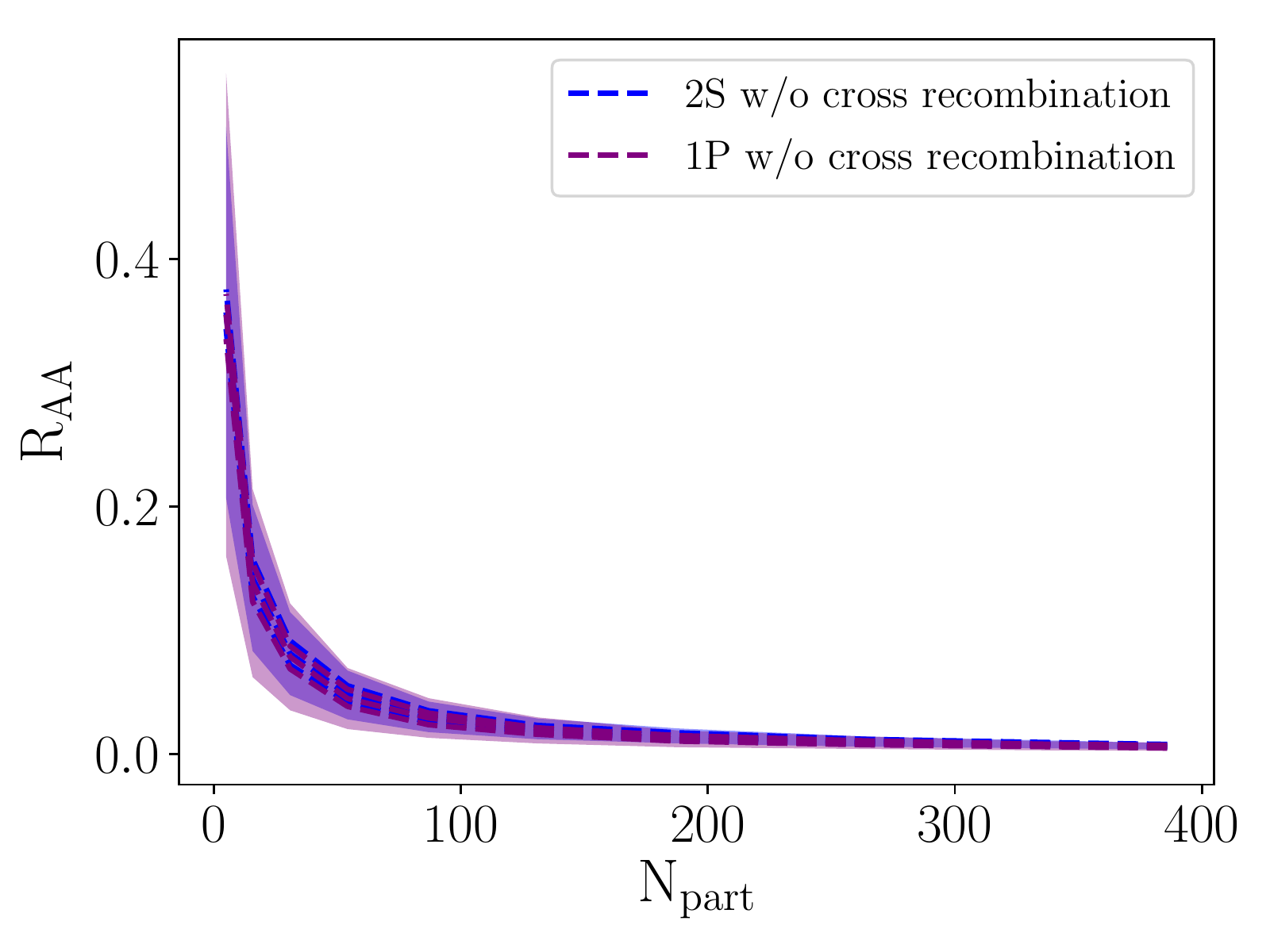}
        \caption{Without cross-talk recombination.}
    \end{subfigure}%
    
    \begin{subfigure}[t]{0.45\textwidth}
        \centering
        \includegraphics[height=2.0in]{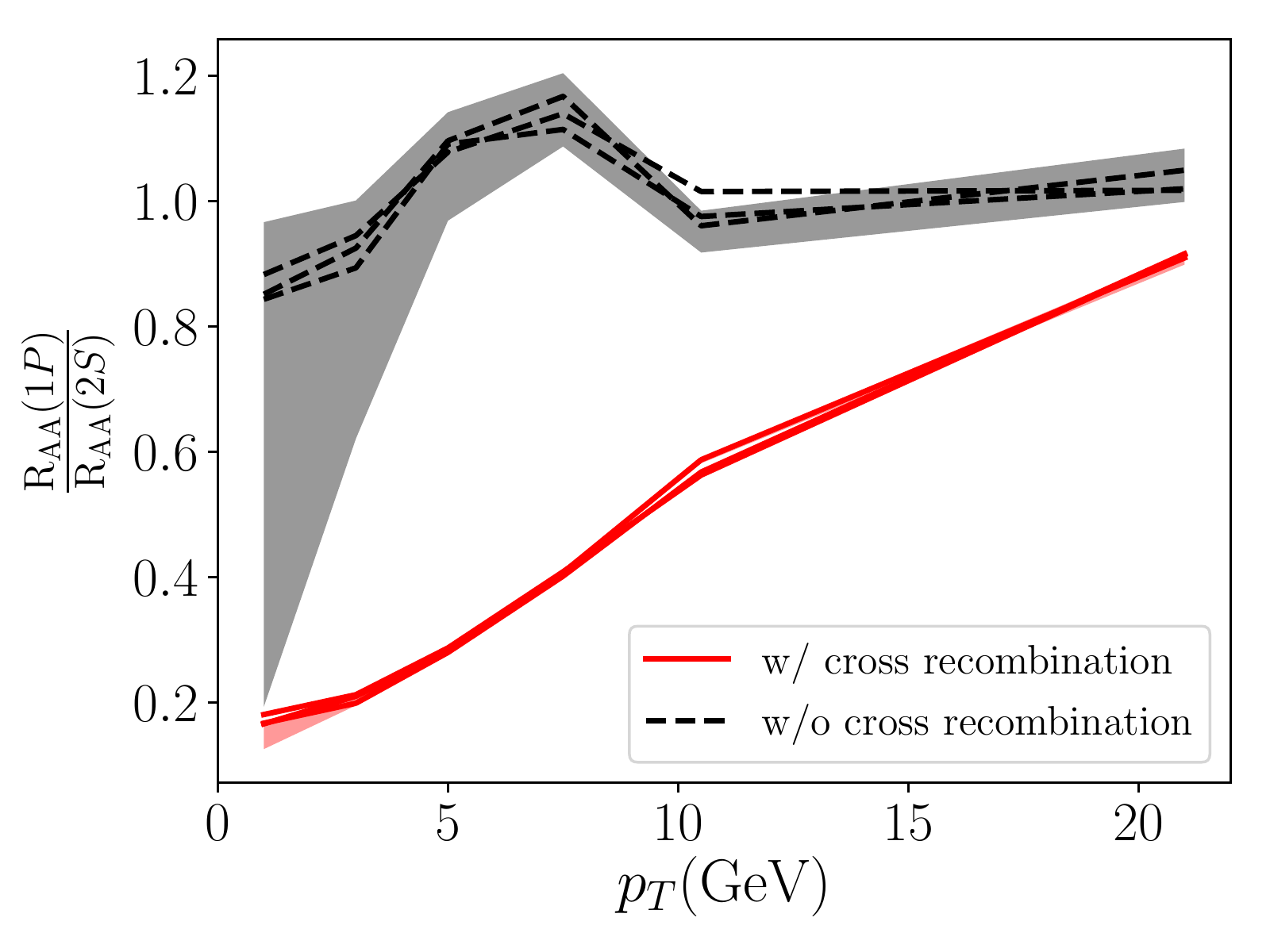}
        \caption{Ratio as a function of $p_T$.}
        \label{fig:ratio}
    \end{subfigure}
\caption{$R_{\ma{AA}}(\chi_b(1P))$ compared with $R_{\ma{AA}}(\Upsilon(2S))$ from calculations with and without cross-talk recombination. The $R_{AA}$'s as functions of centrality are shown in the first row. The ratio of $R_{\ma{AA}}(\chi_b(1P))$ and $R_{\ma{AA}}(\Upsilon(2S))$ as a function of $p_T$ is shown in the second row. Different curves correspond to difference choices of parameters and the band indicates the nPDF uncertainty. The double ratio observable has huge discriminatory power to distinguish calculations with and without correlated recombination.}
\label{fig:chi_b}
\end{figure}

To demonstrate this idea, we plot $R_{\ma{AA}}(\chi_b(1P))$ and $R_{\ma{AA}}(\Upsilon(2S))$ in Fig.~\ref{fig:chi_b} \footnote{All results in Fig.~\ref{fig:chi_b} are calculated from $N^{\text{init}} = 10^5$ simulation events rather than $30000$ as in other figures.} for two cases: with cross-talk recombination and without it. It can be clearly seen that in the former case, $\Upsilon(2S)$ is less suppressed while in the latter case, $\Upsilon(2S)$ and $\chi_b(1P)$ have similar $R_{\ma{AA}}$'s. So we propose a new measurement to test experimentally the importance of correlated recombination, which is the ratio of $R_{\ma{AA}}(\chi_b(1P))$ and $R_{\ma{AA}}(\Upsilon(2S))$. We plot this ratio as a function of transverse momentum in Fig.~\ref{fig:ratio} for both cases. The difference is dramatic: in the case without cross-talk recombination, the double ratio is approximately unity, consistent with the argument based on the quarkonium binding energy and size given above, while in the case with cross-talk recombination, the double ratio can be as small as $0.2$ at low $p_T$ and gradually increases at higher $p_T$. The uncertainties due to the choice of parameters are relatively small in both cases, as can be seen from the variation of the three curves. In the case without cross-talk recombination, the nPDF uncertainty does not cancel out in the ratio, and is very large at low $p_T$, which reflects the difference of the nPDF effects on $2S$ and $1P$ states as shown in Fig.~\ref{fig:cnm}. In the case with cross-talk recombination, the nPDF uncertainty band becomes much narrower. The reason of the narrowness is two-fold. First, the CNM effect and the hot medium effect are multiplicative (see Eq.~(\ref{eqn:final_num})). When the central value of the ratio is smaller, as in the case with cross-talk recombination, the nPDF uncertainty band would look narrower in a plot with a linear scale in the vertical axis. Furthermore, because of cross-talk recombination, different bottomonium states can turn to each other during the evolution and thus the difference of the CNM effects in the initial production is gradually washed out. In a nutshell, after accounting for the calculation uncertainties, the suppression of $\chi_b(1P)$ relative to $\Upsilon(2S)$ is still manifest, especially at low and intermediate $p_T$. 

As can be seen from Fig.~\ref{fig:ratio}, this double ratio observable is of huge model discriminatory power.
However, measurements on the double ratio may be challenging at the moment. Reconstructing the $\chi_b(1P)$ state requires detecting the low energy photon emitted in its decay to the $\Upsilon(1S)$ state. But a large number of photons at low $p_T$ are produced in heavy-ion collisions due to the neutral pion decay, which leads to a large combinatorial background in the $\chi_b(1P)$ reconstruction. Furthermore, the photon resolution in the calorimeter is limited. To overcome the resolution problem, one can reconstruct the photon from its conversion to an electron positron pair due to the interaction with detector materials, but the conversion reduces the signal by a substantial factor of order $10$~\cite{MW}. Moving to intermediate and high $p_T$ would help detecting the emitted photon, but at the same time the interesting signal in the double ratio dies away as $p_T$ increases, needless to say the available experimental statistics at high $p_T$. A tradeoff must be made between the interesting physics, detector resolution and experimental statistics. Considering these factors and the large nPDF uncertainty at low $p_T$ in the case without cross-talk recombination, an optimal $p_T$ window for such a double ratio measurement might be roughly between $5$ and $15$ GeV. Detailed experimental analysis on the $p_T$ searching window is needed. With higher luminosity and improved detector efficiencies, Run 3 of LHC and the sPHENIX program at RHIC may be able to measure this double ratio observable.

\section{Conclusions}
\label{sect:conclusions}
In this paper, we develop a set of coupled transport equations for open heavy quark-antiquark pairs and quarkonia. Our framework is capable of calculating observables for both open and hidden heavy flavors. Dissociation and recombination rates are calculated from pNRQCD via a systematic weak-coupling and nonrelativistic expansion, with the assumption of a weakly-coupled QGP. Recombination rates depend on the real-time heavy quark-antiquark distribution function, which is solved by the transport equation for open heavy flavors. Our framework can handle both uncorrelated and correlated recombination. By numerically solving the coupled transport equations via Monte Carlo techniques, we demonstrate how the ground and excited bottomonia states approach thermal equilibrium inside a QGP box with a constant temperature. Furthermore, we study bottomonia production in heavy-ion collisions using our framework. The initial phase space distributions are calculated from \textsc{Pythia} with the nPDF EPPS16 and the \trento\ model. The medium description utilizes a $2+1$ dimensional viscous hydrodynamic model. The parameters of the \trento\ model and the hydrodynamic equations have been calibrated to experimental observables on light hadrons. We include $\Upsilon(1S)$, $\Upsilon(2S)$, $\Upsilon(3S)$, $\chi_b(1P)$ and $\chi_b(2P)$ states in our reaction network. Our calculations demonstrate the importance of correlated cross-talk recombination in bottomonium phenomenology and can describe most of the experimental data. Discrepancies are seen at mid and large transverse momenta, where omitted higher order terms in the nonrelativistic expansion become gradually more important. We propose a new measurement on the ratio of nuclear modification factors of $\chi_b(1P)$ and $\Upsilon(2S)$ to test the importance of correlated recombination in experiments.

The current formalism can be improved in several ways. First, one can include the effect of the running coupling constant and higher order corrections in the nonrelativistic expansion. One may also consider extending the calculation of the reaction rates to the case of a strongly-coupled QGP, which will be more relevant at low temperature. Second, the simple Coulomb potential can be replaced by a more realistic nonperturbative potential model. The nonperturbative potential may be parametrized and the parameters can be calibrated to experimental observables such as $R_{\ma{AA}}$ and $v_2$, for example, by using the recently developed Bayesian analysis techniques \cite{Bernhard:2019bmu}. A simultaneous description of both open and hidden heavy flavor observables is also possible in our framework and is worth exploring. With a nonperturbative potential, the technical challenge would be to develop a fast numerical algorithm to sample the momenta of the outgoing $Q\bar{Q}$ pair (in dissociation) or quarkonium (in recombination) from the differential reaction rates. No previous studies have done this, but it is crucial for a consistent microscopic treatment like our calculation here. With a Coulomb potential, reaction rates have analytic expressions which are easier to handle in inverse transform and importance samplings (For the case of how to sample efficiently with a Coulomb potential, see Chapter 4.2 of Ref.~\cite{Yao:2019rcb}). For a nonperturbative parametrization of the screened potential, one has to include its dependence on the relative velocity between the quarkonium state and the hydro-cell. We use Coulomb potential here so we do not need to consider the velocity dependence in the potential. Also, the CNM effect at $200$ Au-Au collisions should be better understood and constrained by the measurements in p-Au collisions. Furthermore, the transport in the pre-thermalization stage should be investigated. For open heavy flavors, this has been done in Ref.~\cite{Mrowczynski:2017kso}. If one includes the transport of heavy flavors in the pre-thermalization stage, for consistency, one should also take into account the thermalization process of the medium between the initial hard collision and the hydrodynamics starting time. Parameter calibrations should be re-analyzed with free streaming replaced by a more realistic thermalization process. Finally, we will extend the current study of bottomonium production to charmonium production in heavy-ion collisions. There, we must also initialize unbound $c\bar{c}$ pairs, since many of them are produced primordially and uncorrelated recombination is enhanced. 

\begin{acknowledgments}
XY would like to thank the organizers of the workshop EMMI Rapid Reaction Task Force: Suppression and (re)generation of quarkonium in heavy-ion collisions at the LHC and the organizers of the workshop Quarkonia as Tools 2020, at both of which great discussions on quarkonium in-medium transport came out. This work is supported by U.S. Department of Energy (DOE) research grants DE-FG02-05ER41367. KW acknowledges support from the DOE grant DE-AC02-05CH11231 and National Science Foundation under the grant ACI-1550228 within the JETSCAPE Collaboration. XY acknowledges support from the DOE grant DE-SC0011090, Brookhaven National Laboratory and Department of Physics, Massachusetts Institute of Technology.
\end{acknowledgments}

\appendix
\section{Dissociation and Recombination Terms in Transport Equations}
\label{app:rates}
We will follow most of the notations used in Ref.~\cite{Yao:2018sgn}. We will use ${\bs k}$ to label the momentum of the quarkonium involved in a scattering process. ${\bs q}$ will be used to indicate the momentum of the gluon absorbed or emitted by the quarkonium. This gluon can be real in the process $g+H\leftrightarrow Q+\bar{Q}$ or virtual in inelastic scattering. If the gluon is on-shell, $q=|{\bs q}|$ will be used to represent its energy. For the process $q+H \leftrightarrow q+Q+\bar{Q}$, ${\bs p}_1$ and ${\bs p}_2$ are used indicate the momenta of the light quarks on the left and right respectively. Similarly for the process $g+H \leftrightarrow g+Q+\bar{Q}$, we will use ${\bs q}_1$ and ${\bs q}_2$ to represent the momenta of the gluons on the left and right respectively. In the quarkonium rest frame, the energy of the quarkonium is given by $-|E_{nl}|$ where $E_{nl}$ is the binding energy. In the rest frame of the unbound $Q\bar{Q}$ pair, its energy is given by $\frac{{\bs p}^2_\ma{rel}}{M}$ where ${\bs p}_\ma{rel}$ is their relative momentum. The expressions shown below are valid in the rest frame of quarkonium for dissociation or the center-of-mass frame of a $Q\bar{Q}$ pair for recombination. These two frames are not equivalent, but their difference is negligible (suppressed by $\frac{T}{M}$). In real simulations, the gluon distribution is boosted from the local rest frame of the hydro-cell (where temperature is defined) to these rest frames of the heavy particles. After calculating the reaction rates in the rest frames of the heavy particles, one has to boost them back to the laboratory frame where we keep track of the phase space distributions. In the following, for simplicity, we will only show expressions for quarkonia or $Q\bar{Q}$ pairs whose center-of-mass motions are at rest with respect to the medium local rest frame. In practice, we account for the Lorentz boost properly, as described above.

For later convenience, we define a ``$\delta-$derivative" symbol, first introduced in Ref.~\cite{Yao:2018zze}
\be \nn
&&\frac{\delta }{\delta{{\bs p}_i}} \int \prod_{j=1}^n \frac{\Diff{3}p_j}{(2\pi)^3} h({\bs p}_1, {\bs p}_2, \cdots, {\bs p}_n)\Big|_{{\bs p}_i = {\bs p}} 
\equiv  \frac{\delta }{\delta{w({\bs p}})} \int \prod_{j=1}^n \frac{\Diff{3}p_j}{(2\pi)^3} h({\bs p}_1, {\bs p}_2, \cdots, {\bs p}_n)w({\bs p}_i) \\ 
&=& \int \prod_{j=1, j\neq i}^n \frac{\Diff{3}p_j}{(2\pi)^3} h({\bs p}_1, {\bs p}_2, \cdots, {\bs p}_{i-1}, {\bs p}, {\bs p}_{i+1}, \cdots, {\bs p}_n)\,,
\ee
where the second $\delta$ denotes the standard functional variation and $h({\bs p}_1, {\bs p}_2, \cdots, {\bs p}_n)$ and $w({\bs p}_i)$ are arbitrary independent smooth functions.

\subsection{$g+H\leftrightarrow Q+\bar{Q}$}
The dissociation and recombination terms in the transport equation of the quarkonium state $nls$ can be written as
\be
\ml{C}_{nls}^{\pm}({\bs x}, {\bs p}, t) &=&  \frac{\delta \ml{F}^\pm_{nls}}{\delta{{\bs k}}}  \Big|_{{\bs k}={\bs p}}\,,
\ee
where we used the ``$\delta-$derivative" symbol defined above.
The functions $\ml{F}^\pm_{nls}$ are defined by
\be\nn
\ml{F}^+_{nls} & \equiv & g_+ \int \frac{\diff^3 k}{(2\pi)^3}  \frac{\diff^3 p_{Q}}{(2\pi)^3}  \frac{\diff^3 p_{\bar{Q}} }{(2\pi)^3} \frac{\diff^3 q}{2q(2\pi)^3} (1+n_B(q)) f_{Q\bar{Q}}({\bs x}_Q, {\bs p}_Q, {\bs x}_{\bar{Q}}, {\bs p}_{\bar{Q}}, t) \\
&& (2\pi)^4\delta^3({\bs k} + {\bs q} - {\bs p}_{\ma{cm}}) \delta(-|E_{nl}|+q-\frac{p^2_{\ma{rel}}}{M})  \sum | \ml{M} |^2\\ \nn
\ml{F}^-_{nls} & \equiv &  \int \frac{\diff^3 k}{(2\pi)^3}  \frac{\diff^3 p_{\ma{cm}}}{(2\pi)^3}  \frac{\diff^3 p_{\ma{rel}}}{(2\pi)^3} \frac{\diff^3 q}{2q(2\pi)^3} n_B(q) f_{nls}({\bs x}, {\bs k}, t) \\
&& (2\pi)^4\delta^3({\bs k} + {\bs q} - {\bs p}_{\ma{cm}}) \delta(-|E_{nl}|+q-\frac{p^2_{\ma{rel}}}{M})   \sum | \ml{M} |^2\,,
\ee
where $n_B$ is the Bose-Einstein distribution function, $g_+ = \frac{1}{N_c^2}g_s$ and $g_s$ is the degeneracy factor for spin: $g_s = \frac{3}{4}$ for a quarkonium state with spin $s=1$ and $\frac{1}{4}$ for spin $s=0$. For an arbitrary $Q\bar{Q}$ pair, its probability of being a spin-1 state is $\frac{3}{4}$. Its probability of being a color octet is $\frac{N_c^2-1}{N_c^2}$. When computing the scattering amplitude squared, we have summed over the relevant quantum numbers (color of the $Q\bar{Q}$, gluon polarization and color), so we only use $\frac{1}{N_c^2}$ to avoid double counting. In the definition of $\ml{F}^-_{nls}$, ${\bs p}_{\ma{cm}} = {\bs p}_Q+{\bs p}_{\bar{Q}},$ and ${\bs p}_{\ma{rel}} = \frac{{\bs p}_Q - {\bs p}_{\bar{Q}}}{2}$ are the center-of-mass and relative momenta of a $Q\bar{Q}$ pair with momenta ${\bs p}_Q$ and 
${\bs p}_{\bar{Q}}$.

After summing over the relevant quantum numbers (see Ref.~\cite{Yao:2018sgn} for details), the scattering amplitude squared is given by 
\be
\sum | \ml{M} |^2 \equiv  \frac{2}{3}g^2C_Fq^2 |  \langle   \Psi_{{\bs p}_\ma{rel}} | {\bs r} |   \psi_{nl}  \rangle |^2\,,
\ee
where $C_F=\frac{N_c^2-1}{2N_c}$, $| \psi_{nl}  \rangle$ is the wavefunction of the quarkonium state $nls$ (states with different spins are degenerate) and $|  \Psi_{{\bs p}_\ma{rel}} \rangle $ is the Coulomb scattering wave for the unbound $Q\bar{Q}$ pair. For non-S wave states, we average over the polarizations:
\be
\nn
&&\frac{1}{2l+1}\sum_{m_l=-l}^l \int \diff^3 p_{\ma{rel}} \langle \psi_{nlm_l} | r^i | \Psi_{{\bs p}_\ma{rel}} \rangle  \langle \Psi_{{\bs p}_\ma{rel}} | r^j |  \psi_{nlm_l}  \rangle \\ 
\label{eqn:average_m} 
&=& \frac{1}{3}\delta^{ij}  \frac{1}{2l+1}\sum_{m_l=-l}^l \int \diff^3 p_{\ma{rel}}    |  \langle \Psi_{{\bs p}_\ma{rel}}  | {\bs r} |   \psi_{nlm_l}  \rangle | ^2
\equiv \frac{1}{3}\delta^{ij}  \int \diff^3 p_{\ma{rel}} | \langle \Psi_{{\bs p}_\ma{rel}}  | {\bs r} |   \psi_{nl} \rangle |^2\,.
\ee

\subsection{$q+H\leftrightarrow q+Q+\bar{Q}$ and $g+H\leftrightarrow g+Q+\bar{Q}$}
For the inelastic scattering channels, we have
\be
\ml{C}_{nls,\ma{inel}}^{\pm}({\bs x}, {\bs p}, t) &=&  \frac{\delta \ml{F}^\pm_{nls,\ma{ineq}}}{\delta{{\bs k}}}  \Big|_{{\bs k}={\bs p}} + \frac{\delta \ml{F}^\pm_{nls,\ma{ineg}}}{\delta{{\bs k}}}  \Big|_{{\bs k}={\bs p}}\,.
\ee
The inelastic $\ml{F}^\pm_{nls}$ functions are defined by
\be \nn
\ml{F}^+_{nls,\ma{ineq}} &\equiv&  g_+ \int \frac{\diff^3 k}{(2\pi)^3}  \frac{\diff^3 p_{Q}}{(2\pi)^3}  \frac{\diff^3 p_{\bar{Q}} }{(2\pi)^3} \frac{\diff^3 p_1}{2p_1(2\pi)^3} \frac{\diff^3 p_2}{2p_2(2\pi)^3} \\[4pt]\nn &&n_F(p_2)(1-n_F(p_1)) f_{Q\bar{Q}}({\bs x}_Q, {\bs p}_Q, {\bs x}_{\bar{Q}}, {\bs p}_{\bar{Q}}, t) \\
&& (2\pi)^4\delta^3({\bs k} + {\bs p}_1 - {\bs p}_{\ma{cm}} - {\bs p}_2) \delta(-|E_{nl}|+p_1-\frac{p^2_{\ma{rel}}}{M}-p_2)  \sum | \ml{M}_\ma{ineq} |^2   \\ \nn
\ml{F}^-_{nls,\ma{ineq}}  &\equiv&  \int \frac{\diff^3 k}{(2\pi)^3}  \frac{\diff^3 p_{\ma{cm}}}{(2\pi)^3}  \frac{\diff^3 p_{\ma{rel}} }{(2\pi)^3} \frac{\diff^3 p_1}{2p_1(2\pi)^3} \frac{\diff^3 p_2}{2p_2(2\pi)^3}  n_F(p_1)(1-n_F(p_2))  f_{nls}({\bs x}, {\bs k}, t) \\
&& (2\pi)^4\delta^3({\bs k} + {\bs p}_1 - {\bs p}_{\ma{cm}} - {\bs p}_2) \delta(-|E_{nl}|+p_1-\frac{p^2_{\ma{rel}}}{M}-p_2)  \sum | \ml{M}_\ma{ineq} |^2 \\ \nn
\ml{F}^+_{nls,\ma{ineg}} &\equiv&  g_+ \int \frac{\diff^3 k}{(2\pi)^3}  \frac{\diff^3 p_{Q}}{(2\pi)^3}  \frac{\diff^3 p_{\bar{Q}} }{(2\pi)^3} \frac{\diff^3 q_1}{2q_1(2\pi)^3} \frac{\diff^3 q_2}{2q_2(2\pi)^3} \\[4pt]\nn &&n_B(q_2)(1+n_B(q_1)) f_{Q\bar{Q}}({\bs x}_Q, {\bs p}_Q, {\bs x}_{\bar{Q}}, {\bs p}_{\bar{Q}}, t) \\
&& (2\pi)^4\delta^3({\bs k} + {\bs q}_1 - {\bs p}_{\ma{cm}} - {\bs q}_2) \delta(-|E_{nl}|+q_1-\frac{p^2_{\ma{rel}}}{M}-q_2)  \sum | \ml{M}_\ma{ineg} |^2   \\ \nn
\ml{F}^-_{nls,\ma{ineg}}  &\equiv&  \int \frac{\diff^3 k}{(2\pi)^3}  \frac{\diff^3 p_{\ma{cm}}}{(2\pi)^3}  \frac{\diff^3 p_{\ma{rel}} }{(2\pi)^3} \frac{\diff^3 q_1}{2q_1(2\pi)^3} \frac{\diff^3 q_2}{2q_2(2\pi)^3}  n_B(q_1)(1+n_B(q_2))  f_{nls}({\bs x}, {\bs k}, t) \\
&& (2\pi)^4\delta^3({\bs k} + {\bs q}_1 - {\bs p}_{\ma{cm}} - {\bs q}_2) \delta(-|E_{nl}|+q_1-\frac{p^2_{\ma{rel}}}{M}-q_2)  \sum | \ml{M}_\ma{ineg} |^2 \,,
\ee
where $n_F$ is the Fermi-Dirac distribution for fermions. The scattering amplitudes squared, with summation over relevant quantum numbers are given by (see Ref.~\cite{Yao:2018sgn} for details)
\be
\sum | \ml{M}_\ma{ineq} |^2  &=& \frac{16}{3}g^4T_FC_F  |\langle \Psi_{{\bs p}_\ma{rel}} | {\bs r} |  \psi_{nl}  \rangle|^2   \frac{p_1p_2 + {\bs p}_1\cdot {\bs p}_2}{{\bs q}^2} \\
\sum | \ml{M}_\ma{ineg} |^2  &=& \frac{1}{3}g^4 N_c C_F | \langle \Psi_{{\bs p}_\ma{rel}} | {\bs r} |  \psi_{nl}  \rangle|^2 
 \frac{1+(\hat{q}_1\cdot\hat{q}_2)^2}{{\bs q}^2} (q_1+q_2)^2\,,
\ee
in which $T_F=\frac{1}{2}$, $\hat{q}_i \equiv \frac{{\bs q}_i}{q_i}$ for $i=1,2$.

We will now list the square of the dipole transition matrix elements for different quarkonium states $| \langle \Psi_{{\bs p}_\ma{rel}} | {\bs r} |  \psi_{nl}  \rangle|^2 $.

\subsection{$| \langle \Psi_{{\bs p}_\ma{rel}} | {\bs r} |  \psi_{nl}  \rangle|^2 $}
For non-S wave, we average over the third component of the orbital angular momentum and $| \langle \Psi_{{\bs p}_\ma{rel}} | {\bs r} |  \psi_{nl}  \rangle|^2 $ is defined in Eq.~(\ref{eqn:average_m}). The partial wave expansion of the Coulomb scattering wave $|\Psi_{{\bs p}_\ma{rel}}\rangle $ is given by
\be
\Psi_{{\bs p}_\ma{rel}}(\bs r) = \langle \bs r| \Psi_{{\bs p}_\ma{rel}}\rangle = 4\pi\sum_{\ell,m} i^\ell e^{i\delta_\ell} \frac{F_\ell(\rho)}{\rho} Y_{\ell m}(\hat{r})Y_{\ell m}^*(\hat{p}_{\ma{rel}})\,,
\ee
in which $Y_{\ell m}$ denotes the spherical harmonics and
\be
\rho &=& p_{\ma{rel}}r \\
\delta_\ell &=& \arg\Gamma(1+\ell+i\eta) \\
\label{eqn:eta}
\eta &=& \frac{\alpha_sM}{4N_cp_\ma{rel}}\\
F_\ell(\rho) &=& \frac{2^\ell e^{-\pi\eta/2} |\Gamma(1+\ell+i\eta)| }{(2\ell+1)!} \rho^{\ell+1} e^{i\rho} \,_1F_1(\ell+1+i\eta;2\ell+2;-2i\rho)\,,
\ee
where $_1F_1(a;b;z)$ is the confluent hypergeometric function. We use the Wigner-Eckart theorem to simplify the calculations.

For $1S$ state, we have
\be
| \langle \Psi_{{\bs p}_\ma{rel}} | {\bs r} |  \psi_{1S}  \rangle|^2 = \frac{ 2^9 \pi^2 \eta a_B^7  p_\ma{rel}^2  (1+\eta^2) (2+\eta  a_B p_\ma{rel})^2 }{(1+a_B^2p_\ma{rel}^2)^6(e^{2\pi\eta}-1)} e^{4\eta \arctan{(a_Bp_\ma{rel})}} \,,
\ee
where $a_B = \frac{2}{\alpha_s C_F M}$ is the Bohr radius and $\eta$ is defined in Eq.~(\ref{eqn:eta}).

For $2S$ state, we have
\be \nn
| \langle \Psi_{{\bs p}_\ma{rel}} | {\bs r} |  \psi_{2S}  \rangle|^2 &=& \frac{2^{18}\pi^2\eta a_B^7 p_\ma{rel}^2 (1+\eta^2)}{(1+4a_B^2 p_\ma{rel}^2)^8 (e^{2\pi\eta}-1)} e^{4\eta \arctan{(2a_Bp_\ma{rel})}} \\
&& (-4 - 9\eta a_B p_\ma{rel} + 8a_B^2p_\ma{rel}^2 - 4\eta^2 a_B^2 p_\ma{rel}^2 + 4\eta a_B^3 p_\ma{rel}^3)^2\,.
\ee
This expression can be further simplified by using $\eta a_B p_\ma{rel} = \frac{1}{N_c^2-1}$. These matrix elements of $1S$ and $2S$ have been reported in Ref.~\cite{Brambilla:2011sg}.

For $1P$ state, we find
\be \nn
| \langle \Psi_{{\bs p}_\ma{rel}} | {\bs r} |  \psi_{1P}  \rangle|^2 &=&
\frac{2^{16} \pi^2 \eta a_B^5}{3(1+4a_B^2 p_\ma{rel}^2)^8 (e^{2\pi\eta}-1)} e^{4\eta \arctan{(2a_Bp_\ma{rel})}} \\ \nn
&& \Big[ \big(  8\eta(\eta^2-2)a_B^3p_\ma{rel}^3 + 12(2\eta^2-1)a_B^2p_\ma{rel}^2 +18\eta a_B p_\ma{rel} + 3 \big)^2 \\ 
&& + 16 a_B^4 p_\ma{rel}^4 (3+2\eta a_B p_\ma{rel})^2 (\eta^4 + 5\eta^2 + 4) \Big]\,.
\ee

We do not need the dipole transition matrix elements for $3S$ and $2P$ states, since we assume they cannot be formed inside the QGP. In other words, if a $3S$ or $2P$ state enters the QGP, it melts immediately. In practice, we force them to dissociate in the code when $T>154$ MeV and do not allow them to be (re)generated inside the QGP.

\section{Details on Feed-Down Contributions}
\label{app:fd}
We will use the notations $\sigma_{nl}^\ma{tot}$ and $\sigma_{nl}$ to denote the total and primordial cross sections in a nucleon-nucleon collision. The former contains the latter and feed-down contributions.

To work out the feed-down contributions from excited states to the ground and lower-excited states, the first thing we need is the branching ratio in vacuum. The most recent results reported by the Particle Data Group \cite{Tanabashi:2018oca} are summarized in Table~\ref{tab:br}. For the P-wave states, we need the averaged branching ratio since our transport equations are degenerate in spin. To this end, we use the experimental result on $\frac{\sigma_{\chi_{b2}}(1P)}{\sigma_{\chi_{b1}}(1P)}$ in Ref.~\cite{Khachatryan:2014ofa} and follow the assumptions made in Ref.~\cite{Du:2017qkv} to write
\be
\frac{\sigma_{\chi_{b2}}}{\sigma_{\chi_{b1}}} &=& 0.85 \\
\frac{\sigma_{\chi_{b0}}}{\sigma_{\chi_{b1}}} &=& 1.5 \,,
\ee
for both $1P$ and $2P$ states.
Then we can work out the averaged branching ratios
\be \nn
&&\ma{Br}[\chi_b(1P)\to\Upsilon(1S)] \\ \nn
&\equiv & \frac{ 
\ma{Br}[\chi_{b0}(1P)\to\Upsilon(1S)]\sigma_{\chi_{b0}}+ \ma{Br}[\chi_{b1}(1P)\to\Upsilon(1S)]\sigma_{\chi_{b1}}+ \ma{Br}[\chi_{b2}(1P)\to\Upsilon(1S)]\sigma_{\chi_{b2}} }{\sigma_{\chi_{b0}} + \sigma_{\chi_{b1}} + \sigma_{\chi_{b2}}} \\ 
&=& 0.159 \pm 0.010\\
&&\ma{Br}[\chi_b(2P)\to\Upsilon(1S)] \nn\\ \nn
&\equiv & \frac{ 
\ma{Br}[\chi_{b0}(2P)\to\Upsilon(1S)]\sigma_{\chi_{b0}}+ \ma{Br}[\chi_{b1}(2P)\to\Upsilon(1S)]\sigma_{\chi_{b1}}+ \ma{Br}[\chi_{b2}(2P)\to\Upsilon(1S)]\sigma_{\chi_{b2}} }{\sigma_{\chi_{b0}} + \sigma_{\chi_{b1}} + \sigma_{\chi_{b2}}} \\
&=& 0.056 \pm 0.008\\
&&\ma{Br}[\chi_b(2P)\to\Upsilon(2S)] \nn\\ \nn
&\equiv & \frac{ 
\ma{Br}[\chi_{b0}(2P)\to\Upsilon(2S)]\sigma_{\chi_{b0}}+ \ma{Br}[\chi_{b1}(2P)\to\Upsilon(2S)]\sigma_{\chi_{b1}}+ \ma{Br}[\chi_{b2}(2P)\to\Upsilon(2S)]\sigma_{\chi_{b2}} }{\sigma_{\chi_{b0}} + \sigma_{\chi_{b1}} + \sigma_{\chi_{b2}}} \\
&=& 0.083 \pm 0.010
\ee

\begin{table}
\centering
\begin{tabular}{ | c | c | } 
\hline
Channel & Branching ratio \\ 
\hline
$\Upsilon(2S) \to \Upsilon(1S)$ & 0.265$\pm$0.007 \\
\hline
$\Upsilon(3S) \to \Upsilon(1S)$ & 0.066$\pm$0.002 \\
\hline
$\chi_{b0}(1P) \to \Upsilon(1S)$ & 0.019$\pm$0.003 \\
\hline
$\chi_{b1}(1P) \to \Upsilon(1S)$ & 0.352$\pm$0.020 \\
\hline
$\chi_{b2}(1P) \to \Upsilon(1S)$ & 0.180$\pm$0.010 \\
\hline
$\chi_{b0}(2P) \to \Upsilon(1S)$ & 0.004$\pm$0.002 \\
\hline
$\chi_{b1}(2P) \to \Upsilon(1S)$ & 0.115$\pm$0.014 \\
\hline
$\chi_{b2}(2P) \to \Upsilon(1S)$ & 0.077$\pm$0.011 \\
\hline
$\Upsilon(3S) \to \Upsilon(2S)$ & 0.106$\pm$0.008 \\
\hline
$\chi_{b0}(2P) \to \Upsilon(2S)$ & 0.014$\pm$0.003 \\
\hline
$\chi_{b1}(2P) \to \Upsilon(2S)$ & 0.181$\pm$0.019 \\
\hline
$\chi_{b2}(2P) \to \Upsilon(2S)$ & 0.089$\pm$0.012 \\
\hline
\end{tabular}
\caption{Branching ratios of different bottomonium states.}
\label{tab:br}
\end{table}

\begin{table}
\centering
\begin{tabular}{ | c | c | } 
\hline
Cross sections in proton-proton collisions & Experimental results (nb) from Ref.~\cite{Sirunyan:2018nsz}\\ 
\hline
$B\times \sigma(\Upsilon(1S))$, $|y|<2.4$, $5.02$ TeV  & 3.353$\pm$0.081(stat)$\pm$0.167(syst) \\
\hline
$B\times \sigma(\Upsilon(2S))$, $|y|<2.4$, $5.02$ TeV  & 0.873$\pm$0.031(stat)$\pm$0.046(syst) \\
\hline
$B\times \sigma(\Upsilon(3S))$, $|y|<2.4$, $5.02$ TeV  & 0.404$\pm$0.017(stat)$\pm$0.022(syst) \\
\hline
\end{tabular}
\caption{Experimental inputs of cross sections in proton-proton collisions, where $B$ indicates the branching ratio of the relevant state to $\mu^+\mu^-$.}
\label{tab:pp}
\end{table}

The next thing we need is the primordial cross section ratio. Using the experimental inputs listed in Table~\ref{tab:pp} and following Ref.~\cite{Du:2017qkv} and references therein, we assume 
\be
\label{eqn:sigma2S}
\sigma_{2S}^{\ma{tot}} &=& 0.33 \sigma_{1S}^{\ma{tot}}\\
\label{eqn:sigma3S}
\sigma_{3S} &=& 0.15 \sigma_{1S}^{\ma{tot}}\\
\label{eqn:sigma1P}
\sigma_{1P} &=& 1.08 \sigma_{1S}^{\ma{tot}}\\
\label{eqn:sigma2P}
\sigma_{2P} &=& 0.86 \sigma_{1S}^{\ma{tot}}\,.
\ee
The total cross sections of $1S$ and $2S$ can be written as
\be \nn
\sigma_{1S}^{\ma{tot}} &=& \sigma_{1S} + \sigma_{2S}^{\ma{tot}} \ma{Br}[\Upsilon(2S)\to\Upsilon(1S)]
+ \sigma_{3S} \ma{Br}[\Upsilon(3S)\to\Upsilon(1S)] \\ 
&&+ \sigma_{1P} \ma{Br}[\chi_b(1P)\to\Upsilon(1S)]
+ \sigma_{2P} \ma{Br}[\chi_b(2P)\to\Upsilon(1S)] \\
\sigma_{2S}^{\ma{tot}} &=& \sigma_{2S} + \sigma_{3S} \ma{Br}[\Upsilon(3S)\to\Upsilon(2S)] + \sigma_{2P} \ma{Br}[\chi_b(2P)\to\Upsilon(2S)] \,.
\ee
Using Eqs.~(\ref{eqn:sigma2S}, \ref{eqn:sigma3S}, \ref{eqn:sigma1P}, \ref{eqn:sigma2P}) and the branching ratios, we find
\be
\sigma_{2S} &=& 0.24 \sigma_{1S}^\ma{tot}\\
\sigma_{1S} &=& 0.68 \sigma_{1S}^\ma{tot}\,.
\ee
Experimentally it is known that the feed-down contributions to $\sigma_{1S}^\ma{tot}$ is about $67\%$ (averaged over the transverse momentum), consistent with our prescription here. Now we are ready to compute the ratios of the primordial cross sections. The results are listed in Table~\ref{tab:sigma}.

\begin{table}
\centering
\begin{tabular}{ | c | c | } 
\hline
Primordial cross section ratio & Value \\ 
\hline
$\sigma_{2S}/\sigma_{1S}$ & 0.35 \\
\hline
$\sigma_{3S}/\sigma_{1S}$ & 0.22 \\
\hline
$\sigma_{1P}/\sigma_{1S}$ & 1.59 \\
\hline
$\sigma_{2P}/\sigma_{1S}$ & 1.26 \\
\hline
$\sigma_{3S}/\sigma_{2S}$ & 0.63 \\
\hline
$\sigma_{1P}/\sigma_{2S}$ & 4.54 \\
\hline
$\sigma_{2P}/\sigma_{2S}$ & 3.58 \\
\hline
$\sigma_{3S}/\sigma_{1P}$ & 0.14 \\
\hline
$\sigma_{2P}/\sigma_{1P}$ & 0.80 \\
\hline
\end{tabular}
\caption{Ratios of primordial cross sections.}
\label{tab:sigma}
\end{table}

\bibliographystyle{apsrev4-1}

\begin{thebibliography}{99}

\bibitem{Quigg:1977dd} 
  C.~Quigg and J.~L.~Rosner,
  Phys.\ Lett.\  {\bf 71B}, 153 (1977).

\bibitem{Matsui:1986dk} 
  T.~Matsui and H.~Satz,
  Phys.\ Lett.\ B {\bf 178}, 416 (1986).

\bibitem{Karsch:1987pv} 
  F.~Karsch, M.~T.~Mehr and H.~Satz,
  Z.\ Phys.\ C {\bf 37}, 617 (1988).

\bibitem{McLerran:1981pb} 
  L.~D.~McLerran and B.~Svetitsky,
  Phys.\ Rev.\ D {\bf 24}, 450 (1981).

\bibitem{Mocsy:2007jz} 
  A.~Mocsy and P.~Petreczky,
  Phys.\ Rev.\ Lett.\  {\bf 99}, 211602 (2007)
  [arXiv:0706.2183 [hep-ph]].

\bibitem{Laine:2006ns} 
  M.~Laine, O.~Philipsen, P.~Romatschke and M.~Tassler,
  JHEP {\bf 0703}, 054 (2007)
  [hep-ph/0611300].

\bibitem{Beraudo:2007ky} 
  A.~Beraudo, J.-P.~Blaizot and C.~Ratti,
  Nucl.\ Phys.\ A {\bf 806}, 312 (2008)
  [arXiv:0712.4394 [nucl-th]].
  
\bibitem{Thews:2000rj} 
  R.~L.~Thews, M.~Schroedter and J.~Rafelski,
  Phys.\ Rev.\ C {\bf 63}, 054905 (2001)
  [hep-ph/0007323].

\bibitem{Andronic:2007bi} 
  A.~Andronic, P.~Braun-Munzinger, K.~Redlich and J.~Stachel,
  Phys.\ Lett.\ B {\bf 652}, 259 (2007)
  [nucl-th/0701079 [NUCL-TH]].

\bibitem{Grandchamp:2003uw} 
  L.~Grandchamp, R.~Rapp and G.~E.~Brown,
  Phys.\ Rev.\ Lett.\  {\bf 92}, 212301 (2004)
  [hep-ph/0306077].
  
\bibitem{Grandchamp:2005yw} 
  L.~Grandchamp, S.~Lumpkins, D.~Sun, H.~van Hees and R.~Rapp,
  Phys.\ Rev.\ C {\bf 73}, 064906 (2006)
  [hep-ph/0507314].

\bibitem{Yan:2006ve} 
  L.~Yan, P.~Zhuang and N.~Xu,
  Phys.\ Rev.\ Lett.\  {\bf 97}, 232301 (2006)
  [nucl-th/0608010].

\bibitem{Zhao:2007hh}
X.~Zhao and R.~Rapp,
Phys. Lett. B \textbf{664}, 253-257 (2008)
[arXiv:0712.2407 [hep-ph]].

\bibitem{Liu:2009nb} 
  Y.~Liu, Z.~Qu, N.~Xu and P.~Zhuang,
  Phys.\ Lett.\ B {\bf 678}, 72 (2009)
  [arXiv:0901.2757 [nucl-th]].

\bibitem{Zhao:2010nk}
X.~Zhao and R.~Rapp,
Phys. Rev. C \textbf{82}, 064905 (2010)
[arXiv:1008.5328 [hep-ph]].


\bibitem{Song:2011xi} 
  T.~Song, K.~C.~Han and C.~M.~Ko,
  Phys.\ Rev.\ C {\bf 84}, 034907 (2011)
  [arXiv:1103.6197 [nucl-th]].

\bibitem{Song:2011nu} 
  T.~Song, K.~C.~Han and C.~M.~Ko,
  Phys.\ Rev.\ C {\bf 85}, 014902 (2012)
  [arXiv:1109.6691 [nucl-th]].

\bibitem{Emerick:2011xu}
A.~Emerick, X.~Zhao and R.~Rapp,
Eur. Phys. J. A \textbf{48}, 72 (2012)
[arXiv:1111.6537 [hep-ph]].

\bibitem{Sharma:2012dy} 
  R.~Sharma and I.~Vitev,
  Phys.\ Rev.\ C {\bf 87}, no. 4, 044905 (2013)
  [arXiv:1203.0329 [hep-ph]].
  
\bibitem{Nendzig:2014qka} 
  F.~Nendzig and G.~Wolschin,
  J.\ Phys.\ G {\bf 41}, 095003 (2014)
  [arXiv:1406.5103 [hep-ph]].
      
\bibitem{Krouppa:2015yoa} 
  B.~Krouppa, R.~Ryblewski and M.~Strickland,
  Phys.\ Rev.\ C {\bf 92}, no. 6, 061901 (2015)
  [arXiv:1507.03951 [hep-ph]].

\bibitem{Chen:2017duy} 
  B.~Chen and J.~Zhao,
  Phys.\ Lett.\ B {\bf 772}, 819 (2017)
  [arXiv:1704.05622 [nucl-th]].
 
\bibitem{Zhao:2017yan} 
  J.~Zhao and B.~Chen,
  Phys.\ Lett.\ B {\bf 776}, 17 (2018)
  [arXiv:1705.04558 [nucl-th]].
  
\bibitem{Du:2017qkv} 
  X.~Du, R.~Rapp and M.~He,
  Phys.\ Rev.\ C {\bf 96}, no. 5, 054901 (2017)
  [arXiv:1706.08670 [hep-ph]].

\bibitem{Aronson:2017ymv} 
  S.~Aronson, E.~Borras, B.~Odegard, R.~Sharma and I.~Vitev,
  Phys.\ Lett.\ B {\bf 778}, 384 (2018)
  [arXiv:1709.02372 [hep-ph]].
  
\bibitem{Ferreiro:2018wbd} 
  E.~G.~Ferreiro and J.~P.~Lansberg,
  JHEP {\bf 1810}, 094 (2018)
  [arXiv:1804.04474 [hep-ph]].

\bibitem{Yao:2018zrg} 
  X.~Yao, W.~Ke, Y.~Xu, S.~Bass and B.~M\"uller,
  Nucl.\ Phys.\ A {\bf 982}, 755 (2019)
  [arXiv:1807.06199 [nucl-th]].

\bibitem{Du:2018wsj}
X.~Du and R.~Rapp,
JHEP \textbf{03}, 015 (2019)
[arXiv:1808.10014 [nucl-th]].

\bibitem{Du:2019tjf}
X.~Du, S.~Y.~F.~Liu and R.~Rapp,
Phys. Lett. B \textbf{796}, 20-25 (2019)
[arXiv:1904.00113 [nucl-th]].

\bibitem{Hong:2019ade}
J.~Hong and S.~H.~Lee,
Phys. Lett. B \textbf{801}, 135147 (2020)
[arXiv:1909.07696 [nucl-th]].

\bibitem{Chen:2019qzx}
B.~Chen, M.~Hu, H.~Zhang and J.~Zhao,
Phys. Lett. B \textbf{802}, 135271 (2020)
[arXiv:1910.08275 [nucl-th]].

\bibitem{Kaczmarek:2002mc} 
  O.~Kaczmarek, F.~Karsch, P.~Petreczky and F.~Zantow,
  Phys.\ Lett.\ B {\bf 543}, 41 (2002)
  [hep-lat/0207002].

\bibitem{Bazavov:2018wmo} 
  A.~Bazavov {\it et al.} [TUMQCD Collaboration],
  Phys.\ Rev.\ D {\bf 98}, no. 5, 054511 (2018)
  [arXiv:1804.10600 [hep-lat]].
  
\bibitem{Burnier:2014ssa} 
  Y.~Burnier, O.~Kaczmarek and A.~Rothkopf,
  Phys.\ Rev.\ Lett.\  {\bf 114}, no. 8, 082001 (2015)
  [arXiv:1410.2546 [hep-lat]].

\bibitem{Peskin:1979va} 
  M.~E.~Peskin,
  Nucl.\ Phys.\ B {\bf 156}, 365 (1979).

\bibitem{Bhanot:1979vb} 
  G.~Bhanot and M.~E.~Peskin,
  Nucl.\ Phys.\ B {\bf 156}, 391 (1979).

\bibitem{Brambilla:1999xf} 
  N.~Brambilla, A.~Pineda, J.~Soto and A.~Vairo,
  Nucl.\ Phys.\ B {\bf 566}, 275 (2000)
  [hep-ph/9907240].

\bibitem{Brambilla:2004jw} 
  N.~Brambilla, A.~Pineda, J.~Soto and A.~Vairo,
  Rev.\ Mod.\ Phys.\  {\bf 77}, 1423 (2005)
  [hep-ph/0410047].
      
\bibitem{Fleming:2005pd} 
  S.~Fleming and T.~Mehen,
  Phys.\ Rev.\ D {\bf 73}, 034502 (2006)
  [hep-ph/0509313].
  
\bibitem{Brambilla:2008cx} 
  N.~Brambilla, J.~Ghiglieri, A.~Vairo and P.~Petreczky,
  Phys.\ Rev.\ D {\bf 78}, 014017 (2008)
  [arXiv:0804.0993 [hep-ph]].

\bibitem{Brambilla:2011sg} 
  N.~Brambilla, M.~A.~Escobedo, J.~Ghiglieri and A.~Vairo,
  JHEP {\bf 1112}, 116 (2011)
  [arXiv:1109.5826 [hep-ph]].

\bibitem{Brambilla:2013dpa} 
  N.~Brambilla, M.~A.~Escobedo, J.~Ghiglieri and A.~Vairo,
  JHEP {\bf 1305}, 130 (2013)
  [arXiv:1303.6097 [hep-ph]].

\bibitem{Biondini:2018ovz} 
  S.~Biondini and S.~Vogl,
  JHEP {\bf 1902}, 016 (2019)
  [arXiv:1811.02581 [hep-ph]].

\bibitem{Biondini:2019int} 
  S.~Biondini and S.~Vogl,
  JHEP {\bf 1911}, 147 (2019)
  [arXiv:1907.05766 [hep-ph]].
  
\bibitem{Binder:2020efn}
T.~Binder, B.~Blobel, J.~Harz and K.~Mukaida,
[arXiv:2002.07145 [hep-ph]].

\bibitem{Dumitru:2007hy} 
  A.~Dumitru, Y.~Guo and M.~Strickland,
  Phys.\ Lett.\ B {\bf 662}, 37 (2008)
  [arXiv:0711.4722 [hep-ph]].
  
\bibitem{Dumitru:2009fy} 
  A.~Dumitru, Y.~Guo and M.~Strickland,
  Phys.\ Rev.\ D {\bf 79}, 114003 (2009)
  [arXiv:0903.4703 [hep-ph]].

\bibitem{Du:2016wdx} 
  Q.~Du, A.~Dumitru, Y.~Guo and M.~Strickland,
  JHEP {\bf 1701}, 123 (2017)
  [arXiv:1611.08379 [hep-ph]].

\bibitem{Liu:2006nn} 
  H.~Liu, K.~Rajagopal and U.~A.~Wiedemann,
  Phys.\ Rev.\ Lett.\  {\bf 98}, 182301 (2007)
  [hep-ph/0607062].

\bibitem{Escobedo:2011ie} 
  M.~A.~Escobedo, J.~Soto and M.~Mannarelli,
  Phys.\ Rev.\ D {\bf 84}, 016008 (2011)
  [arXiv:1105.1249 [hep-ph]].
  
\bibitem{Yao:2018nmy} 
  X.~Yao and T.~Mehen,
  Phys.\ Rev.\ D {\bf 99}, no. 9, 096028 (2019)
  [arXiv:1811.07027 [hep-ph]].

\bibitem{Yao:2020kqy}
X.~Yao, W.~Ke, Y.~Xu, S.~A.~Bass, T.~Mehen and B.~M\"uller,
[arXiv:2002.04079 [hep-ph]].

\bibitem{Young:2010jq} 
  C.~Young and K.~Dusling,
  Phys.\ Rev.\ C {\bf 87}, 065206 (2013)
  [arXiv:1001.0935 [nucl-th]].

\bibitem{Borghini:2011ms} 
  N.~Borghini and C.~Gombeaud,
  Eur.\ Phys.\ J.\ C {\bf 72}, 2000 (2012)
  [arXiv:1109.4271 [nucl-th]].

\bibitem{Akamatsu:2011se} 
  Y.~Akamatsu and A.~Rothkopf,
  Phys.\ Rev.\ D {\bf 85}, 105011 (2012)
  [arXiv:1110.1203 [hep-ph]]

\bibitem{Akamatsu:2014qsa} 
  Y.~Akamatsu,
  Phys.\ Rev.\ D {\bf 91}, 056002 (2015)
  [arXiv:1403.5783 [hep-ph]].
 
\bibitem{Blaizot:2015hya} 
  J.~P.~Blaizot, D.~De Boni, P.~Faccioli and G.~Garberoglio,
  Nucl.\ Phys.\ A {\bf 946}, 49 (2016)
  [arXiv:1503.03857 [nucl-th]].

\bibitem{Katz:2015qja} 
  R.~Katz and P.~B.~Gossiaux,
  Annals Phys.\  {\bf 368}, 267 (2016)
  [arXiv:1504.08087 [quant-ph]].
  
\bibitem{Kajimoto:2017rel} 
  S.~Kajimoto, Y.~Akamatsu, M.~Asakawa and A.~Rothkopf,
  Phys.\ Rev.\ D {\bf 97}, no. 1, 014003 (2018)
  [arXiv:1705.03365 [nucl-th]].

\bibitem{DeBoni:2017ocl} 
  D.~De Boni,
  JHEP {\bf 1708}, 064 (2017)
  [arXiv:1705.03567 [hep-ph]].

\bibitem{Blaizot:2017ypk} 
  J.~P.~Blaizot and M.~A.~Escobedo,
  JHEP {\bf 1806}, 034 (2018)
  [arXiv:1711.10812 [hep-ph]].
  
\bibitem{Blaizot:2018oev} 
  J.~P.~Blaizot and M.~A.~Escobedo,
  Phys.\ Rev.\ D {\bf 98}, no. 7, 074007 (2018)
  [arXiv:1803.07996 [hep-ph]].
 

\bibitem{Akamatsu:2018xim} 
  Y.~Akamatsu, M.~Asakawa, S.~Kajimoto and A.~Rothkopf,
  JHEP {\bf 1807}, 029 (2018)
  [arXiv:1805.00167 [nucl-th]].

\bibitem{Miura:2019ssi} 
  T.~Miura, Y.~Akamatsu, M.~Asakawa and A.~Rothkopf,
  Phys.\ Rev.\ D {\bf 101}, no. 3, 034011 (2020)
  [arXiv:1908.06293 [nucl-th]].

\bibitem{Sharma:2019xum}
R.~Sharma and A.~Tiwari,
Phys. Rev. D \textbf{101}, no.7, 074004 (2020)
[arXiv:1912.07036 [hep-ph]].  
\bibitem{Brambilla:2016wgg} 
  N.~Brambilla, M.~A.~Escobedo, J.~Soto and A.~Vairo,
  Phys.\ Rev.\ D {\bf 96}, no. 3, 034021 (2017)
  [arXiv:1612.07248 [hep-ph]].

\bibitem{Brambilla:2017zei} 
  N.~Brambilla, M.~A.~Escobedo, J.~Soto and A.~Vairo,
  Phys.\ Rev.\ D {\bf 97}, no. 7, 074009 (2018)
  [arXiv:1711.04515 [hep-ph]].
  
\bibitem{Brambilla:2019tpt} 
  N.~Brambilla, M.~A.~Escobedo, A.~Vairo and P.~Vander Griend,
  Phys.\ Rev.\ D {\bf 100}, no. 5, 054025 (2019)
  [arXiv:1903.08063 [hep-ph]].

\bibitem{Gossiaux:2008jv} 
  P.~B.~Gossiaux and J.~Aichelin,
  Phys.\ Rev.\ C {\bf 78}, 014904 (2008)
  [arXiv:0802.2525 [hep-ph]].

\bibitem{Gossiaux:2009mk} 
  P.~B.~Gossiaux, R.~Bierkandt and J.~Aichelin,
  Phys.\ Rev.\ C {\bf 79}, 044906 (2009)
  [arXiv:0901.0946 [hep-ph]].
  
\bibitem{Uphoff:2014hza} 
  J.~Uphoff, O.~Fochler, Z.~Xu and C.~Greiner,
  J.\ Phys.\ G {\bf 42}, no. 11, 115106 (2015)
  [arXiv:1408.2964 [hep-ph]].

\bibitem{Cao:2016gvr} 
  S.~Cao, T.~Luo, G.~Y.~Qin and X.~N.~Wang,
  Phys.\ Rev.\ C {\bf 94}, no. 1, 014909 (2016)
  [arXiv:1605.06447 [nucl-th]].

\bibitem{Ke:2018tsh} 
  W.~Ke, Y.~Xu and S.~A.~Bass,
  Phys.\ Rev.\ C {\bf 98}, no. 6, 064901 (2018)
  [arXiv:1806.08848 [nucl-th]].

\bibitem{Xu:2017obm} 
  Y.~Xu, J.~E.~Bernhard, S.~A.~Bass, M.~Nahrgang and S.~Cao,
  Phys.\ Rev.\ C {\bf 97}, no. 1, 014907 (2018)
  [arXiv:1710.00807 [nucl-th]].

\bibitem{Yao:2018sgn} 
  X.~Yao and B.~M\"uller,
  Phys.\ Rev.\ D {\bf 100}, no. 1, 014008 (2019)
  [arXiv:1811.09644 [hep-ph]].
  
\bibitem{Bodwin:1994jh} 
  G.~T.~Bodwin, E.~Braaten and G.~P.~Lepage,
  Phys.\ Rev.\ D {\bf 51}, 1125 (1995)
  Erratum: [Phys.\ Rev.\ D {\bf 55}, 5853 (1997)]
  [hep-ph/9407339].
  
\bibitem{Yao:2017fuc} 
  X.~Yao and B.~M\"uller,
  Phys.\ Rev.\ C {\bf 97}, no. 1, 014908 (2018)
  Erratum: [Phys.\ Rev.\ C {\bf 97}, no. 4, 049903 (2018)]
  [arXiv:1709.03529 [hep-ph]].
  
\bibitem{Sjostrand:2014zea} 
  T.~Sj\"ostrand {\it et al.},
  Comput.\ Phys.\ Commun.\  {\bf 191}, 159 (2015).

\bibitem{Eskola:2016oht} 
  K.~J.~Eskola, P.~Paakkinen, H.~Paukkunen and C.~A.~Salgado,
  Eur.\ Phys.\ J.\ C {\bf 77}, no. 3, 163 (2017)
  [arXiv:1612.05741 [hep-ph]].

\bibitem{Buckley:2014ana}
A.~Buckley, J.~Ferrando, S.~Lloyd, K.~Nordström, B.~Page, M.~Rüfenacht, M.~Schönherr and G.~Watt,
Eur. Phys. J. C \textbf{75}, 132 (2015)
[arXiv:1412.7420 [hep-ph]].

\bibitem{Wang:2019vau} 
  P.~Wang [STAR Collaboration],
  Nucl.\ Phys.\ A {\bf 982}, 723 (2019).

\bibitem{Moreland:2014oya}
J.~S.~Moreland, J.~E.~Bernhard and S.~A.~Bass,
Phys. Rev. C \textbf{92}, no.1, 011901 (2015)
[arXiv:1412.4708 [nucl-th]].

\bibitem{Song:2007ux} 
  H.~Song and U.~W.~Heinz,
  Phys.\ Rev.\ C {\bf 77}, 064901 (2008)
  [arXiv:0712.3715 [nucl-th]].
        
\bibitem{Shen:2014vra} 
  C.~Shen, Z.~Qiu, H.~Song, J.~Bernhard, S.~Bass and U.~Heinz,
  Comput.\ Phys.\ Commun.\  {\bf 199}, 61 (2016)
  [arXiv:1409.8164 [nucl-th]].
  
\bibitem{Bernhard:2016tnd} 
  J.~E.~Bernhard, J.~S.~Moreland, S.~A.~Bass, J.~Liu and U.~Heinz,
  Phys.\ Rev.\ C {\bf 94}, no. 2, 024907 (2016)
  [arXiv:1605.03954 [nucl-th]].

\bibitem{Sirunyan:2018nsz} 
  A.~M.~Sirunyan {\it et al.} [CMS Collaboration],
  Phys.\ Lett.\ B {\bf 790}, 270 (2019).
  [arXiv:1805.09215 [hep-ex]].

\bibitem{Khachatryan:2016xxp} 
  V.~Khachatryan {\it et al.} [CMS Collaboration],
  Phys.\ Lett.\ B {\bf 770}, 357 (2017)
  [arXiv:1611.01510 [nucl-ex]].

\bibitem{Acharya:2019hlv} 
  S.~Acharya {\it et al.} [ALICE Collaboration],
  Phys.\ Rev.\ Lett.\  {\bf 123}, no. 19, 192301 (2019).
  
\bibitem{Sirunyan:2020qec}
A.~M.~Sirunyan \textit{et al.} [CMS],
[arXiv:2006.07707 [hep-ex]].

\bibitem{MW}
  M.~A.~Winn, private communication.

\bibitem{Bernhard:2019bmu} 
  J.~E.~Bernhard, J.~S.~Moreland and S.~A.~Bass,
  Nature Phys.\  {\bf 15}, no. 11, 1113 (2019).

\bibitem{Yao:2019rcb} 
  X.~Yao,
  ProQuest Number: 13882375 (2019)
  [arXiv:1911.08500 [nucl-th]].

\bibitem{Mrowczynski:2017kso} 
  S.~Mrowczynski,
  Eur.\ Phys.\ J.\ A {\bf 54}, no. 3, 43 (2018)
  [arXiv:1706.03127 [nucl-th]].
  
\bibitem{Yao:2018zze} 
  X.~Yao and B.~M\"uller,
  Phys.\ Rev.\ D {\bf 97}, no. 7, 074003 (2018)
  [arXiv:1801.02652 [hep-ph]].

\bibitem{Tanabashi:2018oca} 
  M.~Tanabashi {\it et al.} [Particle Data Group],
  Phys.\ Rev.\ D {\bf 98}, no. 3, 030001 (2018).

\bibitem{Khachatryan:2014ofa}
V.~Khachatryan \textit{et al.} [CMS],
Phys. Lett. B \textbf{743}, 383-402 (2015)
[arXiv:1409.5761 [hep-ex]].
\end{thebibliography}

\end{document}